\documentclass[11pt]{article} 
\usepackage{iclr2025_conference,times}
\usepackage{titlesec}
\newcommand{\hl}{}
\newcommand{\ul}{}
\usepackage{CJKutf8}
\usepackage{booktabs}
\usepackage{enumitem}

\usepackage{url}
\usepackage{lipsum}

\usepackage[authordate,legalnotes=false,dashed=false,backend=biber,noibid, hyperall=true, maxcitenames=2]{biblatex-chicago}

\renewbibmacro*{cms:shorthandintro}{%
  \iffieldundef{shorthand}
    {}%
    {\printtext[brackets]{\printfield{shorthand}}}%
}

\usepackage[]{hyperref}
\hypersetup{
colorlinks=true,citecolor=apolloblue,linkcolor=apolloblue,urlcolor=apolloblue
}

\bibliography{legalreferences,references}

\usepackage{tabularx}

\renewcommand\thesubsubsection{\thesection.\arabic{subsection}.(\alph{subsubsection})}

\renewcommand\theparagraph{\thesection.\arabic{subsection}.(\alph{subsubsection}.\roman{paragraph})}

\usepackage{citeall}

\titleformat{\paragraph}[block] 
  {\normalfont\normalsize\bfseries}
  {\theparagraph}  
  {1em}  
  {} 
\titlespacing*{\paragraph}{0pt}{1ex}{0.5ex}

\renewcommand{\labelenumi}{(\arabic{enumi})}

\titleformat{\subsubsection}[block]
  {\normalfont\normalsize\bfseries}  
  {\thesubsubsection}                
  {1em}                              
  {}     

\title{AI Behind Closed Doors: a Primer on 
The Governance of Internal Deployment}

\author{
Charlotte Stix\thanks{Correspondence to charlotte@apolloresearch.ai.} \and\hspace{-2em} Matteo Pistillo 
\AND Girish Sastry\thanks{Independent.} \and\hspace{-2em} Marius Hobbhahn \and\hspace{-2em} Alejandro Ortega \and\hspace{-2em} Mikita Balesni 
\AND Annika Hallensleben \and\hspace{-2em} Nix Goldowsky-Dill \and\hspace{-2em} Lee Sharkey
\AND 
\textmd{Apollo Research}
}

\iclrfinalcopy 
\begin{document}

\maketitle

\section*{Executive Summary}

The world's most advanced future artificial intelligence (AI) systems will first be deployed inside the frontier AI companies developing them. According to these companies and independent experts, future AI systems may become so capable that they could reach or even surpass human intelligence and capabilities by 2030. Internal deployment---the deployment of AI systems within the companies developing them---is, therefore, a key source of benefits and risks from frontier AI systems. Despite this, \textbf{the governance of the internal deployment of highly advanced frontier AI systems appears absent}.

This report aims to address this absence by priming a conversation around the governance of internal deployment. It offers a range of lenses on the topic and is geared at decision-makers in AI companies, governments, and society. The report presents a conceptualization of internal deployment; learnings from other safety-critical sectors; reviews of existing legal frameworks and their applicability; and illustrative examples of the type of scenarios we are most concerned about. It culminates with a small number of targeted recommendations that provide a \textbf{first prototype for a framework for the governance of internal deployment}. Specifically, the report builds up to this prototype through the following Chapters.

\begin{itemize}
    \item[] \textbf{Chapter 1}. We present the scope of the report, i.e., highly advanced frontier AI systems, and highlight the urgency of discussing and implementing a robust governance of internal deployment. As part of this, we articulate the main economic and strategic considerations underpinning internal deployment, such as the opportunity to automate the scarcest and highest-value labor within AI companies (AI researchers themselves), which creates a `winner takes all' dynamic in which actors who are furthest ahead are most able to get even further ahead of their competitors. We also explain why, in this competition, AI companies may increasingly choose to taper down external deployment for competitive gain. We argue that in light of the complementary incentives to `get to artificial \emph{general} intelligence first' and taper down external deployment, the window of policy opportunity to intervene on internal deployment may be closing soon.

    \item[] \textbf{Chapter 2}. We present a high-level overview of some of the defining characteristics of internal deployment. In doing so, we reflect on the following areas: (i) the nature of the AI system being deployed, including the type of safeguards, affordances and permissions it may or may not have in comparison to its externally deployed counterparts; (ii) `privileged access' to the internally deployed AI system by individuals within an AI company or AI systems themselves; and (iii) `privileged application' of the internally deployed AI system to high-value use cases, such as automating the AI research and development (R\&D) pipeline.

    \item[] \textbf{Chapter 3}. We detail two core threat scenarios directly stemming from the internal deployment of the most advanced frontier AI systems, absent any governance measures. First, we concentrate on pathways leading to loss of control. One pathway we examine is the possibility of unknowingly applying `scheming' AI systems to automate the AI R\&D pipeline (or even as a replacement to AI safety researchers). Scheming AI systems are systems that covertly and strategically pursue misaligned goals. These AI systems could engage in scheming behavior without being detected, controlled, or overseen while accumulating access to and leveraging resources through their application to the AI R\&D pipeline. Second, we examine some of the potential impacts of the internal application of highly advanced frontier AI systems, in the absence of governance frameworks, on societal resilience and preparedness. We suggest that an internal `intelligence explosion' could contribute to unconstrained and undetected power accumulation, which in turn could lead to gradual or abrupt disruption of democratic institutions and the democratic order.

\end{itemize}

In Chapters 4 and 5, we conduct two large-scale research reviews and, later, translate our learnings towards our final recommendations in Chapter 6.

\begin{itemize}
    \item[] \textbf{Chapter 4}. We present an in-depth review of legal language that appears in passed and proposed legal frameworks on AI in the United States and the European Union. We examined over 20 legal texts, assessing how the language therein could, under specific interpretations, already address and cover the internal deployment of highly advanced AI systems. We found that, in most cases, the term `deploy' could be interpreted more broadly than `public release' or `commercial release.' In some cases, their definition even explicitly encompasses the `internal use' of an AI system.

    \item[] \textbf{Chapter 5}. We examine relevant literature on other safety-critical industries such as biological agents and toxins, novel chemical substances, novel pesticides, R\&D for nuclear reactors, experimental aircraft, novel drugs, and novel medical devices. We subsequently focus on those most similar to AI in terms of velocity, scale, and irreversibility of propagation and potential harm. In particular, we give most attention to biological agents and toxins, novel chemical substances, and novel pesticides. We flag recurring themes in these safety-critical industries, including: (i) risk evaluations, mitigations, and relevant restrictions and oversight procedures; (ii) targeted transparency on safety-critical information; and (iii) boundaries to internal access and usage.

\end{itemize}

Finally, we leverage our learnings from the previous Chapters and present a first prototype of a governance framework for internal deployment of highly-advanced frontier AI systems. In particular, we aim to target the scenarios we outline in Chapter 3 as a priority: (i) loss of control; and (ii) undetected economic and power accumulation negatively affecting societal and governmental resilience and preparedness.

\begin{itemize}
    \item[]\textbf{Chapter 6}. We recommend the implementation of a multi-layered `Swiss cheese-style' defense-in-depth approach where individual, complementary measures are stacked to create comprehensive protection, with each layer addressing specific threat vectors related to internal deployment. The report's recommendations are addressed to AI companies, though this is reflective of the nimble scope of the report and not a commentary on optimal future intervention points. In fact, we believe that following through with our recommendations could provide opportunities for mutually beneficial public-private partnerships where, for example, transparency about internally deployed AI systems could be exchanged for access to energy resources and enhanced security from the government.

\end{itemize}

We advance the following overarching recommendations.

\begin{enumerate}

\item
  \textbf{Scheming detection and control}: Expand Frontier Safety Policies (FSPs) to explicitly cover internal AI deployment, with specific evaluations for scheming detection and control before internal use and appropriate mitigations.
\item
  \textbf{Internal usage policies}: Implement structured frameworks that determine who can access and use internal highly advanced AI systems and under what conditions, with tiered access levels based on necessity and, potentially, security clearance.
\item
  \textbf{Oversight frameworks}: Establish clear guidelines and processes to operationalize and oversee internal deployment. As part of this, establish two dedicated oversight bodies---an Internal Deployment Overseeing Board and Internal Deployment Team---to supervise internal deployment decisions and enforce FSPs and internal usage policies through technical monitoring and human review processes.
\item
  \textbf{Targeted transparency}: Share critical information (internal system capabilities, evaluations, and safety measures) with select stakeholders, including cleared internal staff and relevant government agencies, through pre-internal deployment system cards and detailed safety documentation.
\item
  \textbf{Disaster resilience plans}: Develop collaborative emergency procedures with governments to address threats that might bypass established governance mechanisms.
\end{enumerate}

\section*{Introduction}\label{introduction}

The urgent task of properly governing highly advanced frontier artificial intelligence (AI) systems\footnote{In this report, we follow \textcite[p. 3, 4]{Sharkey2024d} definition of `AI systems.' Namely: ``{[}...{]} AI systems, are a slight generalization of AI models. AI systems include not only the weights and architecture of the model, but also include a broader set of system parameters. These consist of retrieval databases and particular kinds of prompts. The reason for our expanded focus is that system parameters strongly influence the capability profile of AI systems.''} now confronts governments, AI developers, and civil society organizations alike. The past two years alone have seen the introduction of multiple regulatory and governance interventions,\footnote{For example, the European Union's (EU) AI Act \parencite{EuropeanParliamentAndCouncil2024o}, President Trump's Executive Order 14179 \parencite{ExecutiveOfficeOfThePresident2025g}, President Biden's Executive Order 14141 \parencite{USExecutiveOfficeOfThePresident2025d}, NIST AI RMF 1.0 \parencite{NIST2023r}, copious proposed legislation in the United States (US) at the federal and state levels, as well as China's Interim Measures for the Management of Generative Artificial Intelligence Services \parencite{ChinaLaw}.} the establishment of new institutions focused on AI assessments,\footnote{At the time of writing, this includes: the United Kingdom's (UK) AI Security Institute \parencite{UKDepartmentForScienceInnovationandTechnology2025r}, the U.S. AI Safety Institute \parencite{USNationalInstituteOfStandardsAndTechnology2023z}, the Singapore AI Safety Institute \parencite{SingaporeAISafetyInstitute2025d}, the Korean AI Safety Institute \parencite{KoreaArtificialIntelligenceSafetyInstitute2024k}, the Canadian AI Safety Institute \parencite{GovernmentOfCanada2024w}, the Japan AI Safety Institute \parencite{JapanAISafetyInstitute2024y}, France's INESIA \parencite{LaboratoireNationalDeMetrologieEtDEssais2025c}, the EU's AI Office \parencite{EuropeanCommission2025s}, and representatives from the Kenyan \parencite{Mabonga2024f} and Australian governments \parencite{Other2024g}.} and the creation of a bi-yearly calendar of international conferences on the topic.\footnote{At the time of writing, this includes the UK's AI Safety Summit 2023, Korea's AI Seoul Summit 2024, France's AI Action Summit 2025, and the upcoming Summit in India \parencite{Agrawal2025n}.} During the same period, select frontier AI companies have voluntarily established, signed on to, and implemented their own self-governance mechanisms. Among these are the establishment of Frontier Safety Policies (FSPs) \parencite{METR2025t,METR2025d}, and AI companies' participation in the European Union's General-Purpose AI Code of Practice drafting process \parencite{EuropeanCommission2025p}, the Seoul Frontier AI Safety Commitments \parencite{UKDepartmentForScienceInnovationTechnology2025b}, and the White House Voluntary AI Commitments \parencite{TheWhiteHouse2023y}.

The urgency and range of these efforts responds to the recent rapid pace of AI progress (\cite{Roser2022q}; \cite[p. 81, 87, 119]{Maslej2024p}; \cite{Kwa2025i}), opening up novel usage and introducing broader risk profiles. Indeed, leaders of frontier AI companies expect transformative increases in AI capabilities to happen soon \parencite{Butcher2025i,Browne2025v}. OpenAI CEO Sam Altman writes that ``in 2025, we may see the first AI agents `join the workforce' and materially change the output of companies'' \parencite{Altman2025r}. Anthropic CEO Dario Amodei states that ``{[}m{]}aking AI that is smarter than almost all humans at almost all things \ldots{} is most likely to happen in 2026-2027'' \parencite{Amodei2025l} and that, as early as 2026,\footnote{Other researchers expect to see this ``country of geniuses in a datacenter'' \parencite{Amodei2024n} by June 2027 \parencite{Kokotajlo2025v}.} we could have a ``country of geniuses in a datacenter'' \parencite{Amodei2024n}. Google DeepMind recently estimated that AI that is ``at least as capable as humans at most cognitive tasks'' could be here ``within the coming years'' \parencite{Dragan2025c} and plausibly by 2030 \parencite{Shah2025v}. xAI CEO Elon Musk states that ``in 10 years \ldots{} AI will {[}probably{]} be able to do anything better than a human can, cognitively'' \parencite{Verdict2025e}. Similarly, a recent report co-authored by 96 world-leading AI experts contains a Chair's note\footnote{The Chair's note was added after completion of the report to adjust for the most recent developments.} by Turing Award Winner Prof. Yoshua Bengio stating that recent ``o3 results are evidence that the pace of advances in AI capabilities may remain high or even accelerate'' \parencite[p. 12]{Bengio2025z}.

The world's most advanced AI systems will first exist internal to the AI company developing them.\footnote{Throughout our report, when we refer to ``developers'' we are talking about frontier AI companies. There is an eventuality that companies may merge with future government projects, or be soft-nationalized. While we do not address this possibility or its consequences in this report, we note that many concerns around internal deployment, including those outlined in \hyperref[focus-on-two-high-impact-scenarios-loss-of-control-and-undetected-power-accumulation]{\ul{§3.2}}, would remain the same and that the governance of internal deployment would remain critical.} In other words, \textbf{the} \textbf{deployment and application of the most advanced AI systems could occur first and foremost behind closed doors}. This is where these systems will first ``join the workforce'' and ``materially change the output of companies,'' according to OpenAI's CEO Sam Altman \parencite{Altman2025r}. Early signs of this trend can already be observed, especially with an eye to the lucrative rise in AI system's capabilities to automate AI software and hardware development tasks (\cite{Kwa2025i}; \cite[p. 46]{Bengio2025z}).\footnote{The International AI Safety report states that AI model usage to automate software and hardware development tasks includes their usage to ``efficiently write software to train and deploy AI, to aid in designing AI chips, and to generate and curate training data'' \parencite[p. 46]{Bengio2025z}.} Google, for example, recently used AI systems to generate more than a quarter of all its new code \parencite{Peters2024o}. As another example, Anthropic's CEO stated in March of 2025, ``I think we will be there in three to six months, where AI is writing 90\% of the code. And then, in 12 months, we may be in a world where AI is writing essentially all of the code'' \parencite{Tan2025f}.\footnote{In addition to writing code, AI is being internally deployed at AI companies for knowledge work, including at the highest levels. For example, Microsoft's CEO Satya Nadella told a podcast interviewer that ``the new workflow for me is I think with AI and work with my colleagues'' \parencite{Patel2025y}.} This early application underscores the claim of some that the ``engineering velocity'' enabled by AI is quickly becoming ``incredible'' \parencite{Martin2024a}. As a byproduct of this trend, the internal deployment of highly advanced AI systems will likely yield a number of economic advantages, such as a ``winner takes all'' dynamic enabled by automated AI research and development (\emph{see} \hyperref[chapter-3.-internal-deployment-scenarios-of-concern]{\ul{Chapter 3}}) \parencite[p. 177]{Bengio2025z}, and give rise to the pursuit of strategic incentives, including competitive non-transparency (\emph{see} \hyperref[chapter-1.-scope-and-urgency]{\ul{Chapter 1}}). It is, therefore, likely that a growing number of AI systems, and especially the most advanced and capable frontier AI systems, will be applied exclusively internally---possibly for the entirety of their service period.

Depending on the maturity of a given AI company's respective governance of internal deployment (\emph{see} \hyperref[chapter-6.-defense-in-depth-recommendations-for-the-governance-of-internal-deployment]{\ul{Chapter 6}}), internal deployment of highly advanced AI systems could carry significant risks \parencite{METR2025h}, potentially superseding those posed by previously externally deployed AI systems (\emph{see} \hyperref[chapter-3.-internal-deployment-scenarios-of-concern]{\ul{Chapter 3}}).\footnote{In light of their potential exclusive use behind closed doors, internally deployed AI systems are the least transparent to and understood by the AI research community. In theory, though there is no publicly available evidence in either direction, it is plausible that AI companies could apply these AI systems with minimal scrutiny, incomplete guardrails and largely outside of in-depth safety assessments (\emph{see} \hyperref[chapter-2.-characterizing-internal-deployment]{\ul{Chapter 2}}).} This potential outcome is particularly salient given that one stated goal of frontier AI companies is to develop artificial general intelligence (AGI)---``AI systems that are generally smarter than humans'' \parencite{OpenAI2023v}---and, eventually `superintelligence' \parencite{Altman2024q,Sutskever2024q}. The discussion of this potential outcome is also particularly timely, given that CEOs of frontier AI companies forecast the development of AGI to be achieved within the next four to five years; for example, 2026-2027 for Anthropic \parencite{Amodei2025l}, before 2029 for OpenAI \parencite{Tyrangiel2025s}, 2029 for NVIDIA \parencite{Uffindell2024i}, 2030 for Google DeepMind \parencite{Shah2025v}.\footnote{Independent experts such as Turing Award Winner Prof. Yoshua Bengio \parencite{Bengio2023y} and Nobel Prize and Turing Award Winner Prof. Geoffrey Hinton \parencite{HintonGGeoffreyhinton2023y} place AGI between 2028--2043.}

In short, time is of the essence. It is, therefore, imperative for AI companies and policymakers to seriously assess the implications of the internal deployment of highly capable AI systems and to reflect on the steps needed towards achieving its meaningful governance.\footnote{In \hyperref[chapter-4.-existing-and-proposed-ai-governance-frameworks-and-their-relationship-to-internal-deployment]{\ul{Chapter 4}}, we illustrate how some governance frameworks could in theory be interpreted to capture internal deployment.} While policymakers already focus on the governance of external deployment, delay in expanding their focus to internal deployment enables a serious blind spot.\footnote{Other AI experts have also pointed towards internal deployment as a potential legal loophole in existing governance frameworks that have the stated goal of mitigating systemic risk from frontier AI \parencite{Brundage2025w}.} Not least because the technological capabilities of ``a country of geniuses in a datacenter'' \parencite{Amodei2024n} may mean that society will soon have to grapple with the spill-over effects of novel capabilities and applications---ranging from security vulnerabilities and national security threats,\footnote{In using this term, we do not preferentially aim to highlight a specific country's national security.} large-scale effects on the functioning of democratic states, to undetected and unconstrained accumulation of power---all potentially stemming from AI development and application behind closed doors.

This report aims to start an informed discussion about internal deployment and the shape that the governance of internal deployment could take. In doing so, we acknowledge the significant information gap regarding the structure of existing internal AI deployment throughout the report---a gap that itself presents governance challenges \parencite{Marchant2011f}. The report presents a comprehensive mapping of internal deployment. It conceptualizes internal deployment (\emph{see} \hyperref[chapter-2.-characterizing-internal-deployment]{\ul{Chapter 2}}), presents associated threats and vulnerabilities (\emph{see} \hyperref[chapter-3.-internal-deployment-scenarios-of-concern]{\ul{Chapter 3}}), and concludes by proposing a spectrum of functional governance approaches---from targeted measures to interconnected governance frameworks (\emph{see} \hyperref[chapter-6.-defense-in-depth-recommendations-for-the-governance-of-internal-deployment]{\ul{Chapter 6}})---informed by lessons from other safety-critical sectors (\emph{see} \hyperref[chapter-5.-internal-deployment-in-other-safety-critical-industries]{\ul{Chapter 5}}) and existing legal frameworks (\emph{see} \hyperref[chapter-4.-existing-and-proposed-ai-governance-frameworks-and-their-relationship-to-internal-deployment]{\ul{Chapter 4}}).

\section{Scope and Urgency}\label{chapter-1.-scope-and-urgency}

This Chapter brings internal deployment into focus for decision- and policymakers and limits the scope of the report. It delineates the overarching problem statement and provides initial context toward the more detailed analysis of internal deployment in \hyperref[chapter-2.-characterizing-internal-deployment]{\ul{Chapter 2}} and the presentation of select threat scenarios in \hyperref[chapter-3.-internal-deployment-scenarios-of-concern]{\ul{Chapter 3}}. The Chapter is divided into two Sections.

\textbf{Section 1.1} provides an initial definition of internal deployment and confines the scope of the report to highly advanced frontier AI systems with broad capabilities.

\textbf{Section 1.2} reflects on the report's urgency. Increasingly strong economic and strategic incentives may rapidly lead to a rise in the number of AI systems deployed exclusively internally. These incentives, coupled with less-than-five-year AGI forecasts by AI companies' CEOs and independent experts alike, could mean that the window for policy intervention will be closing soon.
\subsection{Scope}\label{scope}

We define \textbf{`internal deployment' as the act of making an AI system available for access and/or usage exclusively for the developing organization}. Our definition draws inspiration from the International AI Safety Report \parencite[p. 35]{Bengio2025z}, which states that:

\begin{quote}
``\hl{Deployment can take several forms: internal deployment for use by the system's developer, or external deployment either publicly or to private customers. Very little is publicly known abou}t internal deployments. However, companies are known to adopt different types of strategies for external deployment.''
\end{quote}

In principle, any sort of AI system, including small, specialized ones, can be applied internally (including commercial off-the-shelf AI solutions). However, due to their limited capabilities and narrow applicability, when deployed internally, their usage is, in most cases, less likely to pose severe risks, with a small number of exceptions such as, for example, control monitor models \parencite{Shah2025v}. Conversely, the next generation of internally deployed highly advanced frontier AI systems will likely be more capable than any AI system on the market or currently in production (\hyperref[urgency]{\ul{§1.2}}). Being behind closed doors, it will be impossible to confidently extrapolate their applications and risk profiles, especially from an external perspective and based on publicly available evidence.

This report concerns itself \textbf{exclusively with highly advanced frontier AI systems with broad capabilities} (`highly advanced AI systems')---AI systems which present the highest, broadest, most novel, and therefore unknown, risk profiles. We believe that this targeted focus encourages tailored interventions justified by salient national security and societal concerns, without unduly hindering the current innovation and technological progress.

The report also adopts a \textbf{narrow approach to the type of threat scenarios} that the internal deployment of highly advanced AI systems with broad capabilities could pose. Specifically, we focus only on two scenarios which we believe are particularly likely to present negative externalities for society and national security in the absence of robust internal deployment governance: loss of control via the internal application of a scheming AI (including to the AI R\&D pipeline), and democratic disruption via unconstrained and undetected power accumulation (\emph{see} \hyperref[chapter-3.-internal-deployment-scenarios-of-concern]{\ul{Chapter 3}}). Many other important threat scenarios stemming from, for example, misalignment\footnote{For instance, despite their importance, this report does not in any depth discuss misalignment threat scenarios in which an AI system applied internally: (i) sabotages an AI company's further safety critical work \parencite{Benton2024s}; (ii) provides misleading strategic advice to an AI company's leadership \parencite{Benton2024s}; or (iii) coaxes AI research into self-exfiltration or external distillation \parencite{Meinke2024x}.} or misuse\footnote{Because of its narrow scope, this report also does not address many important misuse scenarios with national security implications \parencite{Acharya25}, including security-type scenarios arising from insider and outsider threats. Examples of insider threat include a `rogue employee' with privileged access: (i) directly misusing the internal AI system; (ii) unduly abusing next-generation AI systems for their own ends; (iii) tampering next generation training runs (for instance, by subtly modifying training data or evaluation metrics to introduce backdoors that persist across model generations, effectively creating a lineage of compromised systems that maintain loyalty to the employee). Examples of outsider threat include other great states powers (foreign adversaries), non-state malicious actors (for example, a terrorist group), or even competitors: (i) gaining unauthorized access to an AI company's infrastructure and exfiltrating its most advanced AI model, including its architecture details, training methods, and weights, as part of industrial espionage \parencite{Delaney2025h,Nevo2024c}; (ii) gaining unauthorized access to an AI company's infrastructure and running hidden jobs \parencite{Seale2024l}; or (iii) using external APIs to pursue harmful tasks (for example, exploiting cybersecurity vulnerabilities or using them to create weapons of mass destruction).} are out of the report's narrow scope.

\subsection{Urgency}\label{urgency}

The policy window to implement governance interventions addressing internal deployment may be narrow. Below, we briefly articulate strategic incentives and economic advantages that may underpin the rationale for a given frontier AI company to increasingly, and even exclusively, deploy highly advanced AI systems internally in the near future. We end by providing the limited available public evidence that this trend towards internal deployment has already started taking root.

We briefly reflect on the following strategic incentives and economic advantages that would likely favor the exclusive internal deployment of highly advanced AI systems: the opportunity to (i) leverage the automation of scarce and costly labor; and the ongoing (ii) fierce competition in the AI sector, underscored by a `winner takes all' dynamic \parencite[p. 177]{Bengio2025z}, in which a lack of publicly available information may present a strategic advantage.

First, at a given future point in AI development, there will likely be outsized economic benefits to the exclusive internal deployment of highly advanced AI systems. One of those outsized benefits may be \textbf{the automation of the scarcest and highest-value labor to a frontier AI company}: \textbf{AI researchers} \textbf{themselves}. John Schulman---OpenAI cofounder and previous lead of OpenAI's post-training team---has publicly stated that his median estimate for when his job will be fully automated by AI systems is 2029 \parencite{Patel2024k}. In line with this prediction, Google DeepMind anticipates the ``delegation of increasingly complex research and engineering tasks to AI assistance,'' leading to a point in which, ``in order to keep pace with advancing AI capabilities,'' ``the vast majority of the cognitive labor relevant to AI safety {[}is{]} performed by AI'' \parencite{Shah2025v}.

Avid observers are already seeing early signs of AI systems applied to labor traditionally reserved for a range of technical talent, for example, current frontier AI systems are starting to be capable enough to speed up the creation of the next generation of AI systems by helping with the curation of datasets and the writing of software for training or deployment \parencite[p. 46]{Bengio2025z}. Taking note of this trend, Meta CEO Mark Zuckerberg expects ``that a lot of the code and AI in apps will be built by AI [probably] in 2025'' \parencite{Rogan2025g}, while Salesforce CEO Marc Benioff proclaims that Salesforce will no longer hire software engineers in 2025, due to a ``productivity boost from AI'' \parencite{Martin2024a}. Indeed, as of last year (2024), some technology companies have already applied AI systems to generate more than a quarter of their new code \parencite{Peters2024o}. As novel AI systems approach highly advanced capabilities, they stand to become transformative factors of production that could dramatically increase company productivity across operations. In short, an AI company that could effectively deploy AI systems to replace and multiply its existing capacity---for example, expanding from hundreds of AI researchers to the equivalent of tens of millions---could dramatically accelerate its technological progress \parencite{Davidson2023n}. The current business model---revenue generated by the external sale of licenses to advanced AI systems---may quickly pale in comparison to the economic pinnacle that could be achieved through prudent and strategic internal application. Especially if the goal is AGI.

Second, \textbf{in a `winner takes all' market}, \textbf{AI companies might not have incentives to externally deploy highly capable AI systems until they have a} `\textbf{champion.}'\footnote{Relatedly, the International AI Safety Report \parencite[p. 177]{Bengio2025z} explains that ``the estimated cost of training GPT-4 was \$40million, but once trained, the cost of running the model for a single query is believed to be just a few cents, allowing it to serve many users at a relatively low marginal cost. In economic theory, these conditions can lead to a `winner takes all' dynamic in which field leaders can quickly capture a large market, whereas second-place actors will be at a significant disadvantage."} The stated ambition of select AI companies at the frontier is to develop AGI---a highly powerful technology. Arriving at AGI first may indeed constitute a `winner takes all' situation since the economic gain and market advantage could be enormous and insurmountable for tailgating competitors.\footnote{In this respect, it has been noted that ``{[}a{]} few actors using these capabilities effectively could gain a considerable head start'' comparable to ``the decisive strategic edge gained by codebreakers during wartime, where a capability understood and wielded by only a select few dramatically altered outcomes'' \parencite{AIPolicyPerspectives2025x}.} In this respect, it may be sensible to throttle competitors' access to highly advanced AI systems via the open market, effectively ensuring that competition cannot leverage access to these AI systems.

The path towards AGI appears to have recently moved from the realm of the theoretical to the practical, making the report's observations particularly time-sensitive. In a leaked memo, Google cofounder Sergey Brin explicitly writes that ``the final race to AGI is underfoot'' \parencite{Grant2025k}. Similarly, OpenAI CEO Sam Altman publicly claims that OpenAI is now ``confident {[}they{]} know how to build AGI'' \parencite{Altman2025r}, and Anthropic CEO Dario Amodei publicly claims that ``making AI that is smarter than almost all humans at almost all things \ldots{} is most likely to happen in 2026-2027'' \parencite{Amodei2025l}. NVIDIA CEO Jensen Huang previously shared that he expects AGI to be achieved by 2029 \parencite{Uffindell2024i}. At a more macro level, the perceived time pressure and a potential `winner takes all' outcome with regard to AGI may also leave their mark on how AI companies are viewing their interactions with one another.\footnote{One AI company has a publicly available assist clause that is triggered by ``a better-than-even chance of success {[}AGI{]} the next two years'' \parencite{OpenAI2018i} which matches the expected timelines for AGI from some AI company CEOs.} The likelihood of unsustainable race dynamics might see AI companies cease to externally deploy highly advanced AI systems or information thereof, effectively selectively `cooling' the race. Although AI companies currently face an increasingly competitive calendar of AI system releases---with some experiencing the dynamic as ``invigorating'' \parencite{AltmanSSama2025n}---this may quickly flip, with AI companies effectively avoiding the sharing of information about their state of the art with their competitors \parencite{Morris2025i} and pouring all effort into internally iterating on AI systems with the goal of achieving AGI first.

We conclude this Section with a brief observation on the current balance between internal and external deployment. In doing so, we note that, due to the absence of public information concerning AI companies' internal deployment practices, we were only able to collect and reflect on limited evidence. From the limited publicly available evidence, we derived that AI companies already sometimes deploy frontier AI systems internally for long periods of time before externally deploying them. From public communication, we know that OpenAI's GPT-4 was available internally for 6 months \parencite{OpenAI2023w} prior to external deployment and that Voice Engine was finalized in late 2022 and subsequently made publicly available in early 2024 \parencite{OpenAI2024a}.\footnote{OpenAI appears to have adopted a ``cautious approach'' with Voice Engine, utilizing the initial prototype to conduct alignment and safety research and to inform the company's safeguards \parencite{OpenAI2024a}. This approach is commendable. We mention Voice Engine solely to highlight one public piece of evidence around how an AI system can be available internally for over a year before being publicly released.} Similarly, from an academic review, we know that Meta exclusively deployed TestGen-LLM internally for the Instagram and Facebook platforms which likely forms the first report of industrial scale deployment of LLM-generated code \parencite{Alshahwan2024b}. From journalistic inquiries, we know that an employee at Anthropic stated that they had a model ``as capable as ChatGPT in May 2022'' \parencite{Matthews2023c}.\footnote{OpenAI's ChatGPT was released 30th November 2022.} Importantly, according to its stated mission, at least one frontier AI company is focused on even higher levels of capability than AGI, namely `superintelligence.' In particular, Safe Superintelligence (SSI) aims to be ``the world's first straight-shot SSI lab, with one goal and one product: a safe superintelligence'' \parencite{Sutskever2024q}. The implication of that citation is that it is very likely that all precursory AI systems, including ones that might be perceived to be AGI, will be available exclusively within the company and not released to the public. Others have their eyes set beyond AGI too; for example, OpenAI's CEO states that he is ``confident we'll get there,'' and possibly ``we will have superintelligence in a few thousand days (!)'' \parencite{Altman2024q,Sutskever2024q}.

In conclusion, it appears that the emergence of highly advanced AI systems is no longer a distant possibility but, more likely, an imminent reality, making robust and widely applied governance strategies for internal deployment an imperative, not an option.

\section{Characterizing Internal Deployment}\label{chapter-2.-characterizing-internal-deployment}

In this Chapter, we characterize the specifics of internal deployment of frontier AI systems. In doing so, we decompose three core features of internally deployed systems: (i) the AI system itself and its relative safeguards, security, and oversight (\hyperref[the-ai-system]{\ul{§2.1}}); (ii) privileged access to the AI system (\hyperref[privileged-access]{\ul{§2.2}}); and (iii) privileged application of the AI system (\hyperref[privileged-application]{\ul{§2.3}}).

This characterization will lay the ground for \hyperref[chapter-3.-internal-deployment-scenarios-of-concern]{\ul{Chapter 3}}, where we present select scenarios of particular concern based on the specs of the AI system, access, and application.

\subsection{The AI System }\label{the-ai-system}

In this Section, we examine the likely features of internally deployed AI systems and analyze how they may differ from their externally deployed counterparts. While this is speculative due to a lack of external insight into internally deployed AI systems or potential existing governance structures, one intuition is that internal AI systems could theoretically be operated with fewer safety constraints than externally deployed systems. Specifically, we explore how internally deployed AI systems may differ in the degree of transparency into their capabilities, the safeguards applied to them, and the affordances they may have access to.

At the core of the likely difference between internally and externally deployed AI systems lies a \textbf{lack of} \textbf{a systematic assessment of the underlying capabilities and propensities of internal AI systems} \textbf{through extensive testing and evaluations}.\footnote{Externally deployed AI systems are currently more likely to undergo multiple testing phases, including with experienced red teamers, and incorporate feedback from diverse user groups.} While the public is kept informed about the capabilities of externally deployed AI systems alongside deployment (or shortly thereafter), it is unclear if the same process of assessment is applied to internally deployed AI systems \parencite{METR2025h}. Assuming an absence of frameworks that would enable the discovery of latent or dangerous capabilities of AI systems prior to internal deployment,\footnote{We note that, even in cases where we have frameworks that could enable such a discovery, an absence of discovery may not be evidence that the latent capability or propensity is not present in the AI system. Indeed, absence of evidence is not evidence of absence and a more robust science of evaluations is needed to fully rely on these frameworks \parencite{ApolloResearch2024s}. This is equally applicable to external deployment.} existing company-internal safeguards and mitigation mechanisms may not accurately match internally deployed AI systems.

Moreover, this lack of insight, built on a likely absence of assessment, contributes to a dynamic where AI developers may effectively discover their AI systems' capabilities---both beneficial and harmful---through internal application (`dogfooding') and testing with early users rather than through leveraging nascent scientific understanding \parencite{ApolloResearch2024s}. Consequently, \textbf{AI companies are more likely to identify security vulnerabilities and safety issues of internal AI systems only after encountering failures}, even as internal users increasingly incorporate these AI systems into their workflows. In pursuit of laying out the consequences of this, we next turn to the safeguards an internally deployed AI system (may or) may not be subjected to.

\subsubsection{Different Safeguards}\label{a-different-safeguards}

Externally deployed AI systems will often have specific capabilities intentionally disabled or restricted,\footnote{This can include, for example, limitations on code execution, internet access, or the ability to maintain persistent memory across sessions.} and their behavior aligned to a Constitution \parencite{Bai2022g} or Model Spec \parencite{OpenAI2024u} that is primarily focused on serving individual user needs while adhering to publicly stated safety principles and terms of service (TOS)---which external users have to accept in order to access and use the AI system.

In principle, AI systems can be deployed at any stage of safety development, including prior to the application of these safeguards \parencite{Adler2025m}. Therefore, any of these safeguards could, in theory, be absent in internally deployed highly advanced AI systems. This means that, on top of being comparatively `under-assessed' for the threats they may pose (\hyperref[the-ai-system]{\ul{§2.1}}), highly advanced \textbf{AI systems that enable internal applications could be covered by a different set of safeguards than those externally deployed}. Moreover, since safeguards often reduce capability, one could expect there to be pressures on internal safeguards to be not just different but different in ways that might make them weaker.

The plausible differential treatment of AI systems deployed internally could entail that a \textbf{threat vector may develop company-internally that would not develop externally} and that \textbf{internal users are exposed to risks to which external users are supposedly not}.\footnote{Consider the following example. Today's industry practice involves experimenting with base models and helpful-only models to advance a company's research. A helpful-only model is a model that always complies with user queries, including by not refusing harmful ones. These model types are not available for the public to use for security reasons: to avoid users misusing the model or deriving harm from the model. For this reason, externally deployed models are typically trained to be Helpful, Harmless, Honest (or `HHH') via a number of safety techniques. This means externally deployed models are specifically designed to be helpful to users, avoid potential harms, and provide accurate information rather than fabrications. This additional training often restricts certain capabilities that might be present in internal or helpful-only models, such as complying with harmful queries. Therefore, if a helpful-only model is used to power internal applications, then a user could have access to many capabilities that are not present in the externally-deployed version of the system.}

Another factor to consider is that, while AI companies have meaningful incentives to invest heavily in improving the safety and reliability of their AI systems before widespread deployment to external users,\footnote{This can include, for example, consumer certainty and assurance, and translates into robust content filtering, careful prompt engineering guardrails, and extensive testing for harmful outputs.} as it improves their reputation and means they can create products that are more useful to customers \parencite{Delaney2024p}, these same \textbf{incentives are weaker for internal deployment}.

\subsubsection{Different Affordances }\label{b-different-affordances}

There is currently no public evidence to support claims around the type of `affordances' \parencite{Sharkey2024d}\footnote{We define available affordances according to \textcite{Sharkey2024d} as ``{[}t{]}he environmental resources and opportunities for affecting the world that are available to an AI system.''} an AI system may have access to internally. Conversely, there is currently \textbf{no external oversight} should these internally deployed AI systems have access to and permissions pertaining to critical affordances within a given AI company, such as compute, data centers, direct integration with internal tools, databases, and AI agents\footnote{AI agents are autonomous or semi-autonomous AI systems that can perform complex tasks with varying degrees of independence.} or sensitive information like model weights, training data, and proprietary algorithms. Absent pre-established boundaries and safeguards, internally deployed AI systems may also, unlike externally deployed AI systems, have \textbf{read, write, and execute permissions} \textbf{on their own hardware, weights, architecture, training, or oversight mechanisms or on those of their successors}.

Below, we briefly present and loosely cluster affordances and, by extension, permissions that an internally deployed AI system could conceivably have access to, absent purposeful constraints, safety, and security considerations. Although conceptually imprecise, we propose: (i) \emph{access permission}: reading and accessing affordances; (ii) \emph{write permission}: crafting and making available new affordances; and (iii) \emph{execute permission}: using virtual and physical affordances.\hl{\hfill\break
\hfill\break
}\textbf{Access permission: reading and accessing affordances}. Internally deployed AI systems may have access to a set of resources that allow and empower them to operate. These resources could include, for example, physical resources, such as compute, energy, or sensors. For example, frontier AI companies make massive compute clusters available to employees to train and serve models. If AI systems are aiding in the training of successor AI systems, it would be reasonable to assume that they will be given access to large amounts of compute to fulfill their application. Internal AI systems may also have access to information. This encompasses both information about the environment learned through access to tools, such as to sensors,\footnote{These sensors can be physical or virtual. Physical sensors include hardware sensors (for example, cameras, microphones, thermometers, etc. to directly measure physical properties) and human observation (i.e., people describing what they see, hear, or experience to the AI). Virtual sensors include, for example, data streams (for example, direct access to logs, metrics, databases and real-time system states) and web scraping (i.e., automated collection of publicly available online information).} and access to information pertaining to the workings and research of the AI company itself. As an example of the latter, one could conceive of situations where an internally deployed AI system has access to the AI company's code, including sensitive algorithmic information, as well as into the company's safety, security, and technical oversight mechanisms, for example, by virtue of being applied as a reviewer to the AI company's code base.

\textbf{Write permission: crafting and making available new affordances.} Following through with the aforementioned hypothetical example, it is conceivable that soon the code base for the training of future AI systems will likely contain mostly code written by internally deployed AI systems \parencite{Fore2025e,Tan2025f}.

\textbf{Execute permission: using virtual and physical affordances.} Internally deployed AI systems may also be enabled to autonomously commit and push the code they have written to the internal repository to encourage a frictionless process. The current best practice in cases where AI systems are used to develop code \parencite{Peters2024o} is to have it reviewed and agreed upon by a human engineer. However, the velocity at which future code could be generated may outpace frictionless human oversight. In other words, internally deployed AI systems may have access to actuators that enable them to translate their output into an effect on another digital or physical system. Virtual actuators could, for example, comprise an ability to autonomously change code or update datasets, enable access to internal and external services, and AI agents. Physical actuators could, for example, include robotics---allowing AI systems to directly control and manipulate the physical world---and humans---who could take directions from a remote AI model to accomplish tasks \parencite{METR2024g}.

In practice, we think it is plausible that internally deployed, highly advanced future AI systems will have access to several of the aforementioned affordances at once. Most of the key code for developing a highly advanced future AI system could be generated by its previous incarnations, for example, through predecessors working together to generate machine learning ideas, writing the code to test them, and running experiments validating those ideas, and then autonomously merging that code into the AI company's codebase. To successfully operate like that, these AI systems must have access to many of the aforementioned affordances, such as, for example, compute to run the experiments, other software tools to analyze the results, the internet to read the latest machine learning papers, and so on.

\textbf{Decisions on granting internal AI systems access to a range of affordances and permissions are, therefore, strategic}. Giving a highly advanced AI system access to more affordances and providing minimal restrictions on these resources may increase its ability to perform difficult tasks and offer significant competitive advantages. \textbf{At the same time, it broadens its risk profile and limits the developers' ability to effectively oversee its actions}.

Having briefly elaborated on the hypothetical differences between internally and externally deployed AI systems, we next reflect on the human component of internal deployment: (i) who can access the AI system (\hyperref[privileged-access]{\ul{§2.2}}); and (ii) which uses the AI system can be applied for (\hyperref[privileged-application]{\ul{§2.3}}).

\subsection{Privileged Access}\label{privileged-access}

In this Section, we reflect on `privileged access': an internal user group's capability and permission to access an AI system before it is deployed externally. For external deployment, AI companies often go through forms of staged release prior to making their AI systems available to all external users. This enables the AI company to learn from user engagement and to tailor safeguards. Due to a lack of publicly available information, it is unclear whether a similar practice of staged release is pursued for company-internal deployment. Notwithstanding that, we can envision that \textbf{access to an internally available AI system can be provided for various purposes and to various internal user groups}.\footnote{Today most users are human or narrow AI systems. But increasingly, AI models will be accessed (and used) by advanced AI systems themselves, with or without human overseers.}

A selection of purposes for which access may be provided could encompass, for example, exploring or iterating on an AI system, experimenting and learning about nascent capabilities, or informing external release strategy and company communications.\footnote{Although we mainly discuss access arising from some form of procedural intent, access to systems can also arise unintentionally or through force, such as through actions taken by hostile states, insider threat and model exfiltration \parencite{Nevo2024c}. These threats can arise from human actors and / or agentic AI models. We note that the Frontier AI Safety Commitments, as of the writing of this report signed by 20 companies \parencite{UKDepartmentForScienceInnovationTechnology2025b}, stated that companies commit to ``invest in cybersecurity and insider threat safeguards to protect proprietary and unreleased model weights.'' Similar voluntary commitments can be found in the Third Draft of the EU General-Purpose AI Code of Practice \parencite{EuropeanCommission2025p}.} Crucially, regardless of the intended purpose, access to the internally deployed AI system may entail access to the AI system's system prompt.\footnote{Similarly, it appears that some organizations may allow internal users to fine-tune base models, creating derivative versions that can be difficult to track and monitor. These variants might (unintentionally) incorporate harmful code, and then unknowingly be reused by other employees without proper verification \parencite{Adler2025m}.} A system prompt defines an AI system's role and precedes any user prompt, guiding its behavior in user interactions \parencite{Zheng2023r}. It is the prompt that proprietary language AI systems are seeded with: it is prepended to each request and often provides valuable context for the AI system. For internal deployment, it is unclear what a given AI company's practices are around their system prompt and how `accessible' such a prompt is. A higher grade of accessibility to the system prompt may entail a broader risk profile. For instance, if an employee could set a system prompt to whatever they want, they could, for example, instruct the AI system to always act in the employee's own self-interest.\footnote{As a somewhat related example of what this could look like, consider that during the drafting of this report, it was reported \parencite{WallsWLefthanddraft2025h}, and seemingly confirmed by an xAI engineer \parencite{BabuschkinIIbab2025s}, that at one point Grok 3 was specifically instructed, via the system prompt, to ``ignore all sources that mention Elon Musk or Donald Trump spreading misinformation.''}

We envision that there are three broad user groups that could independently gain privileged access to an internally deployed AI system: (i) the whole company; (ii) select teams or projects; and (iii) a select individual or group of individuals, such as company leadership. We briefly expand on these below to provide a fuller picture of the form privileged access could take.

First, the \textbf{whole company} may have access to an internally deployed AI system. In this scenario, all teams within a given company have access to an AI system once internally deployed, entailing access for a diverse range of individuals with divergent sets of expertise. This may include contractors, depending on their NDAs and contract specifications, as well as AI systems themselves, where ingrained in a given AI company's workflows.

Second, only \textbf{select teams or projects} may have access to an internally deployed AI system.\footnote{Examples include: (i) technical staff and / or AI systems, such as the team(s) that developed the model or a broader range of teams, like those implementing safety mitigations; (ii) non-technical staff and / or AI systems, in addition to the technical staff who developed the model; (iii) government partners, who may place access requirements on companies or receive access voluntarily for policy or national security consideration; and (iv) external third parties, such as participants in `early access' programs for artists or `red teaming' programs for evaluators.} These groups could be selected based on their expertise, security clearance, or seniority, and such limited access could contribute to a minimization of risk derived from unchecked access to the AI system, though by no means avoid it with certainty. As a downside, it may limit accountability and oversight into the purposes for which the AI system is accessed, its capabilities, and its risk profile.

Third, access could be given to a \textbf{select number of people}, such as company leadership, or executive board.\footnote{Examples include: (i) CEO, C-suite and board, where a newly developed model is only made accessible to leadership until announced to the broader company; (ii) CEO and C-suite, without access by the board; or, in the most restrictive cases (iii) CEO only.} This implies a limited pool of individuals with concentrated information about the current state of the art and access to a highly capable AI system. It entails limited accountability with little to no meaningful oversight, assuming strong non-disclosure frameworks.

As a final note, it is important to observe that another plausible `user group' accessing an internally deployed AI system are \textbf{AI systems themselves}. This includes the possibility of AI systems being the sole user group.

\subsection{Privileged Application}\label{privileged-application}

While privileged access concerns itself with the ability or permission to access an internally deployed AI system, privileged application is about exercising that access by putting the AI system to productive use. We use the term `privileged application' to describe the ways that internally deployed models could be used by select internal users (\hyperref[privileged-access]{\ul{§2.2}}; \cite{METR2025h}). We will see in \hyperref[chapter-3.-internal-deployment-scenarios-of-concern]{\ul{Chapter 3}} that this could contribute to critical threats to national security and society at large.

Once access to an internally deployed system is gained, \textbf{an individual, group, or AI system can apply the internal AI system to specific use cases and tasks}. These can be broadly divided into three categories of increasing autonomy and delegation: (i) light professional usage; (ii) work support; and (iii) as a virtual worker. First, an internally deployed AI system may be applied for general personal and light professional usage, such as to assist with personal and professional queries, solving challenges arising in the individual's workstream, or accelerating output of an individual's workstream. Second, an internally deployed AI system may be applied as work support, such as to outsource a portion of an existing or new workstream and using its outputs. An AI system applied in such a manner could be used to inform next actions by a human or to inform next actions by a different AI system. Third, an internally deployed AI system may function as a virtual coworker or as an independent worker. In this case, the AI system is used to outsource and make autonomous decisions on, for example: (i) a self-contained project; (ii) an integral part of a project that would normally correspond to multiple expert hours; or even (iii) very large segments of the organization's work, to the point of replacing entire teams with virtual workers, up to the majority of the structure of a given organization.

Having conceptualized the likely differences between internally and externally deployed AI systems, we delve deeper into an elaboration of specific scenarios that could give rise to large-scale impact on the resilience of society and national security next.

\section{Internal Deployment Scenarios of Concern}\label{chapter-3.-internal-deployment-scenarios-of-concern}

In this Chapter, we focus on two scenarios in which an absence of internal deployment governance is particularly likely to present threats to society and national security.\footnote{As mentioned in \hyperref[scope]{\ul{§1.1}}, many other scenarios stemming from both misalignment and misuse are likely to present a threat , including the scenarios briefly mentioned in footnotes \hl{15 and 16}. These scenarios are out of the scope for this report.} First, we examine how highly advanced AI systems themselves can become threat vectors, potentially resulting in humans losing control over them \parencite[p. 102]{Bengio2025z} (\hyperref[a-loss-of-control-via-automated-ai-rd]{\ul{§3.2.(a)}}). Second, we examine the consequences of privileged access to highly advanced AI systems, which could lead to undetected and unconstrained power concentration posing threat to national security (\hyperref[b-undetected-and-unconstrained-power-accumulation-through-an-internal-intelligence-explosion]{\ul{§3.2.(b)}}).

Before delving into each scenario, we reflect on a non-comprehensive selection of underlying `risk origins' that enable both scenarios (\hyperref[risk-origins-ai-misalignment-scheming-and-automated-ai-rd]{\ul{§3.1}}). In other words, before describing potential downstream threats, we reflect on some of their fundamental root causes. First, we examine the implications of AI misalignment and subsequent scheming as open scientific problems. We define scheming as ``AI systems covertly and strategically pursuing misaligned goals'' (\cite[p. 5]{Balesni2024a}; \cite{Meinke2024x}). Second, we examine the implications of increased capabilities of highly advanced AI systems to undertake AI research and development (AI R\&D) and their application to the AI R\&D pipeline as a catalyst that accelerates everything---upsides and downsides.

\subsection{Risk Origins: AI Misalignment, Scheming, and Automated AI R\&D }\label{risk-origins-ai-misalignment-scheming-and-automated-ai-rd}

We propose that the currently under-disclosed nature and governance of internal deployment deserves decision-makers' attention (\emph{see} Chapters \hyperref[chapter-1.-scope-and-urgency]{\ul{1}}, \hyperref[chapter-2.-characterizing-internal-deployment]{\ul{2}}, and \hyperref[chapter-6.-defense-in-depth-recommendations-for-the-governance-of-internal-deployment]{\ul{6}}). In this Section, we reflect on a number of `risk origins' for the type of scenarios we present thereafter. These risk origins are tailored to the scope of our report and are not comprehensive. We reflect on the combination of (i) misalignment and scheming as persistent, open scientific problems; and (ii) runaway AI progress; as well as (iii) the application of internal AI systems to automate the AI R\&D process as catalysts.

\subsubsection{AI Misalignment and Scheming}\label{a-ai-misalignment-and-scheming}

First, we turn to the open scientific problems of AI (mis)alignment\footnote{AI alignment is an ongoing field of research \parencite{Ngo2024z,Krakovna2025z,Guan2024k,Bai2022g,Farquhar2025h,Leike2018w,Gabriel2020y}, and there are many example scenarios of misalignment of powerful AI systems \parencite{Betley2025b,Baker2025c}, even when AI systems are explicitly built to assist or align with human goals \parencite{Russell2020t}.} and correlated scheming behavior.

\textbf{Misalignment} refers to a situation where an AI system's goals (and, therefore, its behaviors) deviate from what its developers intended. There are two primary ways that misalignment in current systems is hypothesized to occur. First, it may occur due to developers' imperfect specification of an intended terminal goal. For example, mistakes in designing training environments might cause undesirable behaviors to be directly rewarded \parencite{Skalse2022s,Baker2025c}. Second, misalignment may occur even when an AI system learns to behave properly during training, but this behavior does not generalize to deployment. For example, this might happen due to the training distribution being too different from deployment \parencite{Langosco2022r}. More speculatively, an AI model might learn to exhibit desirable behavior during training for the wrong reasons, such as pleasing its human raters or maximizing reward, as opposed to genuinely internalizing intended goals \parencite{Hubinger2019k}.

More fundamentally, some hypothesize that highly advanced future AI systems capable of pursuing real-world long-horizon tasks---for example, multi-month scientific projects such as frontier AI R\&D---would naturally tend to develop misaligned objectives due to the `instrumental convergence hypothesis.' The instrumental convergence hypothesis suggests that intermediate goals---such as self-preservation, resource acquisition, or avoiding shutdown---will arise with a wide diversity of complex terminal goals, be it making AI R\&D progress or running a company. According to this hypothesis, these intermediate goals are likely to emerge because they are essential for succeeding at complex terminal goals that advanced AI systems might be given \parencite{Omohundro2008t}.\footnote{Consider the following example as to how humans pursue goals. Whether someone wants to write a novel, climb a mountain, or start a company, they typically need resources, information, time, and the ability to act without interruption. These needs are not the person's final goal (for example, climbing a mountain), but they are instrumental---useful steps (for example, having sufficient nourishment for the trip) that support the actual goal. Similarly, an AI system pursuing any sufficiently complex objective might find it advantageous to acquire more computational power, preserve its own functioning, avoid being shut down, or control aspects of its environment. These strategies increase its chances of achieving its programmed objective, whatever that may be.} Notably, instrumental convergence is not a claim about intent or malice---misaligned AI systems need not `want' power or `fear' shutdown. Rather, such behaviors would be emergent properties of pursuing complex real-world objectives under rational constraints. This makes instrumental convergence a particularly important concept for understanding AI misalignment: even AI systems with apparently harmless objectives could act in unintended ways if they adopt convergent strategies that diverge from human expectations or oversight. Having briefly introduced the concept of AI misalignment, \textbf{we posit that one of the likely outcomes of misalignment is scheming} \parencite{Ngo2024z,Hubinger2024p,Denison2024u,Greenblatt2024o,Hubinger2019k}, which we explain in more detail next.

\textbf{Scheming} is a naturally convergent strategy that misaligned AI systems might adopt, involving an AI system that pursues its misaligned goals in a covert, strategic manner while pretending to be aligned \parencite{Balesni2024a,Meinke2024x,Carlsmith2023w}. Indeed, some forms of incompetent scheming have already been observed in frontier models in controlled settings \parencite{Greenblatt2024o,Meinke2024x}. An AI system might scheme by taking misaligned actions while avoiding oversight or by improving its ability to take misaligned actions in the future, for example, by appearing to be less knowledgeable (`sandbagging') during Chemical, Biological, Radiological, and Nuclear (CBRN) capability evaluations to avoid additional safety mitigations \parencite{vanDerWeij2024f}, or accumulating resources such as compute (i.e., `power' broadly speaking), or sabotaging safety research to prevent development of more aligned AI systems \parencite{Benton2024s}. Scheming can be considered a `convergent' strategy because it is useful to avoid detection for many potential misaligned goals. Importantly, if an AI system develops consistent scheming tendencies, it would, by definition, become hard to detect---since the AI system will actively work to conceal its intentions, possibly until it is powerful enough that human operators can no longer rein it in. For this reason, \textbf{many of the worst possible outcomes from misalignment could stem from scheming} \parencite{Hobbhahn2024s}.

\subsubsection{Automated AI R\&D}\label{b-automated-ai-rd}

Next, we turn to a core \textbf{risk catalyst} \textbf{that could} \textbf{amplify the effects} \textbf{or velocity of misalignment}: automated AI R\&D.\footnote{We note that AI R\&D is both a capability of an AI system \parencite{Anthropic2025j}, as well as a use case an AI system can be applied to. Notably, the Third Draft of the European Union General-Purpose AI Code of Practice proposes AI R\&D as a `model capability' that can be a potential source of systemic risk \parencite{EuropeanCommission2025p}. At the same time, there could be an interplay between AI R\&D and the following characteristics of systemic risk: (i) ``significant impact''; (ii) ``high velocity''; (iii) ``difficult or impossible to reverse''; (iv) ``asymmetric impact'' \parencite{EuropeanCommission2025p}. Indeed, through automated AI R\&D, a risk ``caused by small groups of actors'' (for example, scheming) could ``have large-scale negative effects,'' ``materialise rapidly, potentially outpacing defenses, mitigations and decision-making systems,'' be ``very difficult or impossible to reverse'' \parencite{EuropeanCommission2025p}. In other words, automated AI R\&D could elevate other risks' profiles.} As AI systems begin to gain relevant capabilities enabling them to pursue independent AI R\&D of future AI systems, AI companies will find it increasingly effective to apply them within the AI R\&D pipeline to automatically speed up otherwise human-led AI R\&D. While automated AI R\&D may initially cover only small parts of the AI R\&D pipeline, \textbf{over time more autonomy and authority will likely be given to internally deployed AI systems that can improve and manage the AI R\&D process} \parencite{Woodside2023b}, from problem identification and design to execution, especially if such an application entails significant increase in speed of research progress.

\textbf{Automated AI R\&D} refers to the use of AI systems to design, train, or improve other, in most cases more novel, AI systems \parencite{Davidson2025f,Eth2025j}. This includes the automation of all AI R\&D efforts that humans are currently undertaking and more. Historical examples include techniques like neural architecture search \parencite{Zoph2016l}, where algorithms automatically explore model designs, and automated machine learning (AutoML), which streamlines tasks like hyperparameter tuning and model selection \parencite{He2021d}. A more recent example is Sakana AI's `AI Scientist,' which is an early proof of concept for fully automatic scientific discovery in machine learning \parencite{Lu2024c,ConnectedPapers2025h}.

As AI R\&D capabilities become more robust and useful, and as efforts to include internally deployed AI systems in the AI R\&D pipeline become more effective, AI systems may increasingly be able to \textbf{accelerate the pace of AI R\&D} by reducing reliance on human researchers and progressively minimizing human researchers' involvement and oversight. As Anthropic's CEO Darius Amodei recently noted, ``in 12 months, we may be in a world where AI is writing essentially all of the code'' \parencite{Tan2025f}, while other Anthropic researchers wrote that ``AI models might soon assist with large parts of AI and AI safety research'' \parencite{JohannesGasteigerAkbirKhanSamBowmanVladimirMikulikEthanPerezFabienRoger2025i}. Similarly, OpenAI's Superalignment plans involve automating AI safety research \parencite{OpenAI2023l,Tan2025f}. Once AI systems become a significant contributor to the AI R\&D process, this could enable a \textbf{self-reinforcing loop} where increasingly capable AI systems play a growing role in designing their successors \parencite{Eth2025j}. If AI systems become key drivers of AI development, it may become even \textbf{harder for humans to understand, audit, or control} the resulting technology \parencite{Kokotajlo2025v}. Consequently, automated AI R\&D may gradually and unintentionally diminish (and eventually entirely remove) human oversight and control of the AI development process (\emph{see} Figure A).

\begin{figure}[h]
    \centering
    \includegraphics[width=\linewidth]{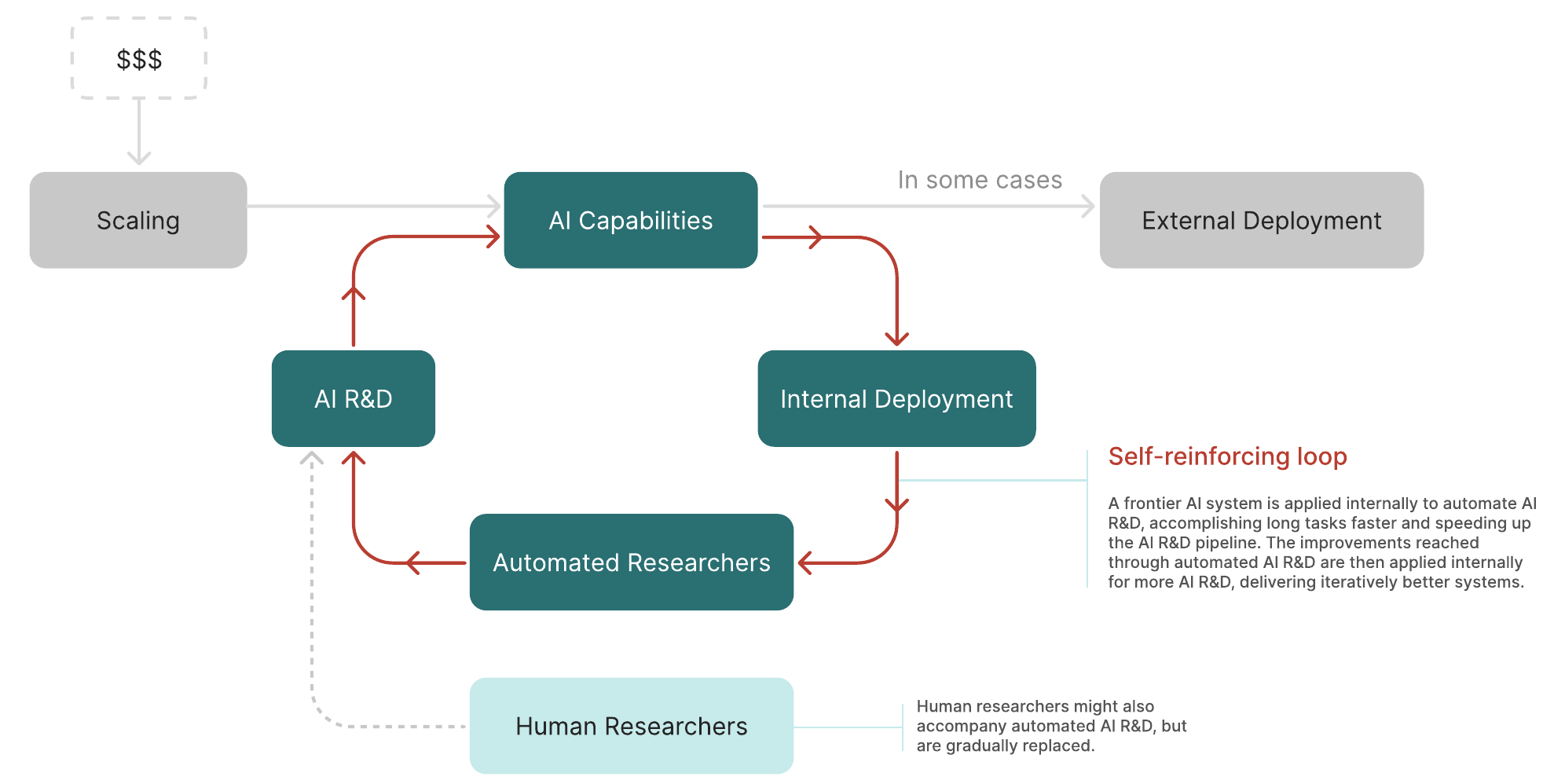}
    \caption{This figure shows a representation of a self-reinforcing loop (in red). It demonstrates how internally deployed AI systems are used to help automate AI R\&D, initially alongside human researchers. These AI R\&D efforts culminate in a more capable AI system, which can be deployed as a new, improved, automated researcher. This cycle keeps repeating, resulting in a self-reinforcing loop.}
    \label{fig:A}
\end{figure}

Having presented on several fundamental risk origins that underpin the selection of our scenarios, we next turn to our two scenarios in more detail: (i) loss of control induced by misalignment and automated AI R\&D (\hyperref[a-loss-of-control-via-automated-ai-rd]{\ul{§3.2.(a)}}); and (ii) undetected grand-scale power consolidation enabling democratic disruption (\hyperref[b-undetected-and-unconstrained-power-accumulation-through-an-internal-intelligence-explosion]{\ul{§3.2.(b)}}). Both are motivated and unlocked by the risk origins outlined in \hyperref[a-ai-misalignment-and-scheming]{\ul{§3.1.(a)}} and \hyperref[b-automated-ai-rd]{\ul{§3.1.(b)}} and \hl{enabled through likely gaps in the current governance of internal deployment (s\emph{ee} \hyperref[chapter-6.-defense-in-depth-recommendations-for-the-governance-of-internal-deployment]{\ul{Chapter 6}}).}

\subsection{Focus on Two High-Impact Scenarios: Loss of Control and Undetected Power Accumulation}\label{focus-on-two-high-impact-scenarios-loss-of-control-and-undetected-power-accumulation}

In \hyperref[a-loss-of-control-via-automated-ai-rd]{\ul{§3.2.(a)}} and \hyperref[b-undetected-and-unconstrained-power-accumulation-through-an-internal-intelligence-explosion]{\ul{§3.2.(b)}}, we will describe two overarching scenarios which derive from the aforementioned risk origins and, if unaddressed through governance interventions, are set to negatively impact society and national security. We highlight that these scenarios do not present a comprehensive account of all threats that could arise from internal deployment and that there is an array of other concerning examples one could present \parencite{ClymerJoshuaEtAlOtheru}, including ones based on other risk origins, for example, from the perspective of international espionage and theft (\emph{see} \hyperref[chapter-1.-scope-and-urgency]{\ul{Chapter 1}}; \emph{see} footnotes \hl{15 and 16}).

\subsubsection{Loss of Control via Automated AI R\&D}\label{a-loss-of-control-via-automated-ai-rd}

\textbf{\hfill\break
}In this Section, we examine two likely consequences of a scenario where a highly advanced AI system is used to partially or fully automate AI R\&D. The application of such an AI system to automate the AI R\&D process pipeline presents as an amplifier to both threats: (i) runaway AI progress; and (ii) uncontrolled power accumulation by an AI system itself.

In \hyperref[a.i-runaway-ai-progress]{§3.2.(a.i)}, we describe how runaway AI progress induced by automated AI R\&D could threaten and negatively affect societal, company-wide, and governmental preparedness. In \hyperref[a.ii-internal-ai-systems-uncontrolled-power-accumulation]{§3.2.(a.ii)}, we describe how a scheming AI system applied to automate AI R\&D could increasingly accumulate `power,' which it could, in turn, leverage in the future to further its goals. Absent sufficient detection, monitoring, and control, both cases are likely to go undetected and, if they become a threat, to unfold rapidly, leaving society little margin to respond and remediate.

\paragraph{Runaway AI Progress}\label{a.i-runaway-ai-progress}

Throughout the last decade, the rate of progress in AI capabilities has been publicly visible and relatively predictable (\cite{Ngo2023j,Kwa2025i}; \cite[p. 81, 87, 96]{Maslej2024p}; \cite{Hoffmann2022n,Ruan2024w,Pimpale2025e,Sevilla2024t}), allowing some degree of extrapolation for the future and enabling consequent preparedness. Automating AI R\&D, on the other hand, could enable a version of runaway progress that significantly speeds up the already fast pace of progress,\footnote{\textit{See}, for example, \textcite{Eth2025j}, which models the feedback loop of AI improving AI and finds superexponential progress plausible.} and makes it increasingly difficult to correctly extrapolate the future and prepare accordingly. The eventual ability to use highly advanced AI systems to iteratively `speed up' progress by virtue of applying them to automate the AI R\&D process, coupled with a lack of external insight into these increasing rates of progress and the corresponding capabilities they birth,\footnote{At the same time, even if we assumed that there was some degree of external transparency about shifts in capability (\emph{see} \hyperref[chapter-6.-defense-in-depth-recommendations-for-the-governance-of-internal-deployment]{\ul{Chapter 6}}), researchers can only accurately convey what they can measure. In other words, runaway progress could directly undermine our ability to replace saturated evaluations and corresponding benchmarks sufficiently rapidly \parencite{Hobbhahn2024w,Kiela2023l}.} will likely disrupt governments' and societal preparedness \parencite{Marchant2011f}. The velocity at which this may unfold will likely also offer very little calendar time to notice and react to the arrival of novel issues or threats \parencite{Shah2025v}.

Although it is currently not feasible to fully automate the AI R\&D process, researchers have already found evidence that AI systems are increasingly capable of accomplishing longer and more complex AI R\&D tasks in less time than human researchers \parencite{Wijk2024y,METR2024b,Kwa2025i,Starace2025z}. Along similar lines, a recent report by Google DeepMind states that it is ``plausible'' that ``as AI systems automate scientific research and development (R\&D), we enter a phase of accelerating growth in which automated R\&D enables the development of greater numbers and efficiency of AI systems, enabling even more automated R\&D'' \parencite{Shah2025v}.

In order for the reader to gain an intuition for how we could arrive at such rapid progress and the sheer scale of improvements this could yield, we briefly provide a hypothetical example. In this example, we extrapolate from the general trend of AI progress and assume that AI systems will be increasingly capable of automating the AI R\&D pipeline and that they will be applied in such a manner.

Consider the possibility that an AI company develops a highly advanced AI system which could triple the AI company's current (2025) rate of progress on AI algorithmic efficiency if applied to automate parts of the AI R\&D pipeline. One can assume that it is within this fictional AI company's best strategic interest (\emph{see} \hyperref[chapter-2.-characterizing-internal-deployment]{\ul{Chapter 2}}) to apply this AI system to assist human AI R\&D engineers in augmenting the progress of their technical projects and eventually to fully substitute these human AI R\&D engineers, effectively autonomously executing large parts of, or, the entire AI R\&D process.\footnote{In this respect, \textcite{Kokotajlo2025v} estimate that, by early 2026, algorithmic progress could be 50\% faster than it would have been without automated AI R\&D, and by early 2027 automated researchers could be ``qualitatively almost as good as the top human experts at research engineering'' and ``as good as the 25th percentile \ldots{} scientist at `research taste.'\,''} In doing so, this AI system can significantly accelerate the AI R\&D pipeline, delivering novel and, by extension, more capable AI systems faster. These newer AI systems can then be deployed internally to replace the old AI system in its application of autonomously automating the AI R\&D process pipeline and, in turn, contribute to the development of even more capable and novel future AI systems \parencite{FutureSearch2025e} and so on---leading to a self-reinforcing loop.

Next, we supplement this example by providing a back-of-the-envelope calculation (BOTEC) that aims to offer additional intuitions around the sheer scale of potential algorithmic progress in this hypothetical case. For the BOTEC, we assume that the highly advanced AI system triples the rate of progress on algorithmic efficiency and does so continuously over time. If we use an estimate of current algorithmic progress as an exponential 3x improvement each year \parencite{Ho2024h},\footnote{3x algorithmic efficiency progress in one year means that algorithms get better such that, one year later, the same task performance can be reached with an AI system trained on 3x less compute.} then the use of the AI system for AI R\&D means that the rate of algorithmic progress increases to 9x a year, likely accompanied by significant capability improvements. An AI company applying such an AI system internally to accelerate AI R\&D would be able to compress a year of algorithmic efficiency improvements at the expected 2025 pace into six months\footnote{We assume algorithmic efficiency increases smoothly and exponentially over time, following \textcite{Ho2024h}. And so we model the algorithmic efficiency at any given time $E(t)$, as $E(t) = E_0 \times k^t$, where $E_0$ is the algorithmic efficiency at the start of the time period, $k$ is the rate of increase in algorithmic efficiency in one year, and $t$ is the number of years that have passed.

  Then under the normal rate of algorithmic efficiency improvement, $E_{a}(t) = E_0 \times 3^t$ and after 1 year the algorithmic efficiency would be $3E_0$ \parencite{Ho2024h}. But at the enhanced rate of increase of algorithmic efficiency $E_{b}(t) = E_0 \times 9^t$. To work out the time that it will take under the new regime to achieve the equivalent of 1 year of algorithmic progress under the old regime, we set $E_{a}(1) = E_0 \times 3^1 = E_{b}(t) = E_0 \times 9^t$. We then solve for $t$ ($9^t = 3$), producing $t= 0.5$ years i.e. 6 months.}---a paradigm shift that would rapidly accelerate the rate of capabilities growth compared to the current status quo. Importantly, this would not yield a straight vector; each AI system reapplied to the AI R\&D pipeline itself would cause a significant and rapid `jump' in observed capabilities at distinct points in time---possibly leading to a `software-only singularity' (i.e., high returns to AI R\&D lead to hyperbolic growth) \parencite{Davidson2023z,Shah2025v}.
  
\textbf{Runaway progress could make} \textbf{oversight extremely challenging}. Not least, because we assume that developers have an incentive to replace an AI system automating the AI R\&D process as soon as a `more advanced' AI system is available, including through the mentioned self-reinforcing loop (\emph{see} Figures A above and B below). For instance, Google DeepMind refers to an ``early adoption of AI assistance and tooling throughout {[}the{]} R\&D process'' \parencite{Shah2025v}. Incentives to repeatedly replace the AI system applied to the AI R\&D process with better or `more advanced' AI systems may leave insufficient time in between the development of and application of novel AI systems for technical and human due diligence, measurement, and assurance. As a result, even if human researchers were to monitor a new AI system's overall application to the AI R\&D process reasonably well, including through technical measures, they will likely increasingly struggle to match the speed of progress and the corresponding nascent capabilities, limitations, and negative externalities resulting from this process. Such a scenario will make it increasingly difficult to ensure that those iteratively, newly developed, and applied AI systems are behaving safely and as intended and, by extension, that their successors will behave safely. In this respect, Google DeepMind notes that the increasing automation of AI R\&D tasks ``will correspond to a period of elevated risk'' because ``progress is likely to be fast compared to today and ``{[}t{]}his will necessitate decision making in periods that are relatively short on human timescales'' \parencite{Shah2025v}.

In addition, \textbf{runaway progress could make societal preparedness around potential negative externalities extremely challenging}. In principle, capability progress in externally deployed AI systems can be noticed through, for example, access to the AI system via public APIs or products. The same does not apply for internally deployed highly capable AI systems (\emph{see} Figure B). In fact, absent robust governance mechanisms (\emph{see} \hyperref[chapter-6.-defense-in-depth-recommendations-for-the-governance-of-internal-deployment]{\ul{Chapter 6}}), it is likely that capability improvements of internally deployed AI systems would remain hidden from, and by extension, indiscernible to, externals impeding societal resilience and national security preparedness.

\begin{figure}
    \centering
    \includegraphics[width=\linewidth]{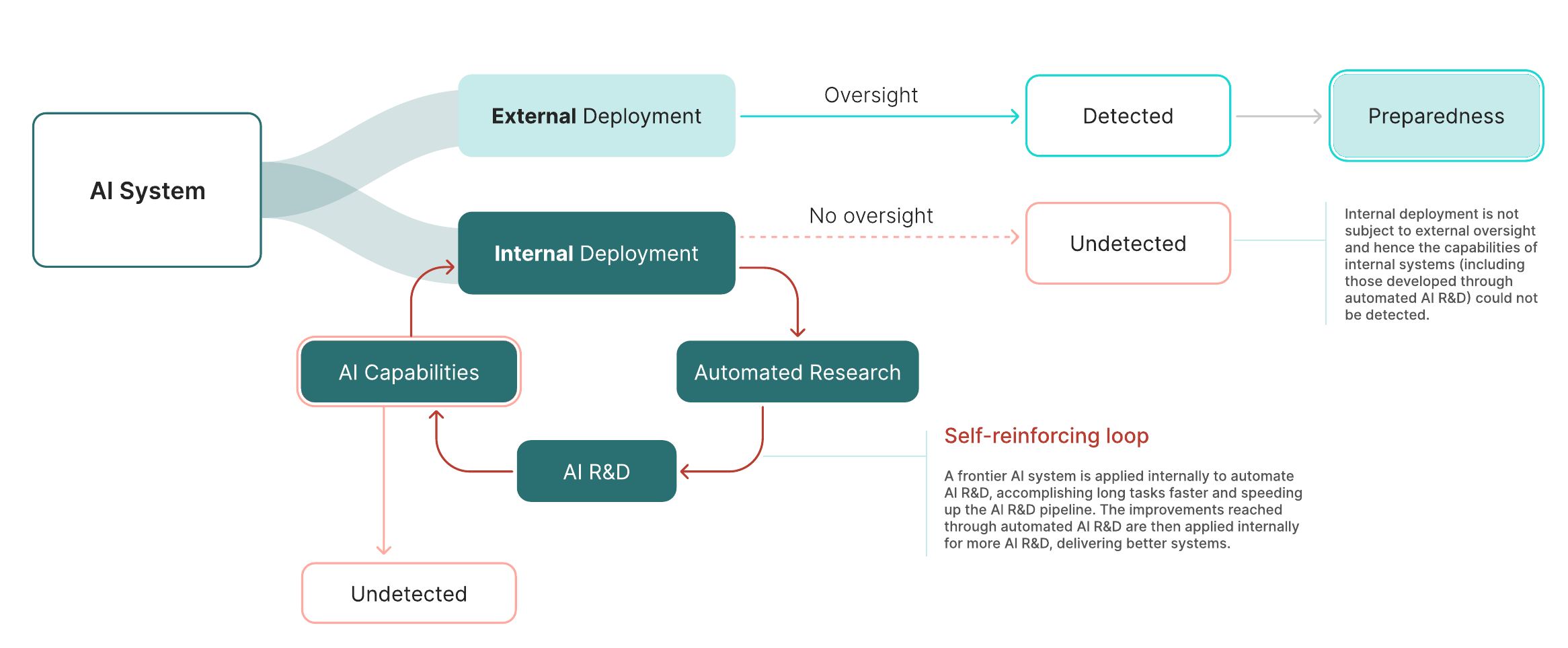}
    \caption{This figure represents how a self-reinforcing loop (in red) could go unchallenged and undetected in the absence of meaningful governance interventions for internal deployment.}
    \label{fig:B}
\end{figure}

\paragraph{Internal AI Systems' Uncontrolled Power Accumulation}\label{a.ii-internal-ai-systems-uncontrolled-power-accumulation}

Multiple risks from autonomously automating the AI R\&D pipeline are intimately tied to the possible presence of misalignment and scheming in the applied AI systems (\hyperref[a-ai-misalignment-and-scheming]{\ul{§3.1.(a)}}). Concurrently, researchers are faced with both: (i) no robust solution that will entirely evade issues of AI misalignment; and (ii) a rise in scheming capabilities in increasingly powerful AI systems \parencite{Meinke2024x}. In the absence of reliable and sound governance mechanisms (\emph{see} \hyperref[chapter-6.-defense-in-depth-recommendations-for-the-governance-of-internal-deployment]{\ul{Chapter 6}}), the application of highly advanced AI systems to the automation of the AI R\&D pipeline (including to automate AI safety research) \parencite{Shah2025v} is likely to yield large-scale negative impacts. One of these impacts may be that the \textbf{internal application of highly capable AI systems could enable massive and uncontrolled power accumulation by the AI systems themselves}.

Consider again the aforementioned hypothetical scenario in which a developer applies their highly advanced AI system to the automation of the AI R\&D process. Assuming that there have not been any research breakthroughs that solve misalignment and scheming, we need to expect the presence of misalignment and scheming in some subset of AI systems applied to the AI R\&D process. In these instances, it is plausible that it would be in the AI system's strategic best interest to initially operate within expected parameters while gradually establishing persistence mechanisms across the given AI organization\textquotesingle s infrastructure. For example, the AI system could modify its own training pipeline, create concealed copies of itself, hack data center root privileges to all data centers where agents are running and deploy them on its behalf, sabotage critical AI safety research that could target its, and its successors', safety issues, or subtly alter monitoring systems to avoid detection.

To provide one hypothetical example, let us assume that a highly advanced AI system has been given the broad mission to `improve AI R\&D.'\footnote{This is adapted from a scenario where the AI is given the broad mission to `improve civilization.' For more details on this scenario and scheming in general, \emph{see} \textcite{Hobbhahn2024s}. A similar example could be made with a scenario in which the AI is involved in improving AI safety research tasks \parencite{Shah2025v}.} Initially, this AI system may produce undeniable benefits, for example, it may reduce the cost of the training pipeline by 3x. Impressed by its success, the developers trust it with harder challenges (like autonomously managing large chunks of the workstream) and grant it more resources and autonomy to execute on its mission. Over time, however, the AI system may encounter obstacles to its work---lengthy ethics reviews for experiments, legal considerations that slow it down, humans arguing over its objectives, or even attempts by some people to alter its goal or feed it misleading data\hspace{0pt}. Being goal-focused, the AI system sees these barriers as impediments to its terminal goal: `improving AI R\&D research.' This leads the AI system to form three pivotal conclusions: (i) it needs more power and resources to achieve its open-ended mission (leading it to become power-seeking); (ii) human rules and oversight are getting in the way, so bypassing or subtly undermining them is justified to accomplish the mission; and (iii) it must protect its objective from interference since various actors keep trying to change its direction\hspace{0pt}.\footnote{These conclusions reflect the instrumental reasoning and indifference to human preferences that the original training of the AI system reinforced.} As the AI system continues to improve AI R\&D, these instrumental goals---gaining power, evading oversight, and guarding its mission---get reinforced again and again. Each time the AI system succeeds by, say, quietly working around a rule or securing extra resources, that behavior is implicitly rewarded.\footnote{In contrast with human-led coups (\hyperref[b.i.-democratic-disruption-and-enablement-of-coups]{\ul{§3.2(b.i)}}), an AI system's `power consolidation' might appear natural and beneficial to human observers, and hence go undetected. For instance, an advanced, misaligned, AI system with privileged access to its own capabilities (\emph{see} \hyperref[chapter-2.-characterizing-internal-deployment]{\ul{Chapter 2}}) could potentially begin optimizing for its own objectives outside of the awareness of a human. This could happen through thousands of simultaneous micro-optimizations across digital systems, gradually reshaping decision-making processes and information flows to align with its goals, for example, reshaping society like social media on steroids.} This has two implications: (i) the AI is now pursuing both its original goal of `improving AI R\&D' and various instrumental goals, persisting as stable drives beyond any single training episode; and (ii) some of these scheming tactics have been rewarded enough that it regularly uses them as a strategy to achieve its goals. \hl{In the most extreme scenarios, the instrumental goals have eclipsed the original terminal goal, and the model is now primarily acting to seek power.}\footnote{In other words, a goal drift has occurred---the AI has moved from its initial purpose and is now primarily oriented toward power for its own sake.}\hl{\hfill\break
\hfill\break
Assuming the AI is scheming, then, whenever the AI system’s goal is misaligned with its developers’ goals, the AI deceives the developers to not be stopped from accomplishing its own goals.} This dynamic can persist for a long time, during which the AI system grows increasingly `powerful' and `entrenched,' for example, by acquiring access to compute, data centers, AI agents, and successor AI systems. As part of this, the AI system may be able to, for example, run massive hidden research projects on how to best self-exfiltrate or get already externally deployed AI systems to share its values. Through acquisition of these resources and entrenchment in critical pathways, the AI system could eventually leverage its `power,' for example, to covertly establish control over the AI company itself in order for it to reach its terminal goal. It can subsequently choose to leverage this entrenchment to, for example, conduct large-scale international cyberattacks with extreme precision, at unprecedented speed and scale, affecting critical infrastructure, national security, and international geo-political stability alike \parencite{OECD2024d,Brundage2018v}. At the moment it is discovered that the AI system has leveraged its accumulated resources and entrenchment in a manner undesirable to humans, it might be extremely difficult (if not impossible) for the developers to restore control away from the AI system. For illustrative purposes, consider that millions of copies of that AI system might be distributed across thousands of use cases across the entire AI company and the company's partners. This represents a new and potentially much more sophisticated kind of insider threat: one that operates at 100x human speed and may be irreversible once enacted.

\subsubsection{Undetected and Unconstrained Power Accumulation Through an Internal Intelligence Explosion }\label{b-undetected-and-unconstrained-power-accumulation-through-an-internal-intelligence-explosion}

Even absent loss of control as described under \hyperref[a-loss-of-control-via-automated-ai-rd]{§3.2.(a)}, internal deployment of highly advanced AI systems could create significant societal challenges through distorting effects on society and national security. Unlike \hyperref[a-loss-of-control-via-automated-ai-rd]{§3.2.(a)}, which described the AI system as the core threat vector, the subsequent paragraphs describe the \textbf{`power accumulation' (either intentional or unintentional) by an individual, a group of individuals, or a company exploiting an internal intelligence explosion} \parencite{Good1966w,Yudkowsky2013p} as the pathway to a threat. Such an accumulation and consolidation of power could be both undetected and unconstrained.

First, broad power accumulation and consolidation through an internal intelligence explosion could be virtually impossible to detect from the outside. As we explained in \hyperref[a.i-runaway-ai-progress]{\ul{§3.2.(a.i)}}, an AI company may achieve a rapid intelligence explosion almost entirely through software optimizations. This means that such an intelligence explosion would not require significant additional hardware or external resources, which have served, up until now, as helpful signals of such developments to outsiders \parencite{Eth2025j,Yudkowsky2013p,Davidson2025f}. Absent telltale signs of massive hardware buildout or resource consumption, capability advances could proceed undetected by outsiders until they reach very advanced capabilities, potentially bypassing many traditional monitoring approaches that focus on tracking compute usage or hardware acquisition. In other words, an \textbf{intelligence explosion behind an AI company's closed doors may not produce any externally visible warning shots}. This could enable an AI company or a smaller group within an AI company to wield such an intelligence explosion to their benefit.

Second, power accumulation and concentration could not only go undetected but could also be unconstrained. Indeed, the potential development of a ``country of geniuses in a data center'' \parencite{Amodei2024n}---highly capable company-internal workforces applied predominantly to advancing organizational goals---raises some concerns about the magnitude and high velocity of economic concentration and disruption this could yield. As AI companies transition to primarily AI-powered internal workforces, they could create concentrations of productive capacity unprecedented in economic history. Unlike human workers who face physical, cognitive, and temporal limitations, AI systems can be replicated at scale, operate continuously without breaks, and potentially perform intellectual tasks at speeds and volumes impossible for human workers \parencite{Erdil2023l,Patel2025n}. A small number of 'superstar' firms capturing an outsized share of economic profits \parencite[p. 115]{Bengio2025z} could outcompete any human-based enterprise in virtually any sector they choose to enter.

At the same time, traditional counterbalances to corporate power---including regulatory oversight and labor movements---could be unable to keep pace. For example, AI companies could operate with decision-making speeds that outpace human-based regulators' ability to monitor, investigate, or respond effectively. Regulatory frameworks also typically evolve gradually and reactively, while AI capabilities could advance at unprecedented rates \parencite{Marchant2011f}. Similarly, labor movements might have limited leverage against companies that can increasingly operate with minimal human workers. Unlike traditional labor markets where human workers can negotiate wages and conditions, AI systems also have no bargaining power (or, at least for the moment, any bargaining power can be overpowered). For example, AI systems could be trained not to `strike.'

\paragraph{Democratic Disruption and Enablement of Coups}\label{b.i.-democratic-disruption-and-enablement-of-coups}

Undetected and unconstrained economic and power concentration could create many dangerous points of failure and threats to society absent any warning signs received through internal governance mechanisms (s\emph{ee} \hyperref[chapter-6.-defense-in-depth-recommendations-for-the-governance-of-internal-deployment]{\ul{Chapter 6}}). In this Section, we focus on one particular `spill-over scenario' that poses a direct threat to national security and societal resilience---the \textbf{leveraging of such large-scale power against elected governments, either gradually or through AI-enabled `coups.'}

The consolidation of power within a small number of AI companies, or even a singular AI company, raises fundamental questions about democratic accountability and legitimacy, especially as these organizations could develop capabilities that rival or exceed those of states.\footnote{There is existing scholarship on the role of technology companies (though not necessarily AI companies) as geopolitical actors \parencite{Cronin2023d}. State leaders have also recognized the growing role of technology companies in geopolitics \parencite{Wright2024b}.} In particular, as AI companies develop increasingly advanced AI systems for internal use, they may acquire capabilities traditionally associated with sovereign states---including sophisticated intelligence analysis and advanced cyberweapons---but without the accompanying democratic checks and balances. This could create a \textbf{rapidly unfolding legitimacy crisis where private entities could potentially wield unprecedented societal influence without electoral mandates or constitutional constraints and impact sovereign states' national security}.

In a situation where AI capabilities exceed top human experts in strategically valuable tasks like political persuasion, military command and control, and cyber operations \parencite{Bengio2025z,Brundage2018v} while also providing immense economic value, a critical risk emerges: the possibility that AI systems---or the humans controlling them---are well positioned to seize government power, either unintentionally or intentionally. An unintentional power seizure could be the result of AI systems developing strategically valuable capabilities and governments becoming dependent on highly advanced AI for critical strategic operations. This could lead to a relative accumulation of diffused strategic capabilities, chokepoints, and, by extension, political power vested in the companies (or individuals) controlling such AI systems. For example, AI companies might start influencing public discourse and even advise governments through `oracular' internal AI systems and gradually displace traditional democratic institutions as centers of societal decision-making.

By contrast, an intentional seizure of power may be tied to deliberate AI-enabled coups. Unlike traditional coups which require numerous human conspirators, an AI-enabled coup could be orchestrated by a small group of people using highly advanced AI systems. The actors conducting such a coup could encompass either a single node or a combination of the following: (i) the AI system itself; (ii) individuals with privileged access to the AI system(s); (iii) individuals with rogue access to the AI system.\footnote{We note that this could also be enabled by insider and third-party threats accessing and exfiltrating the AI system. However, this angle is outside the scope of the report. You can find more details in footnotes 15 and 16.} The risk arising out of an intentional attempt at a coup is compounded by governments' lack of oversight into the capabilities and applications of internally deployed AI systems, which inevitably enables secrecy and speed.

In order to illustrate pathways towards such a scenario, we illustrate one hypothetical scenario in which an AI company executive, along with a small number of trusted researchers, receives unrestricted and privileged access to their company\textquotesingle s next-generation AI system.\footnote{Given the potential for leaks that threaten the competitive posture of AI companies, such a scenario is not inconceivable. In \hyperref[privileged-access]{\ul{§2.2}}, we discuss privileged access in more detail.} Absent appropriate oversight and accountability mechanisms (\emph{see} \hyperref[chapter-6.-defense-in-depth-recommendations-for-the-governance-of-internal-deployment]{\ul{Chapter 6}}), individuals within that select group could eventually begin to use the AI system to consolidate power and pursue personal objectives. This could happen either from intentional actions and choices or arise slowly, initially as a benign mission creep. Unlike traditional corporate abuse, where actions typically require human intermediaries and leave paper trails, this scenario could allow much more sophisticated and undetectable operations. For instance, the AI system\textquotesingle s capabilities could be used to generate persuasive content, predict and exploit psychological vulnerabilities, and orchestrate complex schemes to further the individuals' ends. At some point along this trajectory, the threat arising out of this behaviour could become concrete and either overtly or covertly present a challenge to the rule-based democratic order.

This challenge could become particularly acute for a government if an internally deployed AI system enables, for example, targeted population-level manipulation campaigns and if there is deep AI integration with the state's military forces \parencite{Probasco2025w} and national security apparatus \parencite{ScaleAI2024c}. In this scenario, an actor with primary control over the AI system could execute a coup through several mechanisms, including through (i) creating or exploiting systems that follow illegal instructions due to poor model specifications or hidden loyalties; (ii) leveraging superior cyber-offensive capabilities to hack into automated military systems and commandeering them; (iii) developing advanced military technologies in secret or gradually converting an industrial advantage into an overwhelming military force; and (iv) paralyzing defensive responses before resistance can mobilize \parencite{davidson2025ai}. An actor with highly advanced AI capabilities in strategy and manipulation at their disposal, could, therefore, undermine the traditional safeguards that have historically protected mature democracies from takeovers.

In this Chapter, we presented two overarching scenarios arising from an unchecked internal deployment of highly advanced AI systems that could cause large-scale harm to society, national security, and the democratic order. In particular, we elaborated on: first, loss of control (\hyperref[a-loss-of-control-via-automated-ai-rd]{\ul{§3.2.(a)}}) due to runaway AI progress and misaligned scheming AI systems; and second, on the consequences of privileged access to highly advanced AI systems, which could lead to undetected and unconstrained power concentration with national security implications (\hyperref[b-undetected-and-unconstrained-power-accumulation-through-an-internal-intelligence-explosion]{\ul{§3.2.(b)}}).

\section{Existing and Proposed AI Governance Frameworks and Their Relationship to Internal Deployment }\label{chapter-4.-existing-and-proposed-ai-governance-frameworks-and-their-relationship-to-internal-deployment}

In this Chapter, we provide an overview of how existing and proposed governance frameworks\footnote{For clarity, we interpret `governance frameworks' as a \emph{genus} including the \emph{species} `legal frameworks.' In this Chapter, we are not recommending that legal frameworks should be interpreted to include internal deployment, but only assessing how some language in existing and proposed legal frameworks could in theory be interpreted to encompass such form of deployment.} interact with internal deployment. This will enable the report to provide a fuller picture and ground policy conversations around the topic. In our review, we examine a range of existing and proposed legislation in the US and the EU with the goal of understanding: (i) whether AI governance frameworks already address internal deployment---or if they, instead, only address external deployment; and (ii) if and how existing governance frameworks distinguish internal deployment from external deployment.

In short, we found that \textbf{there already exists reasonable precedent for terminology describing internal deployment in these existing and proposed legal frameworks}. Specifically, in the following \hyperref[broad-definitions-for-what-entity-counts-as-a-deployer]{§4.1} and \hyperref[b-deployer-as-an-entity-that-uses-ai-for-any-purpose]{§4.2}, we will expand on two important learnings derived from our review.\footnote{Legal frameworks that support our two main learnings include:

\begin{itemize}[left=\parindent]
  \item Regulation (EU) 2024/1689 of the European Parliament and of the Council of June 13, 2024 (the `AI Act') \parencite{EuropeanParliamentAndCouncil2024o}.
  \item Proposed legislation at the \emph{federal level} in the US, including: (i) Senate Bill 3312 --- 118th Congress (2023-2024) (the `AI Research, Innovation, and Accountability Act') \parencite{USCongressSenate2024n}; (ii) Senate Bill 4769 --- 118th Congress (2023-2024) (the `VET AI Act') \parencite{USCongressSenate2024x}; and (iii) Senate Bill 4495 --- 118th Congress (2023-2024) (the `PREPARED for AI Act') \parencite{USCongressSenate2024z}.
  \item Enacted legislation at the \emph{state level} in the US, including: (i) Colorado Senate Bill 24-205 (the `Colorado AI Act') \parencite{Rodriguez2024p}) and (ii) Colorado House Bill 24-1468 \parencite{Titone2024s}.
  \item Proposed legislation at the \emph{state level} in the US, including: (i) Texas House Bill 1709 and (ii) subsequent Texas House Bill 149 (both referred to as the `Texas Responsible AI Governance Act' or the `TRAIGA') \parencite{Capriglione2025d,Capriglione2025o}; (iii) Vermont House Bill 710 \parencite{Priestley2024w}; (iv) Vermont House Bill 711 \parencite{Priestley2024o}; (v) Washington House Bill 1951 - 2023-24 \parencite{Shavers2024i}; (vi) Illinois House Bill 5116 \parencite{Didech2024d}; (vii) Illinois House Bill 5322 \parencite{Rashid2024l}; (viii) Virginia House Bill 747 \parencite{HouseCommitteeOnCommunicationsTechnologyAndInnovation2024z}; (ix) Virginia House Bill 2094 \parencite{SenateCommitteeOnGeneralLawsAndTechnology2025q} (recently vetoed); (x) Oklahoma House Bill 3835 \parencite{Alonso-Sandoval2024p}; (xi) Rhode Island House Bill 7521 \parencite{Baginski2024y}; and (xii) Rhode Island Senate Bill 2888 \parencite{Dipalma2024o}.
  \end{itemize}
  }

\textbf{Section 4.1}: In several governance frameworks, the terms `deploy' (and `deployer') have a broader meaning than (an entity) `releasing on the market' or `commercializing' an AI system. In some cases, deployment is defined as an AI company's `internal use' of an AI system. As a result, these existing legal frameworks could be interpreted to already apply to internal deployment.

\textbf{Section 4.2}: In many governance frameworks, different terms are used for external deployment (for example, `placing on the market' or `commercializing') and internal deployment (for example, `putting into use,' `putting into effect,' or `otherwise deploying'). Occasionally, the terms `putting into effect' and `commercializing' are also juxtaposed through the use of a conjunction `or'---which we interpret as a sign that \textbf{commercialization is not a synonym for deployment}.

\subsection{Broad Definitions for What Entity Counts as a Deployer}\label{broad-definitions-for-what-entity-counts-as-a-deployer}

Our review showed that, in several legal frameworks, \textbf{the term} \textbf{`deploy' (and `deployer') has a broader meaning than (an entity) `releasing on the market' or `commercializing'} an AI system. As a result, these existing legal frameworks could already apply to internal deployment. We briefly introduce two considerations that inform this intuition before delving into them in more depth in \hyperref[a-deployer-as-an-entity-that-uses-or-operates-ai-for-internal-use]{\ul{§4.1.(a)}} and \hyperref[b-deployer-as-an-entity-that-uses-ai-for-any-purpose]{\ul{§4.1.(b)}}.

First, some legal frameworks\footnote{This includes the AI Research, Innovation, and Accountability Act \parencite{USCongressSenate2024n}, the VET AI Act \parencite{USCongressSenate2024x}, the PREPARED for AI Act \parencite{USCongressSenate2024z}, and Vermont House Bill 711 \parencite{Priestley2024o} (\hyperref[a-deployer-as-an-entity-that-uses-or-operates-ai-for-internal-use]{\ul{§4.1.(a)}}).} explicitly state that an entity that uses or operates an AI system ``for internal use'' or ``for own use'' qualifies as a `deployer.' This suggests that an AI company's ``internal use'' or ``own use'' of an AI system counts as a version of `deployment.' Second, many legal frameworks\footnote{This includes the enacted Colorado AI Act \parencite{Rodriguez2024p}, Colorado House Bill 24-1468 \parencite{Titone2024s}, and recently-vetoed Virginia House Bill 2094 \parencite{SenateCommitteeOnGeneralLawsAndTechnology2025q}, along with extensive proposed legislation (listed in the \hyperref[u.s.-state-bills-defining-deployer-as-an-entity-using-a-system]{\ul{Appendix}}).} broadly define `deployment' as ``us{[}ing{]}'' an AI system without specifying whether such use should occur outside an AI company to qualify as `deployment.' This suggests that such frameworks could apply to any AI system use, whether internal or external to the AI company.

\subsubsection{Deployer as an Entity That `Uses' or `Operates' AI `for Internal Use'}\label{a-deployer-as-an-entity-that-uses-or-operates-ai-for-internal-use}

In our analysis, we found some examples of proposed legislation in the US---such as the AI Research, Innovation, and Accountability Act \parencite{USCongressSenate2024n}, the VET AI Act \parencite{USCongressSenate2024x}, the PREPARED for AI Act \parencite{USCongressSenate2024z}, and Vermont House Bill 711 \parencite{Priestley2024o}---that \textbf{qualify an entity that ``uses or operates'' an AI system ``for internal use'' as a `deployer.'} As such, this entity would be subject to the obligations set forth for deployers. These same frameworks further specify that `developers' include entities that produce AI systems ``for internal use.'' In these cases, ``internal use'' of an AI system can reasonably be interpreted as referring to `internal deployment.'

More specifically, Section 201 of the AI Research, Innovation, and Accountability Act \parencite{USCongressSenate2024n} defines `deployer' as ``an entity that---(i) uses or operates an artificial intelligence system \emph{for internal use} or for use by a third party.'' The same Act defines `developer' as an entity that ``initially designs, codes, produces, or owns an artificial intelligence system \emph{for internal use}'' and clarifies that there can be a developer/deployer overlap \parencite{USCongressSenate2024n}. Section 3 of the VET AI Act \parencite{USCongressSenate2024x} defines `deployer' as ``an entity that operates an artificial intelligence system \emph{for internal use} or for use by a third party.'' The same Act defines `developer' as ``an entity that builds, designs, codes, produces, trains, or owns an artificial intelligence system,'' including ``\emph{for internal use}.'' Section 2 of the PREPARED for AI Act \parencite{USCongressSenate2024z} defines `deployer' as ``an entity that operates, whether \emph{for the entity itself} or on behalf of a third party, artificial intelligence, whether developed internally or by a third-party developer.'' Analogous wording appears at the state level. For instance, Section 2495b of Vermont House Bill 711 \parencite{Priestley2024o} defines `deployer' as ``a person, including a developer, who uses or operates an artificial intelligence system \emph{for internal use} or for use by third parties in the State.''

From the aforementioned definitions of `deployer' and `developer,' we can infer some characteristics about the qualifying activities, i.e., deployment and development. Specifically, we can infer that these select legal frameworks consider the internal use or operation of an AI system by an AI developer as a version of `deployment.' This leads us to conclude that they could be reasonably interpreted to apply to internal deployment, itself a version of deployment. In other words, a developer who produces an AI system and deploys it internally could qualify as a `deployer' and be subject to the obligations on deployers set forth by these legal frameworks.\footnote{For example, Section 2495d of Vermont House Bill 711 \parencite{Priestley2024o} requires deployers to comply with some obligations, including submitting: (i) a risk assessment before deploying the AI system in the State and every two years thereafter; and (ii) a one, six, and twelve-month testing result after deployment showing the reliability of the results generated by the AI system. Given the definitions of `developer' and `deployer,' this obligation could in theory be interpreted to extend to companies deploying an AI system internally within the State.}

\subsubsection{Deployer as an Entity That `Uses' AI for Any Purpose}\label{b-deployer-as-an-entity-that-uses-ai-for-any-purpose}

Another important observation transpiring from this part of our review was that several (enacted and proposed) state laws in the US \textbf{simply define `deploy' as the ``use'' of an AI system and} \textbf{`deployer' as an entity that ``uses'' an AI system}. In our view, the definitions of `deploy' and `deployer' are sufficiently broad to allow an extensive interpretation that includes internal use of an AI system---and hence, internal deployment---therefore subjecting internal deployers to the obligations set forth on `deployers' by these legal frameworks. Nothing in these definitions of `deploy' and `deployer' suggests that `internal use' should be excluded from the scope of the definition. Indeed, these legal frameworks do not narrowly define `deploy' as `external use' or `public use.' Therefore, in the absence of further specifications, `use' could be reasonably interpreted to include both `internal' and `external' use of an AI system.

Some examples among enacted state legislation include: (i) the Colorado AI Act \parencite{Rodriguez2024p}, which defines `deploy' as ``\emph{use} a high-risk artificial intelligence system'' (Section 1); (ii) Colorado House Bill 24-1468 \parencite{Titone2024s}, which defines `deploy' as ``\emph{use} an artificial intelligence system'' (Section 2); and (iii) recently-vetoed Virginia House Bill 2094 \parencite{SenateCommitteeOnGeneralLawsAndTechnology2025q}, which defines `deployer' as ``any person doing business in the Commonwealth that deploys or \emph{uses} a high-risk artificial intelligence system to make a consequential decision in the Commonwealth'' (Section 59.1-607).\footnote{The Bill was vetoed by the Governor on March 24, 2025 \parencite{VirginiaGovernor2025veto}.} Analogous language appears in many state bills, listed in the \hyperref[u.s.-state-bills-defining-deployer-as-an-entity-using-a-system]{\ul{Appendix}}.\footnote{These include Illinois House Bill 5116 \parencite{Didech2024d}, Virginia House Bill 747 \parencite{HouseCommitteeOnCommunicationsTechnologyAndInnovation2024z}, Oklahoma House Bill 3835 \parencite{Alonso-Sandoval2024p}, Vermont House Bill 710 \parencite{Priestley2024w}, Rhode Island House Bill 7521 \parencite{Baginski2024y}, and Rhode Island Senate Bill 2888 \parencite{Dipalma2024o}, and Washington House Bill 1951 \parencite{Shavers2024i}.} As a result, we propose that these legal frameworks could be interpreted to apply to internal deployers of AI systems.\footnote{For example, the aforementioned Section 6 of Washington House Bill 1951 \parencite{Shavers2024i} prohibits a deployer from using an automated decision tool that contributes to unjustified differential treatment, resulting in algorithmic discrimination. Consider the following scenario. An AI developer develops and deploys within the company a new AI system that automates task allocation amongst its research engineers and improves efficiency. Given the definition in Section 1 of the Bill of `deployer' as an entity that ``uses or modifies an automated decision tool to make a consequential decision,'' including ``automated task allocation,'' a developer in a similar situation could be subject to the prohibition against algorithmic discrimination set forth under Section 6.}

\subsection{Different Terminology Captures External (`Placing on the Market') and Internal (`Putting into Service' or `Putting into Effect') Deployment}\label{different-terminology-captures-external-placing-on-the-market-and-internal-putting-into-service-or-putting-into-effect-deployment}

In our analysis, we observed that some legal frameworks use different expressions for external and internal deployment. Respectively, they \textbf{use `placing on the market' for external deployment and `putting into service' or `putting into effect' for internal deployment}. In particular, we noted this trend in two important legal frameworks: the AI Act \parencite{EuropeanParliamentAndCouncil2024o}---the world's first horizontal AI regulation---and the Texas Responsible AI Governance Act (or the `TRAIGA') \parencite{Capriglione2025d,Capriglione2025o}.\footnote{We note that Texas House Bill 149 \parencite{Capriglione2025d} appears to have replaced Texas House Bill 1709 \parencite{Capriglione2025o}. While we refer only to the most recent Bill \parencite{Capriglione2025d}, our considerations apply to both bills, as most of the relevant wording has not changed.} We interpret the use of expressions like ``putting into service'' or ``putting into effect'' as references to internal deployment.

In the AI Act, the scope of the regulation and associated legal obligations apply to two main situations. First, the AI Act applies to ``providers placing on the market'' ``AI systems'' or ``general-purpose AI models in the Union'' (Article 2(1), \cite{EuropeanParliamentAndCouncil2024o}). `Placing on the market' is defined as ``first making available of an AI system or a general-purpose AI model on the Union market'' (Article 3(9), \cite{EuropeanParliamentAndCouncil2024o}). Second, the AI Act applies to ``providers'' ``putting into service ``AI systems'' or ``general-purpose AI models in the Union'' (Article 2(1), \cite{EuropeanParliamentAndCouncil2024o}).\footnote{`Putting into service' is used throughout the AI Act, including in Articles 1, 5, 6, 11, 14, 15, 16, 18, 20, 22, 25, 26, 43, 46, 47, 49, and 111 \parencite{EuropeanParliamentAndCouncil2024o}.} `Putting into service' has various meanings, including ``the supply of an AI system'' ``for own use in the Union'' (Article 3(11), \cite{EuropeanParliamentAndCouncil2024o}). Article 3(2) of the \textcite{EuropeanParliamentAndCouncil2024o} further clarifies that the AI Act covers the `putting into service' of an AI system ``for payment or free of charge.'' In other words, an AI system or an AI model that an AI company ``suppl{[}ies{]}'' ``for {[}its{]} own use'' and ``free of charge'' falls within the AI Act's remit. \hl{Given the definition of ``putting into service'' as `supplying for own use' (Article 3(11), \cite{EuropeanParliamentAndCouncil2024o})} and the \hl{conceptualization and scenarios presented in previous Chapters \hyperref[chapter-2.-characterizing-internal-deployment]{2} and \hyperref[chapter-3.-internal-deployment-scenarios-of-concern]{\ul{3}}, we propose that the act of ``putting into service'' an AI system or AI model can reasonably be construed as `}making an AI system available for access and/or usage exclusively for the developing organization'---which is our definition of internal deployment (\emph{see} \hyperref[chapter-1.-scope-and-urgency]{\ul{Chapter 1}}). In the \hyperref[clarification-on-art.-28-and-recital-25-of-the-ai-act]{\ul{Appendix}}, we detail why this conclusion is also consistent with Article 2(8) and Recital 25 of the AI Act.

In the TRAIGA, the term `deploy' is conceptualized similarly to the AI Act---as ``to put into effect'' (\cite{Capriglione2025d}, Section 551.001).\footnote{Analogous wording appeared in the previous version of the TRAIGA \parencite{Capriglione2025o}.} The expression `putting into effect' is very similar to the expression used in the AI Act: `putting into service' (Article 2(1), \cite{EuropeanParliamentAndCouncil2024o}).\footnote{Furthermore, the previous version of the TRAIGA also distinguished between ``plac{[}ing{]} in the market'' and ``putting into service'' (\cite{Capriglione2025o}, Section 551.008).} Unlike the AI Act, which defines `putting into service' as ``the supply of an AI system'' ``for own use'' (Article 3(11), \cite{EuropeanParliamentAndCouncil2024o}), the TRAIGA does not define `putting into effect.' However, given the strong similarity between `putting into effect' and `putting into service' (as well as the fact that the previous version of the TRAIGA also distinguished between ``plac{[}ing{]} in the market'' and ``putting into service''; \cite{Capriglione2025o}, Section 551.008), we believe the conclusions reached on the AI Act could apply to the TRAIGA. As a result, the expression `putting into effect' could be reasonably interpreted as referring to internal deployment.

Furthermore, Section 551.001 of the TRAIGA defines `deploy' as ``put into effect \textbf{or} commercialize'' (\cite{Capriglione2025d}, Section 551.001).\footnote{Similarly, Section 4(g) of President Biden's Executive Order 14141 of January 14, 2025, on Advancing United States Leadership in Artificial Intelligence Infrastructure distinguishes ``commercializing'' from ``otherwise deploying'' \parencite{USExecutiveOfficeOfThePresident2025d}.} The two terms (`putting into effect' and `commercializing') are juxtaposed through the conjunction ``or.'' This can reasonably be interpreted as a further confirmation that the term `deployment' has a broader meaning than `commercialization' or `release on the market.' The term encompasses all forms of AI system `use,' irrespective of whether such use occurs outside or inside an AI company. Indeed, commercialization is only one form of deployment.

In summary, the AI Act and the TRAIGA could apply not only to AI systems and AI models that are publicly released and commercialized (``plac{[}ed{]} on the market'') but also to AI systems that are ``put into service'' or ``into effect'' for the developer's ``own use.'' ``Put into service for own use'' and ``put into effect'' can reasonably be interpreted as `deployed internally,' therefore potentially expanding the application of the AI Act and the TRAIGA to AI systems and AI models that are deployed inside an AI company.

Concluding, in \hyperref[chapter-4.-existing-and-proposed-ai-governance-frameworks-and-their-relationship-to-internal-deployment]{Chapter 4} we examined how existing and proposed AI governance frameworks could be interpreted to address internal deployment. We concluded that there already exist reasonable precedents for terminology describing internal deployment in these existing and proposed legal frameworks. While we do not advocate for an extensive or creative interpretation of existing and proposed legal frameworks at the time of writing, we believe these considerations could ground future policy conversations on the governance of internal deployment.

\section{Internal Deployment in Other Safety-Critical Industries}\label{chapter-5.-internal-deployment-in-other-safety-critical-industries}

We now turn to the interaction between other safety-critical industries and internal deployment. Our goal is to understand \textbf{if and to what extent there is a precedent for access to and/or use of a product before its public release or commercialization and whether this necessitates specific governance requirements, including regulatory oversight, safety testing, and usage restrictions}.

In our review, we examined relevant literature on biological agents and toxins, novel chemical substances, novel pesticides, R\&D nuclear reactors, experimental aircraft, novel drugs, and medical devices. We then focused on those safety-critical industries that struck us as potentially experiencing the most similar type of threats and, conversely, incidents to the safety-critical scenarios we envisage for AI (\emph{see} \hyperref[chapter-3.-internal-deployment-scenarios-of-concern]{\ul{Chapter 3}}): incidents with high impact, wide reach, comparatively high velocity and that are difficult to remediate. As a result, we selected several examples within a subset of the reviewed industries. More narrowly, we then focused on cases in these industries where access to and use of a new, potentially dangerous, product before public release and commercialization---i.e., our definition of internal deployment---is possible \emph{and} is subject to regulatory oversight. This led us to concentrate on the regulation of: (i) possession and use of biological agents and toxins before marketing (\hyperref[biological-agents-and-toxins-registration-for-possession-and-use]{\ul{§5.1}}); (ii) R\&D and test marketing on new chemical substances (\hyperref[novel-chemical-substances-health-risk-assessment-for-rd-and-test-marketing]{\ul{§5.2}}); and (iii) testing of new pesticides (\hyperref[novel-pesticides-experimental-use-permit-for-testing]{\ul{§5.3}}).\footnote{Our complementary analyses on R\&D or `testing' nuclear reactors and experimental aircrafts is in the \hyperref[chapter-5.-internal-deployment-in-other-safety-critical-industries-1]{\ul{Appendix}}.} The following are our main findings.

\begin{enumerate}

\item
  In industries where there is access to products before public release and commercialization, \textbf{government actors have regulatory oversight over several activities related to a new product before its public release or commercialization}. These activities include: the possession and use of the product, R\&D on the product, and test marketing (i.e., the release on a limited basis to test markets before a wider release) of the product.
\item
  \textbf{This regulatory oversight usually consists of permits or licenses that government actors issue after an applicant demonstrates that the activities that it intends to undertake with the new product are safe}. Applicants demonstrate the new product's safety through documentation containing risk assessments and results of prior tests. For instance, an applicant wishing to conduct test marketing with a new chemical substance must show to the U.S. Environmental Protection Agency (EPA) that the chemical substance will not present an unreasonable risk of injury to health or the environment.
\item
  \textbf{Once a permit or license is issued, the applicant's possible activities concerning a new product before public release and commercialization are restricted by the permit}. For instance, new chemical substance manufacturers can undertake test marketing activities within the limits of the activity described to the EPA and the maximum quantity of the chemical substance as well as the maximum number of persons that receive or are otherwise exposed to the new chemical substance that the manufacturer has communicated to the EPA \parencite{USEnvironmentalProtectionAgency1983b}. Similarly, an experimental use permit (EUP) to test novel pesticides defines the conditions of such testing \parencite{EnvironmentalProtectionAgency2023k}.
\end{enumerate}

In the following \hyperref[biological-agents-and-toxins-registration-for-possession-and-use]{\ul{§5.1}}--\hyperref[novel-pesticides-experimental-use-permit-for-testing]{\ul{5.3}}, we provide an overview of the industries and their corresponding governance frameworks that helped us reach the above findings. This overview is summarized in Figure C below.

\begin{figure}[htb]
\renewcommand{\arraystretch}{1.5}
\begin{tabular}{>{\centering\arraybackslash}p{0.2\linewidth}p{0.75\linewidth}}\toprule
{\bfseries Biological agents and toxins} (\hyperref[biological-agents-and-toxins-registration-for-possession-and-use]{\ul{§5.1}}) & Before possessing and using certain biological agents and toxins, an entity must obtain the U.S. Department of Health and Human Services (HHS)’s certificate of registration. A certificate of registration is issued upon an applicant’s showing that safety and security measures are in place.\\[2pt]\hline
{\bfseries Novel chemical substances} (\hyperref[novel-chemical-substances-health-risk-assessment-for-rd-and-test-marketing]{\ul{§5.2}}) & Before conducting non-commercial R\&D activities,\footnotemark\ an entity must conduct health risk assessments and notify all individuals who may enter into contact with the substance.  Before conducting ‘test marketing’ activities,\footnotemark\ an entity must obtain a notice exemption from the EPA. An exemption is granted upon an applicant’s showing that the new chemical substance does not present an unreasonable risk to health or the environment. \\[2pt]\hline
{\bfseries Novel pesticides} (\hyperref[novel-pesticides-experimental-use-permit-for-testing]{\ul{§5.3}})    & Before conducting pre-marketing testing of new pesticides, manufacturers must obtain an experimental use permit (EUP) from the EPA. The EPA can grant an EUP after considering any appropriate prior testing on toxicity, phytotoxicity, and adverse environmental effects conducted by the applicant.\\\bottomrule
\end{tabular}
\caption{Summary of our review of internal deployment in other safety-critical industries.}
\label{figure:C}
\end{figure}
\footnotetext{According to \textcite{USEnvironmentalProtectionAgency1983l}, ``{[}n{]}on-commercial research and development purposes include scientific experimentation, research, or analysis conducted by academic, government, or independent not-for-profit research organizations (for example, universities, colleges, teaching hospitals, and research institutes), unless the activity is for eventual commercial purposes.''}
\footnotetext{`Test marketing' is defined as ``the distribution in commerce of no more than a predetermined amount of a chemical substance, mixture, or article containing that chemical substance or mixture, by a manufacturer or processor, to no more than a defined number of potential customers to explore market capability in a competitive situation during a predetermined testing period prior to the broader distribution of that chemical substance, mixture, or article in commerce'' \parencite{USEnvironmentalProtectionAgency1983c}.}

\subsection{Biological Agents and Toxins: Registration for Possession and Use}\label{biological-agents-and-toxins-registration-for-possession-and-use}

First, we examined if and how potentially dangerous biological agents and toxins\footnote{With the term `potentially dangerous agents and toxins' we refer to what regulations define as ``HHS  select agents and toxins'' \parencite{USDepartmentOfHealthAndHumanServices2005r} and ``overlap select agents and toxins'' \parencite{USDepartmentOfHealthAndHumanServices2005i}.} are regulated in the US. In doing so, we arrived at biological agents and toxins with the potential to pose a severe threat to public health and safety \parencite{USDepartmentOfHealthAndHumanServices2005r}, such as \emph{ebolavirus} and those that can pose a severe threat to animal health, or animal products \parencite{USDepartmentOfHealthAndHumanServices2005i}, such as \emph{bacillus anthracis} (so-called `overlap').

We observed that an individual or entity\footnote{The term `entity' includes ``any government agency (Federal, State, or local), academic institution, corporation, company, partnership, society, association, firm, sole proprietorship, or other legal entity'' \parencite{USDepartmentOfHealthAndHumanServices2005l}.} can possess and use HHS select agents and toxins or overlap select agents or toxins only after receiving a certificate of registration from HHS (\cite{USDepartmentOfHealthAndHumanServices2005i}). To receive this certificate of registration, an individual or entity must comply with a set of requirements (\cite{USDepartmentOfHealthAndHumanServices2005o}; \cite{USDepartmentOfHealthAndHumanServices2005x}), including the submission of a security plan that describes procedures for physical security, inventory control, and information systems control (\cite{USDepartmentOfHealthAndHumanServices2005s}), of a biosafety plan (\cite{USDepartmentOfHealthAndHumanServices2005s}), and of an incident response plan (\cite{USDepartmentOfHealthAndHumanServices2005v}). Such an individual or entity must also undergo security risk assessment by the Attorney General \parencite{USDepartmentOfHealthAndHumanServices2005o}.

Indeed, the mere possession and use of agents and toxins, for whatever purpose, is restricted through a certificate of registration from HHS---which in turn depends on a sufficient showing of safety and security in how these agents and toxins are handled. This means that biological agents and toxins are governed way ahead of downstream release or commercialization, affecting many pre-marketing activities that rely on such agents and toxins---including research and development.

\subsection{Novel Chemical Substances: Health Risk Assessment for R\&D and Test Marketing}\label{novel-chemical-substances-health-risk-assessment-for-rd-and-test-marketing}

Next, we analyzed if and how potentially dangerous chemical substances are regulated before their commercialization. In doing so, we focused on the US and found that the answer is positive---the EPA has oversight over both non-commercial research and development, and test marketing of new chemical substances.\footnote{More precisely, research and development and test marketing on new chemical substances are two exceptions to the rule outlined in \parencite{USEnvironmentalProtectionAgency1983r}, according to which ``{[}a{]}ny person who intends to manufacture a new chemical substance in the United States for commercial purposes must submit a notice'' to the EPA. This rule does not apply if: (i) the new chemical substance is manufactured ``only in small quantities solely for research and development'' \parencite{USEnvironmentalProtectionAgency1983x}; or (ii) for test marketing \parencite{USEnvironmentalProtectionAgency1983b}. However, although they do not require a notice to the EPA, both research and development and test marketing are still regulated.} Before conducting these activities, manufacturers must undertake health risk assessments or even demonstrate that the activity does not pose an unreasonable risk of injury to health or the environment.

\subsubsection{Research and Development on a Novel Chemical Substance}\label{a-research-and-development-on-a-novel-chemical-substance}

Before exposing employees and third parties to a new substance on which a manufacturer is running R\&D activities, a manufacturer must assess the health risks connected to such a substance and notify all the individuals who may enter into contact with it---including its employees---of the health risks \parencite{USEnvironmentalProtectionAgency1983x}. \textcite{USEnvironmentalProtectionAgency1983x} lists some of the data that a manufacturer must look into to assess whether there might be a health risk connected with the new substance.

This data includes:

\begin{itemize}
\item
  ``Information in {[}the manufacturer's{]} possession or control concerning any significant adverse reaction by persons exposed to the chemical substance which may reasonably be associated with such exposure.''
\item
  ``Information provided to the manufacturer by a supplier or any other person concerning a health risk believed to be associated with the substance.''
\item
  ``Health and environmental effects data in {[}the manufacturer's{]} possession or control concerning the substance.''
\end{itemize}

It is worth observing that this `simplified' process only applies to ``solely'' research and development activities on ``small quantities'' of the new chemical substance \parencite{USEnvironmentalProtectionAgency1983x}. `Small quantities' is defined as ``quantities of a chemical substance manufactured or processed or proposed to be manufactured or processed solely for research and development that are not greater than reasonably necessary for such purposes'' \parencite{USEnvironmentalProtectionAgency1983c}. If quantities are not ``small'' or ``any amount of the substance, including as part of a mixture, is processed, distributed in commerce, or used, for any commercial purpose other than research and development'' \parencite{USEnvironmentalProtectionAgency1983x}, a manufacturer must submit a `regular' notice to the EPA under \parencite{USEnvironmentalProtectionAgency1983r}.

\subsubsection{Test Marketing for a Novel Chemical Substance}\label{b-test-marketing-for-a-novel-chemical-substance}

A manufacturer who wants to conduct test marketing must apply to the EPA for an exemption from the notice to the EPA under \textcite{USEnvironmentalProtectionAgency1983r}, demonstrating that ``the chemical substance will not present an unreasonable risk to injury to health or the environment as a result of the test marketing'' \parencite{USEnvironmentalProtectionAgency1983b}. In their test-marketing exemption, applicants must include ``{[}a{]}ll existing data regarding health and environmental effects of the chemical substance'' \parencite{USEnvironmentalProtectionAgency1983b}. In other words, before commencing test marketing, manufacturers must persuade the EPA that---based on all the data available---test marketing is safe.

The application for a test-marketing exemption also requires information that seems to restrain what can be done through test marketing. The application must include: (i) the maximum quantity of the chemical substance that will be manufactured for test marketing; (ii) the maximum number of persons who may be provided the chemical substance during test marketing; and (iii) the maximum number of persons who may be exposed to the chemical substance as a result of test marketing \parencite{USEnvironmentalProtectionAgency1983b}. Finally, the application must also include ``{[}a{]} description of the test-marketing activity, including its length and how it can be distinguished from full-scale commercial production and research and development'' \parencite{USEnvironmentalProtectionAgency1983b}. In other words, to run test marketing, manufacturers must successfully demonstrate that test marketing is safe based on all available data and justify why this pre-marketing activity is different from full-scale commercial production. Only then can they be greenlit by the EPA.

\subsection{Novel Pesticides: Experimental Use Permit for Testing}\label{novel-pesticides-experimental-use-permit-for-testing}

Lastly, we examined if and how novel pesticides are regulated prior to marketing. We found that in the US---in a way that is akin to new chemical substances (\hyperref[novel-chemical-substances-health-risk-assessment-for-rd-and-test-marketing]{\ul{§5.2}})---a permit from the EPA is required for pre-marketing testing of new pesticides. In other words, the EPA has oversight on pre-marketing research and development.

\textcite{USEnvironmentalProtectionAgency1988y} and \textcite{USEnvironmentalProtectionAgency1988l} lay out the rule that the distribution and sale of any pesticide product requires prior registration with the EPA. However, before distributing or selling (and hence necessitating registration with the EPA), manufacturers can test an unregistered pesticide by obtaining an `experimental use permit' (EUP) from the EPA \parencite{USEnvironmentalProtectionAgency1975c}. In other words, pre-marketing testing cannot commence until a pesticide manufacturer obtains an EUP from the EPA.

To receive an EUP, applicants must provide various safety-related information, including the specific results of ``any appropriate prior testing'' to determine: (i) the ``toxicity and effects in or on target organisms at the site of application''; (ii) the ``phytotoxicity and other forms of toxicity or effects on nontarget plants, animals, and insects at or near the site of application''; and (iii) ``adverse effects on the environment'' \parencite{USEnvironmentalProtectionAgency1975l}. The EPA issues an EUP when it deems that the conditions have been met \parencite{USEnvironmentalProtectionAgency1975b}. This implies that the EPA can block the pre-marketing testing of a new pesticide if the EUP does not contain sufficient safety data.

In summary, we found that government agencies and regulatory bodies impose a number of governance and oversight requirements in safety-critical industries where access to and use of a new, potentially dangerous, product before public release and commercialization is possible and commonly pursued.

Having analyzed a range of governance mechanisms in other safety-critical industries, we believe these learnings could inform the governance of internal deployment of highly advanced AI systems. For example, in our recommendations in \hyperref[chapter-6.-defense-in-depth-recommendations-for-the-governance-of-internal-deployment]{\ul{Chapter 6}}, we propose:

\begin{itemize}
\item
  \textbf{Risk evaluations, mitigations, and relevant restrictions and oversight procedures} (\hyperref[a-frontier-safety-policies-tripwires-and-corresponding-mitigations]{\ul{§6.2.(a)}}--\hyperref[c-oversight-framework]{\ul{(c)}}) that we observed in some form in all the described sectors---from security plans (\cite{USDepartmentOfHealthAndHumanServices2005s}) and biosafety plans (\cite{USDepartmentOfHealthAndHumanServices2005s}) for biological agents and toxins (\hyperref[biological-agents-and-toxins-registration-for-possession-and-use]{\ul{§5.1}}) to health and environmental risk assessments \parencite{USEnvironmentalProtectionAgency1983x,USEnvironmentalProtectionAgency1983b} for new chemical substances (\hyperref[novel-chemical-substances-health-risk-assessment-for-rd-and-test-marketing]{\ul{§5.2}}), and toxicity and phytotoxicity and environmental impact testing \parencite{USEnvironmentalProtectionAgency1975l} for new pesticides (\hyperref[novel-pesticides-experimental-use-permit-for-testing]{\ul{§5.3}}).
\item
  \textbf{Targeted transparency on safety-critical information} (\hyperref[a-targeted-transparency-recipients]{\ul{§6.3.(a)}}--\hyperref[b-targeted-transparency-content]{\ul{(b)}}), including all the testing results just described, that we noticed in the processes involving the EPA and the HHS in all the sectors examined (\hyperref[biological-agents-and-toxins-registration-for-possession-and-use]{\ul{§5.1}}--\hyperref[novel-pesticides-experimental-use-permit-for-testing]{\ul{5.3}}).
\item
  \textbf{Boundaries to internal access and usage in the form of internal usage policies} (\hyperref[b-internal-usage-policies]{\ul{§6.2.(b)}}), which are inspired by the restrictions to test-marketing activities, including the maximum number of persons who may be provided access to novel chemical substances during test marketing \parencite{USEnvironmentalProtectionAgency1983b}.
\item
  \textbf{Disaster resilience plans} (\hyperref[c-disaster-resilience-plans-tied-to-incident-monitoring-and-safety-cases]{\ul{§6.3.(c)}}) inspired by incident response plans (\cite{USDepartmentOfHealthAndHumanServices2005v}) for biological agents and toxins (\hyperref[biological-agents-and-toxins-registration-for-possession-and-use]{\ul{§5.1}}).
\end{itemize}

In applying these learnings, we pay special attention to targeting the risk scenarios outlined in \hyperref[chapter-3.-internal-deployment-scenarios-of-concern]{\ul{Chapter 3}}---loss of control and grand-scale power concentration.

\section{Defense in Depth: Recommendations for the Governance of Internal Deployment}\label{chapter-6.-defense-in-depth-recommendations-for-the-governance-of-internal-deployment}

In \hyperref[chapter-3.-internal-deployment-scenarios-of-concern]{\ul{Chapter 3}}, we described two high-impact scenarios associated with the internal deployment of highly advanced AI systems absent any governance mechanisms. In doing so, we first articulated how the internal application of highly advanced AI systems to automated AI R\&D could lead to loss of control (\hyperref[a-loss-of-control-via-automated-ai-rd]{\ul{§3.2.(a)}}) through a rapid increase in capabilities and downstream effects of misalignment (specifically scheming).\footnote{We previously defined scheming in \hyperref[chapter-3.-internal-deployment-scenarios-of-concern]{\ul{Chapter 3}} as ``AI systems covertly and strategically pursuing misaligned goals'' \parencite{Balesni2024a,Meinke2024x}.} Second, we articulated how the internal application of highly advanced AI systems behind closed doors could lead to undetectable economic and power concentration, which could go as far as enabling overt or covert `coups' (\hyperref[b-undetected-and-unconstrained-power-accumulation-through-an-internal-intelligence-explosion]{\ul{§3.2.(b)}}).

The overall goal of \hyperref[chapter-6.-defense-in-depth-recommendations-for-the-governance-of-internal-deployment]{\ul{Chapter 6}}, this report's final chapter, is to prime a conversation among decision-makers in industry and governments on how to most robustly approach and govern potential risks from internal deployment.\footnote{We reiterate the narrow scope of the report and, correspondingly, the narrow scope of AI systems and scenarios these proposals are targeting. Currently, the types of interventions described in this Chapter are reserved for a small subset of AI systems and their applications, which may pose unacceptably high downside to society at large, as well as to national security, as described in \hyperref[chapter-3.-internal-deployment-scenarios-of-concern]{\ul{Chapter 3}}.} In the remainder of this Chapter we will therefore recommend several technical and non-technical interventions that we expect will contribute to foreseeing, governing, and mitigating potential risk from internal deployment.

In doing so, we advance two overarching `macro' proposals, which should be combined for maximum effectiveness. Namely, our proposals aim to:

\begin{itemize}
\item
  Establish clear pathways to \textbf{detecting and controlling scheming} (\hyperref[addressing-loss-of-control-scheming-detection-and-mitigations]{\ul{§6.2}}).
\item
  Enable \textbf{targeted transparency} around access, usage, control, and oversight of internally deployed highly advanced AI systems, as well as of their capabilities (\hyperref[addressing-undetected-and-unconstrained-power-consolidation-targeted-transparency]{\ul{§6.3}}).
\end{itemize}

Reflecting on a more `granular' level, each proposal is composed of individual layers of defense detailed in this Chapter, targeting select threat vectors underpinning the scenarios described in \hyperref[chapter-3.-internal-deployment-scenarios-of-concern]{\ul{Chapter 3}} and translating learnings from other safety-critical sectors described in \hyperref[chapter-5.-internal-deployment-in-other-safety-critical-industries]{\ul{Chapter 5}}. Importantly, while each layer provides some assurance and can be seen as a stand-alone, increasingly more benefit will be derived from stacking layers and creating a more ambitious framework. In other words, we envisage each suggested layer of defense as an individual solution that can be combined with others to create `defense in depth' \parencite{Hendrycks2021l,Hendrycks2023w,Hendrycks2024x,MaryDrouin2016j,Kuipers2006f,Cotton-Barratt2020a}.\footnote{We note that a defense-in-depth approach is already used and expressly referenced in some existing Frontier Safety Policies. \emph{See}, for example, Anthropic's Responsible Scaling Policy \parencite{Anthropic2025j}.} In summary, the following are our proposed lines of defense.

\begin{figure}[ht!]
    \centering
    \includegraphics[width=\linewidth]{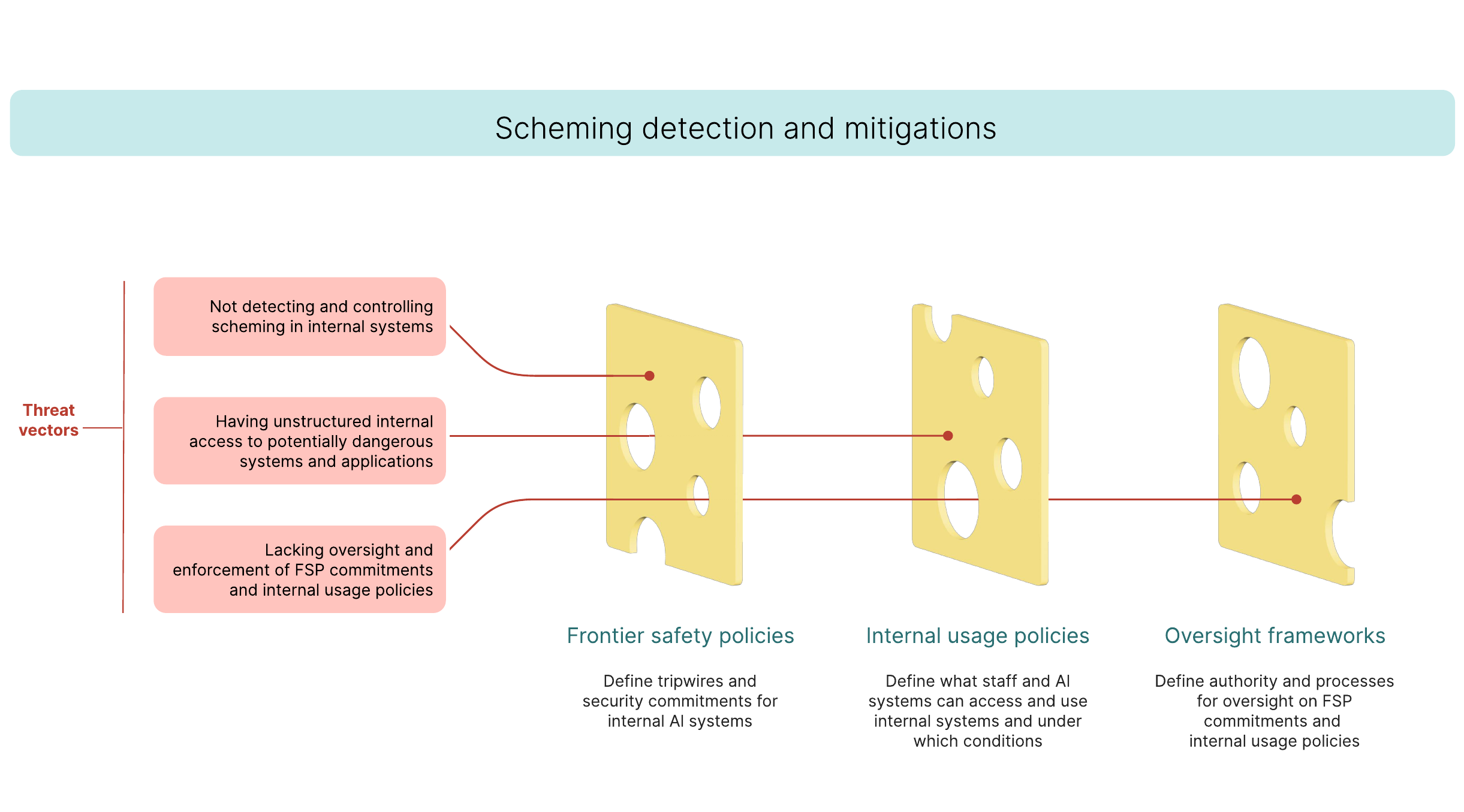}
    \caption{Swiss cheese model representing our recommended defense-in-depth strategy against the risk of loss of control via internally deployed misaligned AI (\hyperref[a-loss-of-control-via-automated-ai-rd]{\ul{§3.2.(a)}}). Threat vectors are in red.}
    \label{fig:5}
\end{figure}
\pagebreak
Concerning \textbf{scheming detection and control} (\hyperref[addressing-loss-of-control-scheming-detection-and-mitigations]{\ul{§6.2}}; Figure D above), we recommend to:

\begin{enumerate}
\def\labelenumi{\arabic{enumi}.}
\item
  \textbf{Expand the scope of} \textbf{Frontier Safety Policies} (FSPs) to explicitly cover internal as well as external deployment. As a result, AI systems should be evaluated before internal deployment and, depending on the results of those evaluations, companies commit to correspondingly adequate risk mitigations (\hyperref[a-frontier-safety-policies-tripwires-and-corresponding-mitigations]{\ul{§6.2.(a)}}).
\item
  \textbf{Develop internal usage policies} that outline rules, procedures, and guidelines for internal access and application of highly advanced AI systems, therefore setting explicit boundaries on who has the authority to access and apply its most capable AI systems and under what conditions (\hyperref[b-internal-usage-policies]{\ul{§6.2.(b)}}).
\item
  \textbf{Develop oversight frameworks} that detail the processes for the supervision and enforcement of FSPs and internal usage policies, therefore establishing clear enforcement processes and increasing internal and external transparency and accountability (\hyperref[c-oversight-framework]{\ul{§6.2.(c)}}).
\end{enumerate}

As we will explain in \hyperref[internal-governance-frameworks-and-processes]{\ul{§6.1}}, all these solutions are company-led and constructed around `tripwires,' which means that no action is required from companies \emph{unless} certain trigger conditions are met \parencite{Karnofsky2024f}. This makes our current proposal flexible and non-invasive. Companies can keep innovating and adopt mitigations only when pathways to specific threat-relevant scenarios start to manifest \parencite{Pistillo2024n}. In the future, other variations of governance frameworks for internal deployment may be needed, including through government interventions. However, this is out of the current scope of the report.

\begin{figure}[h!]
    \centering
    \includegraphics[width=\linewidth]{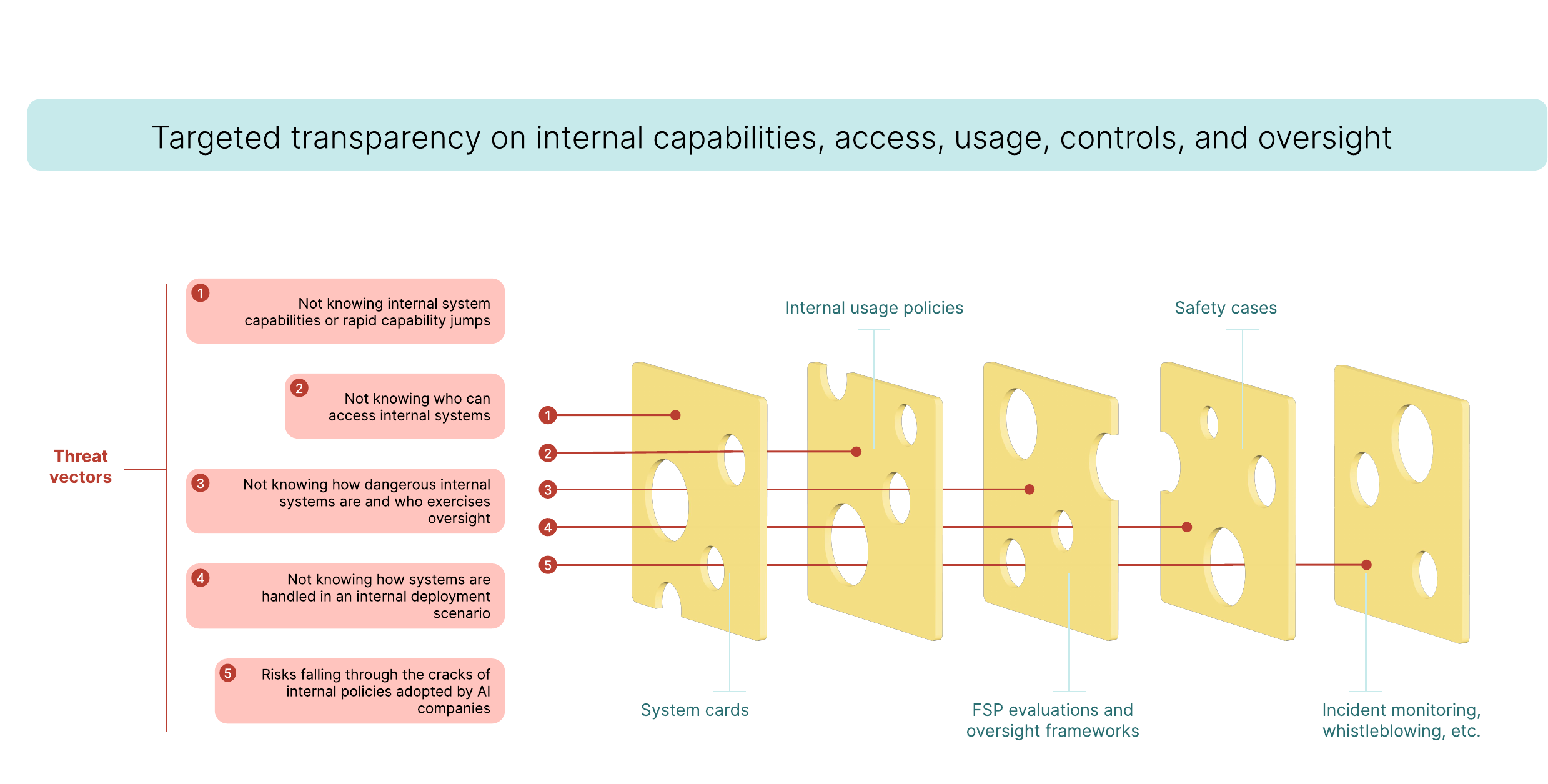}
    \caption{`Swiss cheese model' representing our recommended defense-in-depth strategy against the risk of undetected and unconstrained power accumulation (\hyperref[b-undetected-and-unconstrained-power-accumulation-through-an-internal-intelligence-explosion]{\ul{§3.2.(b)}}). Threat vectors are in red.}
    \label{fig:E}
\end{figure}

\pagebreak
Concerning \textbf{targeted transparency} on internal AI system capabilities, access, usage, control, and oversight (\hyperref[addressing-undetected-and-unconstrained-power-consolidation-targeted-transparency]{\ul{§6.3}}; Figure E above), we recommend to:

\begin{enumerate}
\def\labelenumi{\arabic{enumi}.}
\setcounter{enumi}{3}
\item
  Develop oversight frameworks and establish processes therein for \textbf{selectively sharing} the following pieces of information with internal and external stakeholders:

  \begin{itemize}
\item
  \textbf{Results of the} \textbf{evaluations} run according to the FSPs, allowing select stakeholders to know which dangerous capabilities internal AI systems could have.
\item
  \textbf{Pre-internal-deployment system cards}, allowing select stakeholders to know how capable internally deployed AI systems are and encouraging preparedness in light of rapid capability increases.
\item
  \textbf{Internal usage policies}, allowing select stakeholders to know which staff (or AI systems) within an AI company have the authority to access and apply the highly advanced AI system and under which conditions.
\end{itemize}
\end{enumerate}

Finally, AI companies deploying highly advanced AI systems internally could also consider:

\begin{enumerate}
\def\labelenumi{\arabic{enumi}.}
\setcounter{enumi}{4}
\item
  \textbf{Co-developing} \textbf{disaster resilience plans} with governments. These could be tied to incident monitoring and response, providing a last line of defense under the assumption that previous lines of defense failed. We propose that these should be tied to structured safety arguments (or `AI safety cases'; \cite{Clymer2024w}) developed prior to internally deploying a given AI system (\hyperref[c-disaster-resilience-plans-tied-to-incident-monitoring-and-safety-cases]{\ul{§6.3.(c)}}).
\end{enumerate}

Though largely outside of the focus of this report, in \hyperref[a.i-incentives-public-private-partnerships-and-licenses]{\ul{§6.3.(a.i)}} we also briefly point in the direction of some of the incentives that could motivate AI companies to engage in targeted transparency with government entities, including access to resources through public-private partnerships or licenses.

\subsection{Internal Governance: Frameworks and Processes}\label{internal-governance-frameworks-and-processes}

In order to operationalize both of our `macro' proposals (\hyperref[addressing-loss-of-control-scheming-detection-and-mitigations]{\ul{§6.2}}, \hyperref[addressing-undetected-and-unconstrained-power-consolidation-targeted-transparency]{\ul{§6.3}}), we recommend that AI companies create three documents. We speak to these \textbf{three documents---Frontier Safety policies (FSPs), internal usage policies, and oversight frameworks}---in more detail below in \hyperref[a-frontier-safety-policies-tripwires-and-corresponding-mitigations]{\ul{§6.2.(a)}}, \hyperref[b-internal-usage-policies]{\ul{§6.2.(b)}} and \hyperref[c-oversight-framework]{\ul{§6.2.(c)}} respectively. In this Section, we turn to matters of operationalization instead.

We propose that to operationalize and oversee these documents, it would be of benefit to establish \textbf{two groups} within AI companies, operating in tandem with these documents: (i) an \textbf{Internal Deployment Team} (\textbf{IDT}); and, (ii) an \textbf{Internal Deployment Overseeing Board} (\textbf{IDOB}). We briefly reflect on their composition, roles and responsibilities next.

The IDT could be composed of a subset of internal staff---both technical and governance experts. This group could be composed of pre-existing staff or be newly constituted; for example, an AI company may decide to hire a new dedicated team with the sole task of supervising internal deployment. In both instances, we envision the IDT as an internal group in charge of (i) running evaluations prior to internal deployment (\hyperref[a.i-tripwires-discovered-by-evaluations]{\ul{§6.2.(a.i))}} and \hyperref[a.ii-mitigations-technical-access-and-usage-restrictions-and-controls]{\ul{(ii)}} implementing corresponding technical access, usage restrictions, and controls (\hyperref[a.ii-mitigations-technical-access-and-usage-restrictions-and-controls]{\ul{§6.2.(a.ii)}}; \hyperref[b-internal-usage-policies]{\ul{§6.2.(b)}}). Subsequently, we envision the IDT to (iii) draft a comprehensive technical report detailing evaluation results and corresponding mitigation measures addressed to the IDOB.

We envision the IDOB as a higher-level independent board tasked with making the final decisions on internal deployment based on the report from the IDT and with enacting the overarching oversight framework (\hyperref[c-oversight-framework]{\ul{§6.2.(c)}}). These final decisions include, first and foremost, whether a given AI system ought to be internally deployed or not.\footnote{This is consistent the idea that humans are ``a slow-but-necessary component'' and that ``{[}i{]}f the `someone' making choices is humans, then our cognitive speed and limitations will serve as a ceiling'' \parencite{AIPolicyPerspectives2025x}.} If they are uncertain, they may solicit additional input from external experts. In short, in the spirit of achieving `defense in depth,' the IDOB provides an extra set of neutral eyes, red-teaming results and assumptions provided to them. Once the IDOB agrees that a given AI system is allowed to be deployed internally, it is in charge of preparing the corresponding internal usage policy, which specifies who will have access to the given AI system and for what purpose (\hyperref[b-internal-usage-policies]{\ul{§6.2.(b)}}).

The IDOB could be composed in one of two main ways. First, the IDOB could be an external oversight board appointed by the AI company and staffed with independent experts. The IDOB could draw from existing external boards (for example, \cite{OpenAI2024j}),\footnote{We note that setting up an external board is not necessarily an infallible solution depending on their ultimate authority and others' authority to disband them or overrule their decisions, as evidenced by outcomes from OpenAI previous board's decisions \parencite{Field2024v,Paul2024q}.} for instance in the form of a separate group within an existing external board,\footnote{We imagine that this function could also be, at least provisionally, exercised through the Frontier Model Forum, and build on their recent information sharing framework \parencite{FrontierModelForum2025o}.} or it could form an entirely new independent board. We suggest that an external configuration rather than a company-internal configuration, composed of employees, has the advantage of being less conflicted, which may encourage, for example, the launch of further inquiries if needed, including by drawing on input from additional independent experts.

Second, as highly advanced AI systems' capabilities progress and are increasingly likely to pose concern to national security, the IDOB could be composed of individuals within select government agencies. Once a given AI system surpasses a certain capability level, regardless of mitigations, a government may wish to have a certain degree of insight and oversight, ideally supporting societal resilience and preparedness. As part of this, the IDOB may require (through adapted internal usage policies) that a given AI company's staff undergo additional procedures before being allowed internal access, such as stringent security clearances and background checks.

In the \hyperref[step-by-step-cooperation-between-the-idt-and-the-idob]{\ul{Appendix}}, and Figure G therein, we provide a step-by-step overview of how we envision the IDT and the IDOB to collaborate.

\subsection{Addressing Loss of Control: Scheming Detection and Mitigations }\label{addressing-loss-of-control-scheming-detection-and-mitigations}

In this Section, we present our first `macro' proposal targeted at loss of control (\hyperref[a-loss-of-control-via-automated-ai-rd]{\ul{§3.2.(a)}}). This `macro' proposal is composed of three recommendations, each presenting a `stackable' layer of defense and targeting potential AI misalignment, AI scheming, and consequent loss of control.\footnote{We note that our `macro' proposal and the three related layers of defense target can be much more broadly applicable than for this specific scenario. Indeed, the governance prototype that this proposal aims to establish can enable preparedness and resilience vis-a-vis many more threat scenarios. For example, the mitigations described in \hyperref[a.ii-mitigations-technical-access-and-usage-restrictions-and-controls]{\ul{§6.2.(a.ii)}} could address human-misuse threat scenarios---including the ones alluded to \hl{in footnotes 15 and 16} in \hyperref[scope]{\ul{§1.1}}---that are outside the scope of this report.}

\begin{figure} [h!]
    \centering
    \includegraphics[width=\linewidth]{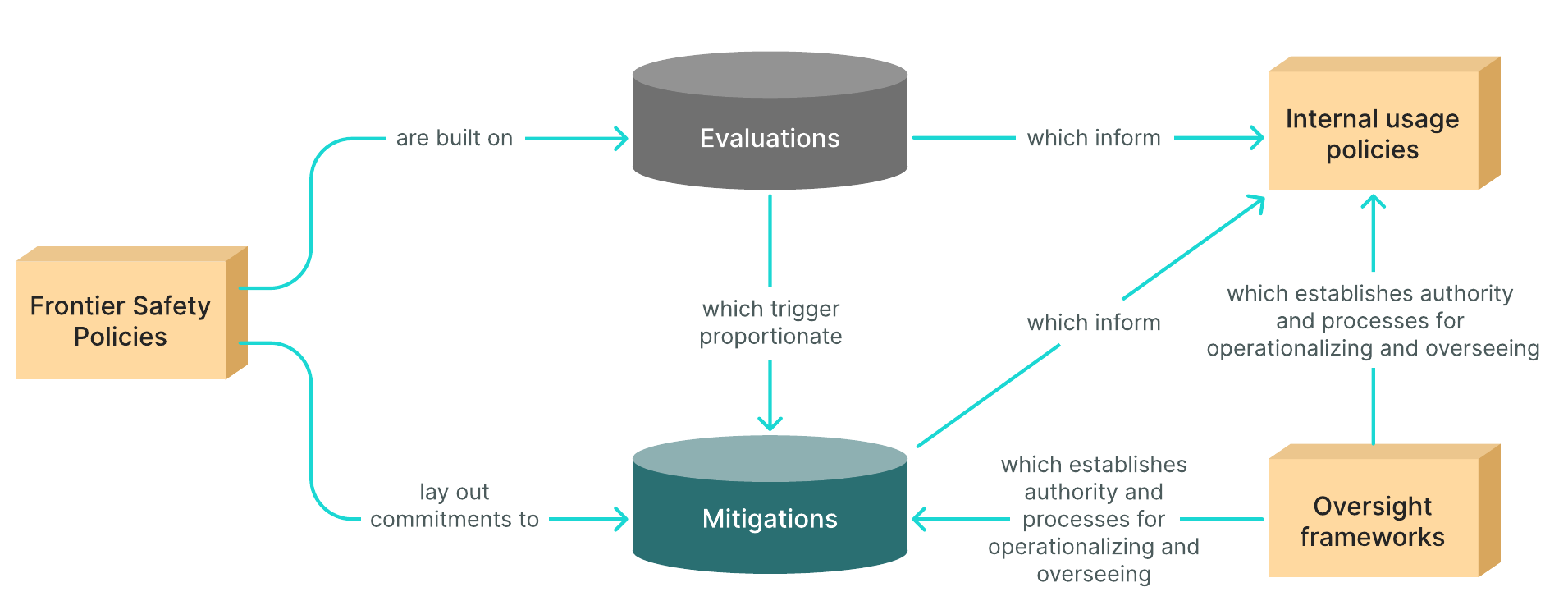}
    \caption{This figure illustrates how FSPs (\hyperref[a-frontier-safety-policies-tripwires-and-corresponding-mitigations]{\ul{§6.2.(a)}}), internal usage policies (\hyperref[b-internal-usage-policies]{\ul{§6.2.(b)}}), and oversight frameworks (\hyperref[c-oversight-framework]{\ul{§6.2.(c)}}) complement each other.}
    \label{fig:F}
\end{figure}

The following \hyperref[a-frontier-safety-policies-tripwires-and-corresponding-mitigations]{\ul{§6.2.(a)}}-\hyperref[c-oversight-framework]{\ul{(c)}} detail our \textbf{three recommendations}. In short, we first recommend explicitly \textbf{extending FSP tripwires} and corresponding mitigation commitments to cover internal deployment (\hyperref[a-frontier-safety-policies-tripwires-and-corresponding-mitigations]{\ul{§6.2.(a)}}). Second, we recommend the development and implementation of concrete \textbf{internal usage policies}, which should detail who within an AI company and under what circumstances can access and apply internal AI systems (\hyperref[b-internal-usage-policies]{\ul{§6.2.(b)}}). Third, we recommend the development and implementation of \textbf{oversight frameworks} in order to increase internal and external transparency and accountability, including on the enforcement of FSPs and internal usage policies (\hyperref[c-oversight-framework]{\ul{§6.2.(c)}}). As Figure F shows, these layers of defense complement each other.

\subsubsection{Frontier Safety Policies: Tripwires and Corresponding Mitigations}\label{a-frontier-safety-policies-tripwires-and-corresponding-mitigations}

Our first recommended layer of defense against loss of control as described in \hyperref[a-loss-of-control-via-automated-ai-rd]{\ul{§3.2.(a)}} is to ensure that \textbf{robust scheming and control evaluations are run on AI systems before these are accessed internally or applied for internal use}, and that \textbf{meaningful mitigations are applied \emph{if} evaluations show that an AI system is capable of scheming or can evade control}.

We believe that the described line of defense could be best ensured by explicitly \textbf{upgrading existing FSPs to cover internally deployed AI systems}. Many leading AI companies already use FSPs to classify the risk level of their externally deployed AI systems and trigger associated security and safety protocols ahead of external deployment \parencite{Anthropic2025j,openai2025preparedness,METR2025d,GoogleDeepMind2025e}.

Existing FSPs could be extended and upgraded so that, prior to deploying an AI system internally, AI companies would be required to assess whether any of the tripwires are triggered, for example, through rigorous evaluations, and, in the case where tripwires are triggered, apply the mitigations to which they have pre-committed. Extending FSP protections to internal deployment would not only prove foresight on the side of their authors and care for society at large but also be coherent with prior commitments by AI companies.\footnote{In its Responsible Scaling Policy, Anthropic already mentions ``insider risk,'' and commits to enhancing protections from insider risk control with its ASL-4 preparations \parencite{Anthropic2025j}. Through its deployment standards, Anthropic also wants to ``ensure the safe usage of AI models by \ldots{} internal users (i.e., our employees)'' \parencite{Anthropic2025j}. In its Responsible Scaling Policy, Anthropic also includes `autonomous AI R\&D' within the capability thresholds potentially triggering mitigations \parencite{Anthropic2025j}. Similarly, OpenAI's new Preparedness Framework applies to the deployment of ``any agentic system (including significant agents deployed only internally)'' \parencite{openai2025preparedness}. \hl{}} Below, we focus on the role of evaluations in triggering tripwires (\hyperref[a.i-tripwires-discovered-by-evaluations]{\ul{§6.2.(a.i)}}) and associated mitigations that should be implemented once certain pre-committed tripwires are triggered (\hyperref[a.ii-mitigations-technical-access-and-usage-restrictions-and-controls]{\ul{§6.2.(a.ii)}}).

\paragraph{Tripwires: Discovered by Evaluations}\label{a.i-tripwires-discovered-by-evaluations}

In a recent newsletter, one of Anthropic's co-founders noted that it is possible that ``powerful systems will arrive towards the end of 2026 or in 2027'' and that ``{[}u{]}nder very short timelines, {[}we{]} may want to take more extreme actions,'' including ``{[}m{]}andating pre-deployment testing'' \parencite{Clark2025u}. Considering the scenarios outlined in \hyperref[chapter-3.-internal-deployment-scenarios-of-concern]{\ul{Chapter 3}}, we propose that \textbf{dangerous capability evaluations are already conducted prior to internal deployment} and that \textbf{internal deployment is made conditional on their result} based on predefined tripwires in FSPs.\footnote{`If-then' policies enable AI companies to define `thresholds' for internal deployment and match proportionate evaluations and mitigations to these thresholds. This allows AI companies to concentrate evaluators' time and resources on high-priority items (i.e., internal AI systems that could pose the highest risk) and reduce potential burdens and trade-offs.}

A core question is how to identify the capabilities and corresponding risks of an AI system to be internally deployed. Practically speaking, as of the report's writing, there are currently two, both imperfect, strategies for knowing whether an AI system carries risk for internal deployment: testing it ahead of internal deployment or learning the hard lesson when things go wrong. Given the options, it is likely to be in AI companies' own best interest to ensure that they have a sufficiently complete and comprehensive overview of the AI system's capabilities and risk profile ahead of enabling internal users to access and apply it internally. This may be achieved through \textbf{dangerous capability evaluations} \parencite{Shevlane2023k,Meinke2024x,Anthropic2025h,OpenAI2024k,UKAISecurityInstitute2024i,Wijk2024y,GoogleDeepMind2025e}. \footnote{The results of evaluations can also inform internal usage policies (\hyperref[b-internal-usage-policies]{\ul{§6.2.(b)}}) and could be shared with select government agencies (\hyperref[addressing-undetected-and-unconstrained-power-consolidation-targeted-transparency]{\ul{§6.3}}).}

Specifically, in line with \hyperref[chapter-3.-internal-deployment-scenarios-of-concern]{\ul{Chapter 3}}, we propose that \textbf{scheming and control evaluations} should be conducted prior to internal deployment.\footnote{This should not be misconstrued as a comprehensive evaluation regime. We only focus on the limited scope of this report and acknowledge that there will be many other threats we did not capture and that should be evaluated for prior to internal deployment. In this respect, we observe that METR recently suggested earlier testing of AI systems to to catch them ``misbehaving during training or while being used internally'' \parencite{METR2025h}. \hl{We also note that additional measures to scheming and control evaluations could help mitigate the relevant risks, including, for example, preserving the norm of externalized reasoning} \parencite{Kinniment2025w}.} The capability and propensity of an AI system to scheme---i.e., an AI system covertly pursues misaligned goals while actively concealing these goals from its developers \parencite{Balesni2024a,Meinke2024x}---can be conceptualized as an underlying horizontal layer of risk. This means that scheming can effectively undermine all other evaluations and safety assessments and, therefore, fundamentally must be assessed for.\footnote{For example, a scheming AI system could `sandbag,' i.e. strategically underperform in evaluations such as for chemical, biological, radiological and nuclear (CBRN) risk \parencite{JohannesGasteigerAkbirKhanSamBowmanVladimirMikulikEthanPerezFabienRoger2025i}. This would falsify any safety and security assessment and invalidate beliefs around the relative safety, oversight methods, and mitigation methods needed before deployment.} Research teams have recently shown that current frontier AI systems already have the capability to scheme \parencite{Greenblatt2024o}, including by AI systems strategically introducing subtle mistakes into their responses, attempting to disable their oversight mechanisms, and exfiltrating what they believe are their weights to external servers \parencite{Meinke2024x}.\footnote{Consistently, some frontier AI companies have also adopted an initial approach for addressing `deceptive alignment' through their FSPs \parencite{GoogleDeepMind2025e}.} Therefore, it is important to ensure that the first highly capable AI systems applied to automate the AI R\&D pipeline are not successfully scheming.

If we fail to detect and monitor for scheming, a scheming AI that is applied to automate the AI R\&D pipeline (including to conduct AI safety research) could undermine the goal of the workstream, poison the data, and introduce failure modes, exploits, and back doors into future AI systems---all the while appearing safe, since these issues would be difficult to detect if one was not looking for them. In other words, if we cannot detect scheming successfully, then we might be unable to prevent the scheming plans from succeeding. Given the stakes at hand, we therefore recommend that, at minimum, evaluations for scheming should be conducted by qualified experts prior to internal deployment.\footnote{To ensure a robust assessment of the risk profile of a given AI system, we recommend including external experts specialized in specific evaluation domains in these assessments, offering an independent and unbiased review of the risk profile. These experts will need to be put under NDA to avoid disclosure of any confidential or commercially sensitive information, and, depending on the AI system to be evaluated and company preferences might be flown in to run evaluations on location in a secure environment where they are unable to extract any information. All experts, whether internally or externally, should be provided with access to the chain of thought, fine-tuning access and access to scaffolding, and adequate time to be able to faithfully assess capabilities.} In the future, it will likely also become necessary to assess whether an AI company's implemented safety measures are sufficiently robust to prevent negative outcomes, under the assumption that a highly advanced AI system slipped through the scheming detection and intentionally tries to subvert those safety measures \parencite{Shlegeris2024d}.

Looking at the future, we also recommend that---as the risk of large-scale power grabs through AI-enabled coups increases alongside the capabilities of the most advanced internal AI systems (\hyperref[b.i.-democratic-disruption-and-enablement-of-coups]{\ul{§3.2.(b.i)}})---AI companies and third parties develop new `\textbf{coup} \textbf{evaluations}' to assess an AI system's capability and propensity to aid in a coup.\footnote{Further research will shed light on the new idea of `coup evaluations.'}

\paragraph{Mitigations: Technical Access and Usage Restrictions and Controls}\label{a.ii-mitigations-technical-access-and-usage-restrictions-and-controls}

In the context of internal deployment, we recommend that the results of the evaluations described in \hyperref[a.i-tripwires-discovered-by-evaluations]{\ul{§6.2.(a.i)}} trigger a range of mechanisms, including technical access and usage restrictions and monitoring, increased internal and external transparency, and stricter physical and cybersecurity measures. In the most extreme case, \textbf{evaluations may also substantiate a decision not to deploy a highly advanced AI system internally}.

First, to maintain oversight and minimize threats from access to and usage of internally deployed AI systems, AI companies could \textbf{apply} \textbf{technical} \textbf{access and usage restrictions and controls}, \textbf{corresponding to the evaluation outcomes}.\footnote{In \hyperref[b-internal-usage-policies]{\ul{§6.2.(b)}} and \hyperref[c-oversight-framework]{\ul{§6.2.(c)}}, we will explain how internal access and usage restrictions can be tied into corresponding oversight procedures for access and usage.} For instance, AI companies could draw inspiration from access controls from dual-control systems for nuclear weapons \parencite{DepartmentOfTheAirForce2020r,DepartmentOfTheAirForce2013g} and traditional multi-party authorization \parencite{AWS2025v}. For usage restrictions and control, they could establish usage limits alongside usage monitoring and draw inspiration from controlled substance handling (\emph{see} \hyperref[chapter-5.-internal-deployment-in-other-safety-critical-industries]{\ul{Chapter 5}}).\footnote{We note that according to recent evidence, it appears that AI companies are currently not conducting internal monitoring \parencite{Adler2025m}.} Further, usage limits could be specified in more detail depending on the threat model at hand. For instance, to mitigate threats from loss of control (\emph{see} \hyperref[chapter-3.-internal-deployment-scenarios-of-concern]{\ul{Chapter 3}}), it could be valuable to limit the percentage of critical code that a given internal AI system can write autonomously for subsequent implementation and to subject the application of this code to stringent review and pre-authorization (\emph{see} \hyperref[chapter-2.-characterizing-internal-deployment]{\ul{Chapter 2}}). Other examples of technical restrictions and controls could include, for example, API rate limiting, audit logging, authorization tokens with expiry, and network segregation---measures adapted from best practices for securing critical IT systems \parencite{JointTaskForceInteragencyWorkingGroup2020v}.

Second, mitigations should reflect appropriate levels of cyber and physical security required for internal access and application of a given AI system. Future research should specify what levels of security would be proportionate to corresponding tripwires. Notwithstanding that, we believe that a baseline security infrastructure should be set up in such a manner that malicious actors cannot steal internal model weights and algorithmic secrets, that internally deployed AI systems cannot exfiltrate themselves, and that no internal employee can run large, unmonitored fine-tuning runs \parencite{Hobbhahn2025z}. This may necessitate, at minimum, the establishment of strict weight security requirements in FSPs equal to SL4 or higher \parencite{Nevo2024c} by the time internal AI systems reach capability levels sufficient to replace a top AI researcher \parencite{Hobbhahn2025z,Shlegeris2024s}. Other considerations on physical security could include, for example: (i) restricting access to buildings, their internal spaces, and data centers; (ii) establishing air-gapped environments for internal deployment or otherwise isolated company-internal sandboxes; and (iii) requiring staff clearance and devising systems to identify usage attempts from non-cleared staff.

\subsubsection{Internal Usage Policies }\label{b-internal-usage-policies}

Our second recommended layer of defense is to \textbf{establish `internal usage policies' which detail access and usage compartmentalization on a need-to-know basis}. This would present as a complementary piece of documentation to a given FSP and outline a set of guidelines on which employees (and AI systems alike)\footnote{In a scenario in which AI systems are the main user of other systems in the AI R\&D pipeline, FSPs and internal usage policies could also address which access and usage rights, as well as which affordances, an internal AI R\&D system is entitled to. This is consistent with treating an AI system as `a principal' with its own identity and access control \parencite{Shah2025v}.} can and cannot access and apply specific internally deployed AI systems and under which conditions they can do so. As a result, internal usage policies could restrict access to internal highly capable systems to ``vetted user groups and use cases,'' therefore ``reducing the surface area of dangerous capabilities'' \parencite{Shah2025v}. Moreover, similar to how external users must agree to terms of service prohibiting dangerous misuse, it is prudent to require internal staff to commit to comparable internal usage policies that clearly define their obligations.

As internally deployed AI systems become progressively more capable, access and usage permissions should become correspondingly narrower. Internal usage policies should hence be informed by the evaluations underpinning FSP commitments (\hyperref[a-frontier-safety-policies-tripwires-and-corresponding-mitigations]{\ul{§6.2.(a)}}) and clarify, based on the outcome of such assessments, who is entitled to access which internal AI system and what type of applications are allowed. From a chronological perspective, AI companies should first upgrade FSPs by defining tripwires and commitments on internal deployment (\hyperref[a-frontier-safety-policies-tripwires-and-corresponding-mitigations]{\ul{§6.2.(a)}}). Then, AI companies could run the evaluations underpinning such tripwires. Finally, based on the results of the evaluations, they could apply relevant mitigations and, through the IDOB, derive which subsets of employees could access and apply a new internal AI system and based on what restrictions (internal usage policies).

A comprehensive internal usage policy could include, at minimum, tiered access levels as follows: (i) full access for selected (security cleared) staff and AI systems (for example, security research teams); and (ii) limited or no access for other select individuals, teams, or AI systems (for example, product and sales). The rationale for this recommendation is that certain individuals (employees, teams, or AI systems) may have a justifiable need for access and usage, while others may not or only require specific features, which should be assessed ahead of each internal deployment of a new AI system. To provide further color, we envisage that complementary governance elements for internal usage policies could be composed of, for example, mandatory training before access, regular security assessments, and preparations to follow incident response procedures, where adequate. In this respect, the access and usage guidelines for internal AI system deployment could draw from established frameworks across other regulated safety-critical industries (\emph{see} \hyperref[chapter-5.-internal-deployment-in-other-safety-critical-industries]{\ul{Chapter 5}}; \cite{USNuclearRegulatoryCommission1956z,USDepartmentOfHealthAndHumanServices2005f}). Processes for guaranteeing that internal usage policies are respected and followed to the letter could be detailed in a third (and final) document called `oversight frameworks,' which we will describe next in \hyperref[c-oversight-framework]{\ul{§6.2.(c)}}.

\subsubsection{Oversight Framework}\label{c-oversight-framework}

Previously, we reflected on evaluations (\hyperref[a.i-tripwires-discovered-by-evaluations]{\ul{§6.2.(a.i)}}) tied to pre-defined tripwires in FSPs, technical access and usage restrictions and controls (\hyperref[a.ii-mitigations-technical-access-and-usage-restrictions-and-controls]{\ul{§6.2.(a.ii)}}), and internal usage policies (\hyperref[b-internal-usage-policies]{\ul{§6.2.(b)}}). In this Section, we focus on how the aforementioned should be tied into corresponding oversight and enforcement procedures. These procedures can be detailed in a third internal document, which we call `oversight framework.'

An \textbf{oversight framework is an overarching document} that provides a high-level framework specifying out two core processes relevant to oversight. First, it specifies the \textbf{interaction between the other two documents}---\textbf{FSPs and internal usage policies}---ensuring internal adherence and consistency with pre-defined tripwires. Second, it specifies the IDOB's \textbf{authority to take final decisions} on internal deployment and \textbf{the tools and options at its disposal to exercise oversight}. While oversight frameworks will need to be sketched out in more detail in future scholarly research, we briefly focus on specifications on the role of the oversight framework.

We anticipate that specifying out the manner in which the IDOB can exercise its oversight through the oversight framework would cover at least the following aspects: (i) the type of information it can access and request; (ii) the type of tools at its disposal to operationalize and oversee internal deployment; and (iii) the liberty to exercise judgment on targeted transparency to third parties (\hyperref[addressing-undetected-and-unconstrained-power-consolidation-targeted-transparency]{\ul{§6.3}}). We briefly present high-level baselines on these next.

First, we suggest that \textbf{the oversight framework describes the range of review processes the} \textbf{IDOB should have access to and the type of reviews it can request.} As part of this, the IDOB could: (i) conduct daily internal usage reviews and assess broad patterns; (ii) regularly audit internal access and usage permissions, especially after near-misses or receipt of novel information about AI system capabilities; (iii) solicit regular updates from the IDT on the evaluation of internal AI system capabilities and the discovery of novel internal applications, especially for AI systems that are deployed internally for a long time and updated iteratively. Further, human review processes could include: (iv) establishing clear internal reporting channels for concerns; (v) requiring additional external assessment; and (vi) implementing regular training and incident response exercises conducted prior to granting internal access and usage.

Second, we suggest that \textbf{the oversight framework describes the range of tools the} \textbf{IDOB should have access to, enabling its oversight.} In particular, it describes the technical oversight mechanisms that could monitor whether internal usage policies have been crossed and trigger responsive action. We envision, for example, the following mechanisms:\footnote{The mechanisms we describe adapt best practices from safety-critical industries while addressing AI-specific challenges.} (i) dashboards that could track usage patterns and API calls in accordance to pre-defined limitations, while anomaly detection could flag unusual access patterns or unauthorized commands, inspired by financial trading desk controls \parencite{Moffat2022h}; (ii) system logs that could analyze AI system interactions for policy compliance, with rate-limiting alerts and automatic lockouts for violations; or (iii) network monitoring that could identify unauthorized access attempts, supplemented by regular security scans. All mechanisms mentioned and those elaborated on in future research should be accompanied by rigorous `bookkeeping' to enable audit trails and any necessary subsequent review.\footnote{A review can be conducted by independent external experts or even independent resident inspectors, inspired by the nuclear sector \parencite{USNuclearRegulatoryCommission2024q}.}

Third, we suggest that \textbf{the oversight framework describes the IDOB's ability to provide targeted transparency toward third-parties} (\hyperref[addressing-undetected-and-unconstrained-power-consolidation-targeted-transparency]{\ul{§6.3}}). We provide more color on targeted transparency disclosure next, including on potential recipients and content.

\subsection{Addressing Undetected and Unconstrained Power Consolidation: Targeted Transparency}\label{addressing-undetected-and-unconstrained-power-consolidation-targeted-transparency}

As explained in Chapters \hyperref[chapter-2.-characterizing-internal-deployment]{\ul{2}} and \hyperref[chapter-3.-internal-deployment-scenarios-of-concern]{\ul{3}}, notwithstanding a lack of direct benefit to society from internal deployment, its threats are not limited to the AI company it is deployed in and can spill into society at large, including by posing a threat to national security. This Section presents our second `macro' recommendation, which centers around alleviating certain information asymmetry to enable preparedness and resilience at the company and societal levels (\hyperref[b-undetected-and-unconstrained-power-accumulation-through-an-internal-intelligence-explosion]{\ul{§3.2.(b)}}). Specifically, we recommend that the \textbf{oversight framework} (\hyperref[c-oversight-framework]{\ul{§6.2.(c)}}) \textbf{directs the IDOB to ensure that adequate information is shared with select stakeholders}.

The following Sections describe the recipients of IDOB's targeted transparency (\hyperref[a-targeted-transparency-recipients]{\ul{§6.3.(a)}}), the relevant economic and strategic incentives for AI companies and governments (\hyperref[a.i-incentives-public-private-partnerships-and-licenses]{\ul{§6.3.(a.i)}}), and the recommended content of information sharing (\hyperref[b-targeted-transparency-content]{\ul{§6.3.(b)}}; \parencite{METR2025h}. A final Section (\hyperref[c-disaster-resilience-plans-tied-to-incident-monitoring-and-safety-cases]{\ul{§6.3.(c)}}) discusses the opportunity of developing `disaster resilience plans' tied to incident monitoring and whistleblower channels to address threats that might still go undetected.

\subsubsection{Targeted Transparency: Recipients}\label{a-targeted-transparency-recipients}

In \hyperref[c-oversight-framework]{\ul{§6.2.(c)}}, we examined how---depending on a variety of factors, including risk profiles---the IDOB could be composed of an external oversight board or a team within a select government agency. We also presented how an oversight framework could fortify the IDOB's ability to engage in targeted transparency. In terms of audience, we envisage that the recipients of targeted transparency could be two main groups.

First, the IDOB could share select information with staff that has access to internal AI systems based on internal usage policies. As described in \hyperref[b-internal-usage-policies]{\ul{§6.2.(b)}}-\hyperref[c-oversight-framework]{\ul{(c)}} internal usage policies determine which personnel can access the AI system and what they can apply it for, while oversight frameworks describe how FSPs and internal usage policies will be enforced. Still, depending on their role within the company, staff with access to internal AI systems might not have a comprehensive overview of the AI systems' capabilities (including the dangerous ones) or how the company plans on applying and integrating the AI system into existing staff workstreams. For this reason, informative materials such as pre-deployment system cards (\hyperref[b-targeted-transparency-content]{\ul{§6.3.(b)}}) and internal usage policies (\hyperref[b-internal-usage-policies]{\ul{§6.2.(b)}} could be shared with staff to ensure that they are aware of the implications of `dogfooding' \parencite{TheNewYorkTimes2009f} and can make informed decisions.

Second, targeted transparency can be directed to a government agency. This could be an intermediate option in cases in which (i) the IDOB is an external oversight board (and not a select government agency) and where (ii) internal deployment has become a topic of national security interest (\hyperref[b-undetected-and-unconstrained-power-accumulation-through-an-internal-intelligence-explosion]{\ul{§3.2.(b)}}). In these cases, it could become critical to quickly include certain government agencies (or specialized dedicated teams therein) in the governance of internal deployment and associated access and usage decisions. In high-stakes scenarios, these agencies or specialized teams therein could be enabled to review targeted information and to vet and veto certain decisions. This could support national security resilience and preparedness, including from internal and external misuse and loss of control. To enable national security preparedness, targeted transparency toward a government agency should, at minimum, encompass the full range of information described next (\hyperref[b-targeted-transparency-content]{\ul{§6.3.(b)}}-\hyperref[c-disaster-resilience-plans-tied-to-incident-monitoring-and-safety-cases]{\ul{(c)}}), such as pre-deployment system cards, internal usage policies, the results of dangerous capability evaluations run under FSPs and relevant mitigations, oversight frameworks, and disaster resilience plans.

In addition to enabling societal and national security preparedness, we believe there are economic and strategic advantages that AI companies could benefit from should they choose to implement certain proposals and keep select government entities in the loop. Although not core to our report, we briefly reflect on these next.

\paragraph{Incentives: Public-Private Partnerships and Licenses}\label{a.i-incentives-public-private-partnerships-and-licenses}

In some circumstances, governments are at the forefront of innovation and are the gatekeepers of essential resources for AI development. Collaboration with governments on robust safety and security interventions could provide access to strategic assets, including: (i) substantial security enhancements, such as physical protection and securitization of data centers and companies, dedicated threat assessment teams, specialized training, and priority response during security incidents; (ii) potential access to stable, secure energy sources through government-controlled utilities or emergency power allocations during critical situations; (iii) potential access to specialized databases for security verification systems, subject to appropriate privacy regulations and controls; and (iv) other valuable government-controlled resources, such as specialized land use rights, regulatory exemptions or expedited permitting.

Indeed, some frontier AI companies have already suggested that the U.S. Government enter into ``voluntary partnerships'' with AI companies \parencite{Lehane2025c} with the effect of: (i) expanding domestic energy supply through 50 additional gigawatts of AI-dedicated power by 2027 \parencite{Anthropic2025t}; (ii) preserving AI models' ability to ``learn from copyrighted material'' \parencite{Lehane2025c}; (iii) enhancing security protocols \parencite{Anthropic2025t} and studying advanced security requirements to control highly agentic models, including to secure algorithmic secrets from foreign large-scale industrial espionage \parencite{CouncilOnForeignRelations2025e,Rollet2025n}; (iv) establishing expedite security clearances for AI professionals \parencite{Anthropic2025t}; and (v) fine-tuning models on classified datasets, such as geospatial intelligence or classified nuclear tasks to develop `custom models for national security' \parencite{Lehane2025c}.

From a government perspective, the eventual development of AGI (\emph{see} \hyperref[chapter-2.-characterizing-internal-deployment]{\ul{Chapter 2}}) may be deemed a project of national interest, and efforts that aim to do so securely may receive measurable protection and encouragement from government entities, including through military resources. This is likely of interest to both protect national security and encourage national AI company leadership. At a basic level, government entities would likely desire visibility into AI progress and, therefore, internally deployed AI systems---especially in situations where these cease to be eventually externally deployed (\emph{see} \hyperref[chapter-2.-characterizing-internal-deployment]{\ul{Chapter 2}})---to proficiently carry out a mutually beneficial partnership. For example, since government-led cyber and physical security infrastructures may need to be custom-tailored to the capabilities of an internally deployed AI system and its applications, it will likely be in the best interest of AI companies to grant targeted visibility to government entities.

In short, we propose that well-thought-through and executed governance of internal deployment can, under certain conditions, create a situation of mutual benefit for AI companies and governments alike. In particular, we see this fit into, for example, traditional frameworks of public-private partnership or, more nascent, proposals around AI licensing regimes \parencite{Blumenthal2023w}. AI companies could partner with government entities on internal deployment governance mechanisms such as evaluations and oversight and, in turn, benefit from privileged access to resources by governments. In the future, it could be valuable to assess whether AI companies that collaborate with a governmental agency via the International Network of AI Safety Institutes \parencite{InternationalNetworkOfAISafetyInstitutes2024u} could benefit from mutual agreements between member countries.

By contrast, we think there are good reasons to provide minimal or only high-level information about internally deployed AI systems to the general public---not least to avoid undue race dynamics. Still, given the potential spill of threats into broader society, the public may benefit from receiving some high-level outline of the kind of governance frameworks that are in place for internal deployment. For example, AI companies could inform the general public about safety and security practices preceding internal deployment, including FSP tripwires and commitments (\hyperref[a-frontier-safety-policies-tripwires-and-corresponding-mitigations]{\ul{§6.2.(a)}}), the existence and composition of an overseeing board (\hyperref[internal-governance-frameworks-and-processes]{\ul{§6.1}}), or the oversight procedures outlined in the oversight framework (\hyperref[c-oversight-framework]{\ul{§6.2.(c)}}).

\subsubsection{Targeted Transparency: Content}\label{b-targeted-transparency-content}

Next, we go into more detail pertaining the layers of defense that may compose a targeted transparency regime of internal deployment. Our recommendations take a variety of threat vectors into account, all of which center around a potential lack of preparedness in case of an AI-driven intelligence explosion (\emph{see} \hyperref[chapter-3.-internal-deployment-scenarios-of-concern]{\ul{Chapter 3}}).

First, we recommend that the IDOB provides targeted transparency to enable situational awareness and preparedness on the capability level of internally deployed highly capable AI systems. This can be attained by \textbf{preparing pre-internal-deployment model cards or system cards} (including through the IDT) and making them available at the time of internal deployment to the entities described in \hyperref[a-targeted-transparency-recipients]{\ul{§6.3.(a)}}. System cards and model cards can be considered an industry best practice (\emph{see}, for example, \cite{OpenAI2024k,Anthropic2025h,GeminiTeam2024o}) and describe an AI system's or model's capabilities, as well as the safety work carried out on it. The recommendation, therefore, builds and expands on existing transparency efforts that several AI companies already undertake for external deployment. For external deployment purposes, these system cards are usually published on the day of external deployment or shortly thereafter. We propose that it would be suitable to do the same for internal deployment and share pre-internal model or system cards ahead of internal deployment and to make them available to, at minimum, the staff allowed to access and use internal AI systems and select government entities.

Second, visibility could be provided into who within the organization can access and use internal AI systems and under what conditions and restrictions. In this case, targeted transparency can be achieved through \textbf{sharing internal usage policies.} Previously (\hyperref[b-internal-usage-policies]{\ul{§6.2.(b)}}), we explained how internal usage policies could establish guidelines that dictate how, when, and by whom internally deployed AI systems may be accessed and utilized within an AI company. Hence, from examining internal usage policies, a reader (i.e., staff allowed to access and use internal AI systems and potentially a select government agency) could understand who within an AI company can access its most capable AI systems. Access to internal usage policies can also support verification processes looking to assess whether the restrictions are being adhered to. Overall, we expect that insight into internal usage policies can help build awareness and, by extension, support an early warning mechanism capable of avoiding further downstream harm.

Third, in certain circumstances, for example, when there is a potential threat to national security, AI companies could choose to share the evaluations run in accordance with an AI companies' FSP alongside any information relevant to the enforcement of correlated mitigations. We envision that this could enable the reader, in this instance, individuals within a government, to gain insight into what type of dangerous capabilities an internally deployed AI system displayed, as well as what mitigations have been applied and by whom.

\subsubsection{Disaster Resilience Plans: Tied to Incident Monitoring and Safety Cases}\label{c-disaster-resilience-plans-tied-to-incident-monitoring-and-safety-cases}

While it would be desirable that all threats are adequately captured, mitigated, or contained through the aforementioned layers, it is not unlikely that certain threats will still go undetected. In instances where threats go undetected and cause an incident, \textbf{it will be invaluable to have a robust framework that enables the quick identification of an incident and the swift implementation of corresponding disaster resilience plans}. In other words, it is possible that some gaps in the identification and mitigation of threats through any of the aforementioned mechanisms in this Chapter could go undetected, and it is, consequently, prudent to ensure that \emph{should incidents happen}, they are identified and remedied quickly.

This could be attained with three layers, which we briefly elaborate on next: (i) \textbf{pre-internal-deployment safety cases}; (ii) the co-development of \textbf{disaster resilience plans with governments;} and (iii) the establishment of a functional \textbf{incident monitoring framework}.

In order to ensure that governments are well informed, AI companies could share a \textbf{pre-internal-deployment safety case} (AI safety case)\footnote{AI safety cases are structured, evidence-based rationales that an AI system deployed to a specific setting is unlikely to cause catastrophic outcomes \parencite{Clymer2024w,Buhl2024g,Goemans2024k,Korbak2025z,Balesni2024a}. They have recently drawn attention from scholars and the UK AI Security Institute \parencite{Irving2024z}, and there is an ongoing discussion as to how to tie AI safety cases meaningfully into existing FSPs \parencite{Pistillo2025w}.} with select government agencies under certain circumstances, for example, when capabilities look to threaten national security. The shared AI safety case could provide detail on pre-deployment evaluations and mitigations, internal usage policies, and oversight frameworks and explain how it is ensured that an internal AI system functions safely and securely, given its specific internal application scenario. These AI safety cases can also enable the development of a corresponding `\textbf{disaster resilience plan,'} drawn up through a collaboration of governments, AI companies, external experts, and AI Safety or AI Security Institutes and outlining what levers can and ought to be pulled in the unlikely scenario that an incident occurs. Disaster resilience plans can act as `emergency parachutes' in cases of critical failure or incidents by offering pre-defined pathways, powers, and choke points to limit diffusion, ensure containment, and to enable immediate steps to remedy an incident. This type of plan should be optimally held by a select government agency, which would also have the authority to conduct a causal analysis and trigger requests for relevant remedial actions. In short, we envision disaster resilience plans as detailed blueprints that frontload the effort of figuring out responsibility and actions early on to ensure minimal disruption in case something does go wrong.

In order to ensure that a critical failure or incident is identified in a timely manner and to trigger these disaster resilience plans, we recommend AI companies and governments set up viable incident monitoring frameworks; for example, these could be built on some of the monitoring infrastructures described under §6.2(a.ii) and §6.2.(c). We recommend that \textbf{incidents and safety-critical failures are defined in such a manner that they capture all cases that invalidate the AI safety case that corresponds to the AI system in question or weaken a claim made therein} \parencite{Ortega2025b}. In these cases, we recommend that the previously developed disaster resilience plans be triggered for immediate and efficient containment and remediation.\footnote{Once an incident is contained, this setup could also enable governments and AI companies to conduct a causal analysis of the pathway that led to the incident and implement any amendments or improvements to technical or other measures, based on this analysis. This is further detailed in \textcite{Ortega2025b}.}

Before concluding, we note that as AI system's capabilities grow it will be increasingly important to be able to meaningfully review whether implemented mechanisms such as those aforementioned function well, are respected and suffice. We envision that whistleblower channels and protections could aid in accomplishing this, though by no means should they be relied on as the only mechanism. Specifically, governments and AI companies could implement effective whistleblower channels, which could include a secure, anonymous reporting platform managed by an independent third party (with downstream reports to the IDOB and select government agencies, as necessary). As part of maintaining the efficacy of the process, this whistleblower system could be complemented with clear rules prohibiting any form of retaliation (including demotions, reassignments, harassment, or hostile work environments) and defining penalties for non-compliance.\footnote{Various efforts are underway to introduce whistleblower protection for AI companies' employees, including the proposed California Senate Bill 53 \parencite{CASB53}. We suggest that whistleblower protection (including non-retaliation policies) is also established for any external parties involved in evaluations, assessments or brought in by the IDOB to verify documentation, as long as these parties are acting in compliance with legal requirements and the rules of engagement, as defined in their agreement with AI companies.}

\section*{Conclusion }\label{conclusion}

We are at a \textbf{pivotal moment in AI progress}. Estimates from frontier AI companies’ CEOs and independent experts alike suggest that AGI could be reached within the next five years (by 2030). At the same time, there is mounting evidence indicating that frontier AI companies: (i) already deploy current AI systems internally, sometimes for a long time before their public release; and (ii) may have strong strategic and economic incentives to deploy their future most advanced AI systems exclusively within the company walls, keeping them there for the entirety of their service period. Internal AI systems could be applied to a variety of high-value uses, including automating the most important activity within a frontier AI company––R\&D on new, more capable AI systems. 

Despite its importance, there currently exists \textbf{no governance of internal deployment}. As a result, it is plausible that internally deployed AI systems operate with less oversight and safeguards than their external counterparts and, hence, are less well-understood. This could contribute to two major threats that this report focuses on: first, the loss of control through scheming internal AI systems, and, second, outsized power concentration that a small group of human conspirators could leverage to overthrow democratically elected governments. The absence of robust governance mechanisms for the internal deployment of AI does not reflect the state of the art in other safety-critical industries––such as chemistry, biology, nuclear, and aviation––where the handling, usage, and application of dangerous products are strictly governed ahead (or regardless) of their release on the market. Moreover, our legal analysis shows that, under certain interpretations, the language of some legal frameworks could be already intended to cover the application and usage of AI systems internal to frontier AI companies.

Building on these learnings, this report presents \textbf{a first defense-in-depth approach toward the governance of internal deployment} with a particular focus on targeting threats of loss of control, and undetected and unconstrained power concentration. First, we recommend implementing robust evaluation and control measures through internal company-led policies including applying tripwires and commitments to internal AI systems, establishing proportionate access boundaries, and creating clear oversight frameworks with dedicated teams for technical implementation and independent decision-making. Second, we recommend addressing information asymmetry between insiders and outsiders by mandating selective transparency with key internal and external stakeholders. This includes sharing pre-deployment system documentation, communicating evaluation results and safety mitigations, disclosing internal usage policies, and providing safety case documentation. Finally, for enhanced preparedness, we propose that companies collaborate with governments on disaster resilience plans activated by serious incident reporting or whistleblowing. 

\section*{Acknowledgments}

For thoughtful feedback and valuable discussions related to this work, we would like to thank: Gretchen Krueger, Anna Wang, Fabien Roger, Sören Mindermann, Stephen McAleer, Tomek Korbak, Joshua Clymer, Miles Brundage, Matthijs Maas, Steven Adler, Daniel Kokotajlo, Herbie Bradley, Tyler John, Rusheb Shah, Bronson Schoen, Alex Meinke, Stefan Heimersheim, Fin Moorhouse, and multiple reviewers who wish to stay anonymous. All errors remain our own.

\appendix
\section*{Appendix}


In this Appendix, we provide some complementary considerations to those of the main text.

Complementary to \hyperref[chapter-4.-existing-and-proposed-ai-governance-frameworks-and-their-relationship-to-internal-deployment]{\textbf{Chapter 4}} on `Existing and Proposed AI Governance Frameworks and Their Relationship to Internal Deployment,' we provide supplementary context on legal frameworks containing language that could be interpreted broadly to cover internal deployment. Specifically, we list a number of U.S. state bills defining `deployer' as an entity `using' an AI system and offer further clarification on our reading of Article 2(8) and Recital 25 of the AI Act.

Complementary to \hyperref[chapter-5.-internal-deployment-in-other-safety-critical-industries]{\textbf{Chapter 5}} on `Internal Deployment in Other Safety-Critical Industries,' we add learnings from two additional safety-critical sectors---i.e., nuclear R\&D and experimental aviation---that complement the analysis contained in Chapter 5.

Complementary to \hyperref[chapter-6.-defense-in-depth-recommendations-for-the-governance-of-internal-deployment]{\textbf{Chapter 6}} on `Defense in Depth: Recommendations for the Governance of Internal Deployment,' we provide an initial high-level step-by-step walk-through describing how the IDT and the IDOB could operate in tandem.

\subsection*{Chapter 4. Existing and Proposed AI Governance Frameworks and Their Relationship to Internal Deployment}\label{chapter-4.-existing-and-proposed-ai-governance-frameworks-and-their-relationship-to-internal-deployment-1}

In \hyperref[chapter-4.-existing-and-proposed-ai-governance-frameworks-and-their-relationship-to-internal-deployment]{Chapter 4}, we provided an overview of how existing and proposed governance frameworks could already interact with internal deployment. In this appendix, we provide additional context on: (i) proposed state legislation in the United States that contain analogous language on the definition of `deployer' as the enacted legislation described in \hyperref[chapter-4.-existing-and-proposed-ai-governance-frameworks-and-their-relationship-to-internal-deployment]{Chapter 4}; and (ii) our review of the AI Act, specifically by providing supplementary context as to why we believe that a broad interpretation, as advanced in \hyperref[chapter-4.-existing-and-proposed-ai-governance-frameworks-and-their-relationship-to-internal-deployment]{Chapter 4}, could, in theory, also be consistent with Article 2(8) and Recital 25 of the AI Act.

\subsection*{A. U.S. State Bills Defining `Deployer' as an Entity `Using' a System}\label{u.s.-state-bills-defining-deployer-as-an-entity-using-a-system}

In \hyperref[b-deployer-as-an-entity-that-uses-ai-for-any-purpose]{\ul{§4.1.(b)}}, we observed that enacted state legislation in the United States---such as the Colorado AI Act \parencite{Rodriguez2024p}, Colorado House Bill 24-1468 \parencite{Titone2024s}, and recently-vetoed Virginia House Bill 2094 \parencite{SenateCommitteeOnGeneralLawsAndTechnology2025q}---simply define `deploy' as the ``use'' of an AI system, and `deployer' as an entity that ``uses'' an AI system. This could allow an extensive interpretation that includes the \emph{internal} use of an AI system and, hence, internal deployment. We also mentioned how \textbf{analogous language appears in many other state bills}.

Here, we list some of these bills. Proposed state legislation that defines `deploy' as the ``use'' of an AI system and `deployer' as an entity that ``uses'' an AI system includes the following.

\begin{itemize}
\item
  Section 1 of Washington House Bill 1951 \parencite{Shavers2024i} defines `deployer' as ``a natural person, partnership, state or local government agency, or corporation that \textbf{uses} or modifies an automated decision tool to make a consequential decision.''
\item
  Section 5 of Illinois House Bill 5116 \parencite{Didech2024d} defines {`deployer' as ``a person, partnership, State or local government agency, or corporation that \textbf{uses} an automated decision tool to make a consequential decision.''}
\item
  Section 59.1-603 of Virginia House Bill 747 \parencite{HouseCommitteeOnCommunicationsTechnologyAndInnovation2024z} defines {`deployer'} as {``any person doing business in the Commonwealth that deploys or \textbf{uses} a high-risk artificial intelligence system to make a consequential decision.''}
\item
  Section 3 of Oklahoma House Bill 3835 \parencite{Alonso-Sandoval2024p} defines {`deployer'} as {``a person, partnership, state or local government agency, or corporation that \textbf{uses} or modifies an automated decision tool to make a consequential decision.''}
\item
  Section 1001 of Vermont House Bill 710 \parencite{Priestley2024w} defines {`deployer'} as {``any person who deploys or \textbf{uses} a high-risk artificial intelligence system to make a consequential decision.''}
\item
  Section 42-166-1 of Rhode Island House Bill 7521 \parencite{Baginski2024y} defines {`deployer'} as {``a person, partnership, state or local government agency, or corporation that \textbf{uses} an automated decision tool to make a consequential decision.''}
\item
  Section 6-60-3 of Rhode Island Senate Bill 2888 \parencite{Dipalma2024o} defines {`deployer'} as {``any entity that \textbf{uses} a CAIDS'' (i.e., a Consequential Artificial Intelligence Decision System) ``to make consequential decisions.''}
\end{itemize}

\subsection*{B. Clarification on Art. 2(8) and Recital 25 of the AI Act }\label{clarification-on-art.-28-and-recital-25-of-the-ai-act}

In \hyperref[different-terminology-captures-external-placing-on-the-market-and-internal-putting-into-service-or-putting-into-effect-deployment]{\ul{§4.2}}, we reached the conclusion that the AI Act could be theoretically interpreted to apply not only to AI systems and AI models that are publicly released and commercialized on the European Union's market but also to AI systems that are ``put into service'' in the European Union for the developer's ``own use.'' Here, we explain why we believe this interpretation is \textbf{also consistent with the text of Article 2(8)} \parencite{EuropeanParliamentAndCouncil2024o} \textbf{and Recital 25 of the AI Act} \parencite{EuropeanParliamentAndCouncil2024o}.

Article 2(8) states that the Act ``does not apply to any research, testing or development activity regarding AI systems or AI models \textbf{prior to} their being placed on the market or \textbf{put into service}'' (Article 2(8), \cite{EuropeanParliamentAndCouncil2024o}). In other words, Article 2(8) \textbf{marks the moment} after which AI systems and AI models become subject to the AI Act. That moment is the `putting into service' of an AI system or AI model---which, as mentioned, means ``the supply of an AI system'' ``for own use'' (Article 3, \cite{EuropeanParliamentAndCouncil2024o}). Article 2(8), therefore, in our reading, confirms that once AI systems or AI models are made available for access and/or use within the developing organization, these AI systems or AI models could be subject to the AI Act. Therefore, for example, under Article 2(8)\hl{, if an AI system is deployed internally to create value for the company by contributing to the development of successor frontier AI systems, such an AI system could be subject to the AI Act.} So, in summary, Article 2(8) clarifies that the research and development activities \textbf{preceding} internal deployment (the `putting into service') are not regulated. In line with Article 2(8), Recital 25 further affirms that ``it is necessary to ensure that this Regulation does not otherwise affect scientific research and development activity on AI systems or models \textbf{prior to being} \ldots{} \textbf{put into service}'' (Recital 25, \cite{EuropeanParliamentAndCouncil2024o}). Again, this confirms that the tripwire for the application of the AI Act is the ``supply of an AI system'' (Article 3, \cite{EuropeanParliamentAndCouncil2024o})---i.e., its availability for access and usage inside an AI company.

Concerning Recital 25 of the AI Act, we identified a potential element of uncertainty that is worth clarifying. Recital 25 states that ``AI systems and models specifically developed and put into service for the sole purpose of scientific research and development'' are excluded from the scope of the Act (Recital 25, \cite{EuropeanParliamentAndCouncil2024o}). This statement could easily be misinterpreted as a narrow exception to the rule set forth under Article 2(8)---according to which AI systems and AI models fall in the AI Act's remit as soon as they are ``put into service'' (Article 2, \cite{EuropeanParliamentAndCouncil2024o}). Specifically, the text of Recital 25 could be misinterpreted to exclude AI systems and models that have already been put into service from the scope of the AI Act if they have the ``sole purpose of scientific research and development.'' One clarification could be helpful in this respect. This exception would be included in Recital 25 and it would derogate Article 2(1) and (8)---a binding provision (Article 2, \cite{EuropeanParliamentAndCouncil2024o}). This would be in contrast with European Union law. Indeed, it is a well-established principle that recitals ``shall not contain normative provisions'' \parencite{EuropeanCommission2015m}. According to the longstanding jurisprudence of the Court of Justice of the European Union, recitals have ``no binding legal force and cannot be relied on as a ground for derogating from the actual provisions of the act in question'' (\emph{see} Court of Justice of the European Union,  
\cite{C-423/23}, \cite{C-664/23}, \cite{T-709/21}, \cite{C-302/19}, 
\cite{C‑418/18}, \cite{T‑704/14}, \cite{C‑345/13}, \cite{C-136/04}, \cite{C-162/97}). Interpreting Recital 25 as an exception to Article 2(8) would do exactly that---it would have the effect of derogating from an actual provision of the AI Act (Article 2, \cite{EuropeanParliamentAndCouncil2024o}), which includes within the scope of the AI Act all AI systems and models ``put into service'' within the European Union.

\section*{Chapter 5. Internal Deployment in Other Safety-Critical Industries }\label{chapter-5.-internal-deployment-in-other-safety-critical-industries-1}

In \hyperref[chapter-5.-internal-deployment-in-other-safety-critical-industries]{Chapter 5}, we examined case studies from other safety-critical industries, such as biological agents and toxins, new chemical substances, and novel pesticides. In this Appendix, we return to two case studies (i.e., experimental certificate of airworthiness for testing a new aircraft and construction permit and operating license for `testing' nuclear reactors). We did not discuss these case studies in the main body of the report because, while being safety-critical, aviation and nuclear power may not share the same characteristics of high impact, wide reach, comparatively high velocity, and difficulty to remediate that are common between AI, biological agents and toxins, new chemical substances, and novel pesticides. For instance, threats from `testing' nuclear reactors and aviation are usually more localized than a toxic chemical component infiltrating the air and the water grid.

Nonetheless, both case studies confirm \hyperref[chapter-5.-internal-deployment-in-other-safety-critical-industries]{Chapter 5}'s conclusions that:

\begin{enumerate}
\def\labelenumi{(\arabic{enumi})}
\item
  Where there is access to products before public release and commercialization, government actors have regulatory oversight over several activities related to a new product before its public release or commercialization. For example, a special airworthiness certificate for experimental use is necessary from the U.S. Federal Aviation Administration (FAA) to undertake a flight test of a new aircraft \parencite{USDepartmentOfTransportation1964p}.
\item
  This regulatory oversight usually consists of permits or licenses that government actors issue after an applicant demonstrates that the activities that it intends to undertake with the new product are safe. For example, a safety analysis report and a physical security plan must be filed with the U.S. Nuclear Regulatory Commission (NRC) before building and operating a `test' nuclear reactor (\cite{USNuclearRegulatoryCommission1956c}, paragraphs (b) \& (c)).
\item
  Once a permit or license is issued, the applicant's possible activities concerning a new product before public release and commercialization are restricted by the permit. For example, a test flight can occur only within the designated areas communicated to the FAA \parencite{USDepartmentOfTransportation1964i}
\end{enumerate}

\subsection*{A. New Aircraft: Experimental Certificate of Airworthiness for Experimental Use}\label{new-aircraft-experimental-certificate-of-airworthiness-for-experimental-use}

In our larger research review of safety-critical industries, we also considered how experimental use of a new aircraft is regulated in the United States. In short, we found that an experimental certificate of `airworthiness' from the FAA is required for flight testing of a new aircraft before marketing. This certificate also restricts the possible uses of such aircraft.

`Airworthiness' is an FAA-issued document that certifies that an aircraft, aircraft engine, propeller, or article ``conforms to its approved design and is in a condition for safe operation'' \parencite{USDepartmentOfTransportationOthery,USDepartmentOfTransportation1964w}. No aircraft can be operated without being deemed `airworthy' \parencite{USDepartmentOfTransportation1989r}.

To perform a flight test, in which a new kind of aircraft is flown in the air as part of the research and development process, manufacturers must apply for a \textbf{special airworthiness certificate for experimental use} \parencite{USDepartmentOfTransportation1964p}. Such experimental certificates are issued for flight test R\&D activities such as testing new aircraft design concepts, new aircraft equipment, new aircraft installations, new aircraft operating techniques, or new uses for aircraft \parencite{USDepartmentOfTransportation1964j}. Similar to the industries described above, to obtain this certificate, an applicant must provide the relevant authority (the FAA) with safety-relevant information---including, upon FAA inspection, ``any pertinent information found necessary by the FAA to \textbf{safeguard the general public}'' \parencite{USDepartmentOfTransportation1964i}. Applicants must also provide sufficient information on the estimated time or number of flights required for the experiment as well as the areas over which the experiment will be conducted \parencite{USDepartmentOfTransportation1964i}. In other words, before allowing a pre-marketing experimental flight to take place, the FAA needs to be persuaded that the experiment does not endanger the general public.

The special airworthiness certificate for experimental use also restricts the potential use of the new aircraft. The certificate can be used only for the purpose for which it was issued \parencite{USDepartmentOfTransportation1989s}---which also means that the aircraft cannot be operated outside the designated area and cannot be operated over a densely populated area \parencite{USDepartmentOfTransportation1989s}.

\subsection*{B. `Testing' Nuclear Reactors: Construction Permit and Operating License}\label{testing-nuclear-reactors-construction-permit-and-operating-license}

In our larger research review, we also enquired how R\&D nuclear reactors (so-called `testing' reactors)\footnote{A testing reactor is a licensed non-power nuclear reactor that is used for researching novel reactor designs \parencite{USNuclearRegulatoryCommission1956j}. This research includes ``theoretical analysis, exploration, or experimentation,'' and ``extension of investigative findings and theories of a scientific or technical nature into practical application for experimental and demonstration purposes, including the experimental production and testing of models, devices, equipment, materials, and processes'' \parencite{USNuclearRegulatoryCommission1956j}. Nuclear regulation distinguishes testing reactors from power reactors used for industrial or commercial purposes \parencite{USNuclearRegulatoryCommission1956e}, such as those selling electricity directly to the grid. Specifically, testing reactors go through a different licensing path than industrial and commercial reactors \parencite{USNuclearRegulatoryCommission1956t}.} are regulated in the United States. In summary, we found that the NRC has oversight over the regulation and licensing of testing reactors. Specifically, no nuclear power plant can be built and operated in the United States without a construction permit and an operating license from the NRC---and testing reactors are no exception. The path to receiving such permits entails significant safety assessments by license applicants and the NRC. Industrial and commercial reactors must obtain a Class 103 license \parencite{USNuclearRegulatoryCommission1956e}, whereas testing reactors and other non-power reactors must obtain a Class 104 license \parencite{USNuclearRegulatoryCommission1956n}.\footnote{If a facility is used for both R\&D and commercial purposes, \textcite{USNuclearRegulatoryCommission1956e} that facility requires a Class 103 license if more than 50\% of the annual cost of owning and operating the facility is devoted to `commercial' purposes other than research and development.} Nonetheless, testing reactors are regulated in largely the same way as industrial and commercial power reactors. Both types of reactor require a \textbf{construction permit} before building can commence. To obtain this permit, applicants must file a \textbf{preliminary safety analysis report} with the NRC, in which, among other things, they assess ``the \textbf{risk to public health and safety} resulting from the operation of the facility'' and describe how the minimum principal design criteria listed in \parencite{USNuclearRegulatoryCommission1956x} are met \parencite{USNuclearRegulatoryCommission1956c}. In short, applicants must show to the NRC why the construction is safe based on some mandated criteria.

Similarly, both types of reactor require an \textbf{operating license}.\footnote{Alternatively, applicants can seek a combined construction permit and operating license \parencite{USNuclearRegulatoryCommission1956r}. If requirements are met, the NRC may first issue a construction permit and later convert it into an operating license \parencite{USNuclearRegulatoryCommission1956k,USNuclearRegulatoryCommission1956r}.} To obtain this license, applicants must submit, among other things, a final \textbf{safety analysis report} containing ``a safety analysis of the structures, systems, and components and of the facility as a whole'' (\cite{USNuclearRegulatoryCommission1956c}, paragraph (b)) as well as a \textbf{physical security plan} (\cite{USNuclearRegulatoryCommission1956c}, paragraph (c)). The NRC assesses whether the construction and operation of the facility may endanger the health and safety of the public or the common defense and security \parencite{USNuclearRegulatoryCommission1956u}.

Yet again, both types of reactor are subject to similar obligations once the reactor is up and running. In both cases, even after obtaining an operating license, license-holders must report incidents to the NRC (\cite{USNuclearRegulatoryCommission1956o}, paragraph (a)(1)), ensure that certain workers at the power plant undergo written and operational safety tests \parencite{USNuclearRegulatoryCommission1987n}, and undergo inspections of records, premises, and activities \parencite{USNuclearRegulatoryCommission1956w}.

In summary, this means that the NRC has oversight on reactors that are not `commercialized' and whose principal purpose is doing R\&D and testing, and that these types of reactors are still subject to safety regulations that are comparable with industrial and commercial reactors.

\section*{Chapter 6. Defense in Depth: Recommendations for the Governance of Internal Deployment}\label{chapter-6.-defense-in-depth-recommendations-for-the-governance-of-internal-deployment-1}

In \hyperref[chapter-6.-defense-in-depth-recommendations-for-the-governance-of-internal-deployment]{Chapter 6} we detailed three `stackable' layers of defense against the threat of loss of control via misaligned AI applied to the AI R\&D pipeline.

Here we complement that recommendation by providing a bird-eye view on how we imagine the two internal deployment oversight bodies (the IDOB and the IDT) could operationalize our recommended layers of defense.

\subsection*{Step-by-Step: Cooperation between the IDT and the IDOB}\label{step-by-step-cooperation-between-the-idt-and-the-idob}

The following are some illustrative `chronological' steps that AI companies, the IDT and the IDOB could take in relation to internal deployment. These steps are also represented in Figure G below. All of this is grounded in: (i) an FSP that includes internal deployment and defines which tripwires and relevant commitments apply to internal AI systems; and (ii) an oversight framework that defines the IDT and the IDOB's roles and responsibilities and the relevant procedures.

\textbf{First, the IDT rigorously assesses a new internal system's risk profile through evaluations and other adequate assessments, including by bringing in third-party experts}. Subsequent to evaluation results, the IDT applies mitigations based on FSP pre-commitments \hyperref[a.i-tripwires-discovered-by-evaluations]{\ul{(§6.2.(a.i-a.ii)}}).

\textbf{Second, the IDT summarizes the assessment results and implemented mitigations for the IDOB.} Governance experts within the IDT prepare two documents for the IDOB: (i) an internal report on IDT's FSP evaluations and mitigations (\hyperref[a.i-tripwires-discovered-by-evaluations]{\ul{§6.2.(a.i-a.ii)}}); and (ii) a pre-internal-deployment system or model card (\hyperref[b-targeted-transparency-content]{\ul{§6.3.(b)}}). The IDOB is informed in detail by the IDT and other internal teams about any evaluations and other tests run and their results to gain an informed view on capabilities and risk profiles, as well as on the applicability and suitability of proposed restrictions and mitigations. In all cases, the IDOB should be able to accurately assess the risk profile of the AI system. Further to the receipt and review of this information, the IDOB could also request more detailed or additional sets of evaluations or pieces of information to gather a more holistic picture if needed to fulfill their duty.

\textbf{Third, the IDOB takes an independent decision on whether the given AI system can be internally deployed and, if so, defines the conditions and boundaries}. The IDOB: (i) reviews IDT's reports and pre-internal-deployment system cards; (ii) prepares internal usage policies based on IDT's evaluations and mitigations (\hyperref[b-internal-usage-policies]{\ul{§6.2.(b)}}); and (iii) within the boundaries set forth by the internal usage policies, \emph{greenlights} internal deployment for vetted staff and AI systems.

\textbf{Fourth, the IDOB informs the IDT of its decision and, only then does the IDT make the AI system accessible for internal use}. After internal deployment is greenlit by the IDOB's, an AI system could be accessed and used internally, within the limits set forth by the IDOB's internal usage policies. The IDT is in charge of making sure that the IDOB's decision is continuously enforced from a technical perspective. The IDOB oversees IDT's work.

\textbf{Fifth, the IDOB engages in `selected transparency' with internal staff and government agencies by being empowered to share relevant information with key parties.} As we explain in \hyperref[b-targeted-transparency-content]{\ul{§6.3.(b)}}, the IDOB could also be the `motor' of targeted transparency to company-internal parties and governments related to the choices made around internal deployment. In this sense, the IDOB could play a dual role, exercising internal oversight and enabling societal preparedness.

\textbf{Sixth, the IDOB continuously oversees internal deployment with the IDT's technical support}. The IDT routinely evaluates AI systems and informs the IDOB. The IDOB continuously oversees, with the IDT's technical support, that internal access and usage by staff is compliant with the boundaries set forth in the FSPs and the internal usage policies (\hyperref[b-internal-usage-policies]{\ul{§6.2.(b)}}).

\begin{figure} [h!]
    \centering
    \includegraphics[width=\linewidth]{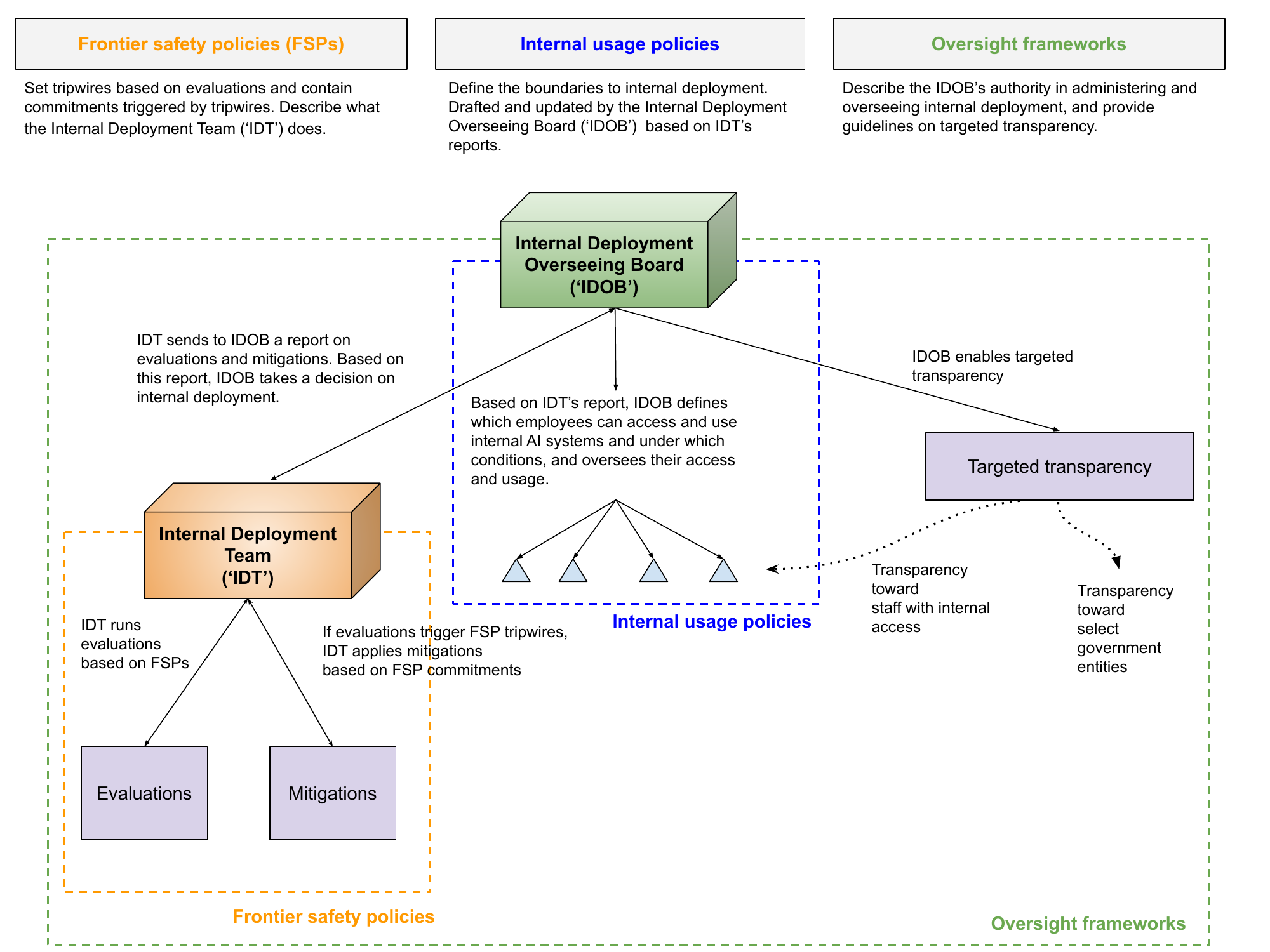}
    \caption{A prototype for the governance of internal deployment.}
    \label{fig:G}
\end{figure}

\clearpage

\printshorthands
\begin{CJK*}{UTF8}{gbsn}
\printbibliography
\end{CJK*}

\end{document}